\documentclass[]{article}
\usepackage{natbib}
\usepackage{emulateapj}
\usepackage{graphicx}
\usepackage{multirow}
\usepackage{color}

\def\msun{{\rm ~M}_{\odot}}
\def\rsun{{\rm ~R}_{\odot}}
\def\myr{{\rm ~Myr }}

\def\pmyr{Myr$^{-1}$ }
\def\mpy{{\rm ~M}_{\odot} {\rm ~yr}^{-1}}
\def\zsun{{\rm ~Z}_{\odot}}

\usepackage{color}

\begin{document}

\title{Double compact objects I: the significance of the common envelope on
merger rates}

 \author{Michal Dominik\altaffilmark{1}, 
         Krzysztof Belczynski\altaffilmark{1,2}, 
         Christopher Fryer\altaffilmark{3}, 
         Daniel E. Holz\altaffilmark{4,5}, 
         Emanuele Berti\altaffilmark{6,7}, 
         Tomasz Bulik\altaffilmark{1}, 
         Ilya Mandel\altaffilmark{8}, 
         Richard O'Shaughnessy\altaffilmark{9}, 
 }

 \affil{
     $^{1}$ Astronomical Observatory, University of Warsaw, Al.
            Ujazdowskie 4, 00-478 Warsaw, Poland  \\
     $^{2}$ Center for Gravitational Wave Astronomy, University of Texas at
            Brownsville, Brownsville, TX 78520\\
     $^{3}$ CCS-2, MSD409, Los Alamos National Laboratory, Los Alamos, NM 87545
            \\
     $^{4}$ Enrico Fermi Institute, Department of Physics, and Kavli Institute
for Cosmological Physics\\University of Chicago, Chicago, IL 60637 \\
     $^{5}$ Theoretical Division, Los Alamos National Laboratory, Los Alamos, NM 87545 \\
     $^{6}$ Department of Physics and Astronomy, The University of Mississippi, 
	    University, MS 38677, USA \\
     $^{7}$ California Institute of Technology, Pasadena, CA 91109, USA \\
     $^{8}$ School of Physics and Astronomy, University of Birmingham, Edgbaston, Birmingham, B15 2TT, UK  \\
     $^{9}$ Center for Gravitation and Cosmology, University of Wisconsin-Milwaukee,Milwaukee, WI 53211, USA
 }

\begin{abstract}

The last decade of observational and theoretical developments
in stellar and binary evolution provides an opportunity to incorporate 
major improvements to the predictions from populations synthesis models. 
We compute the Galactic merger rates for NS-NS,                      
BH-NS, and BH-BH mergers with the StarTrack code.      
The most important revisions include: updated wind mass loss rates 
(allowing for stellar mass
black holes up to $80 \msun$), a realistic treatment of the common
envelope phase
(a process that can affect merger rates by 2--3 orders of magnitude), and a 
qualitatively
new neutron star/black hole mass distribution (consistent with the observed 
``mass gap'').
Our findings include: (i) The binding energy                                 
of the envelope plays a pivotal role in determining whether a binary merges
within a Hubble time.
(ii) Our description of natal kicks from supernovae plays an            
important role, especially for the formation of BH-BH systems. 
(iii) The      
masses of BH-BH systems can be substantially increased in the case of low
metallicities or weak winds. (iv) Certain combinations of parameters      
underpredict the Galactic NS-NS merger rate, and can be ruled out. {\em (v)}
Models incorporating delayed supernovae do not agree with the observed NS/BH
``mass gap'', in accordance with our previous work. 
This is the first in a series of three papers.
The second paper will study the merger rates of double compact
objects as a function of redshift, star formation rate, and             
metallicity. In
the third paper we will present the detection rates for gravitational       
wave observatories, using up-to-date signal waveforms and sensitivity
curves.

\end{abstract}


\section{Introduction}

We investigate the evolution of binary systems that leads to the formation of close
double compact objects (DCOs) merging within a Hubble time: double neutron star
(NS-NS), black hole--neutron star (BH-NS), and double black hole (BH-BH)
systems. Only isolated evolution (i.e., in field populations) is
considered.
Several different groups have provided similar estimates and studies in the
past decade \citep{nele2001,voss,dewi,nutzman,pfahl,danny}. The exploration of the relevant physical
processes and rate estimates were 
obtained with the {\tt StarTrack} population synthesis code \citep{comprehensive,
B1,B3,rarity,startrack,onthemax,nasza,B10,rich1,rich2,rich3}.  In this paper we
revisit these rate estimates, incorporating recent improvements to
the input physics within the {\tt StarTrack} code. Similar rate estimates
performed for dense populations in which dynamical interactions between
stars are important (i.e., globular clusters) have been presented elsewhere
\citep{gultekin,oleary,grin2006,sadowski,ivan,downing,millerlau}. 

To describe the formation of a double compact object one needs to 
know the history of both progenitors, as well as the interactions between them. 
However, the uncertainties associated with a number of evolutionary processes inhibit 
an accurate description. Among the most important unknowns are the details of the common 
envelope (CE) phase, the unknown maximum mass of the NS, the physics
of the supernova explosions that form compact objects, and the wind mass loss rates of their 
progenitors. In this paper we approach the problem by calculating
and investigating a suite of population synthesis models. 
The uncertainties are assessed by altering the parameters and input physics within 
defined limits. This allows us to ``bracket" our ignorance of the
physical processes that are crucial for understanding double compact object formation.

In this paper we investigate the physics associated with the CE
phase, focusing on 
the binding energy parameter, $\lambda$, and the structure of the donor star.  
We introduce a new treatment of the NS/BH formation processes
in core collapse supernovae, and assess the role of natal kicks, the maximum
NS mass, and wind mass loss on the rates. To show the basic dependence of 
our results on metallicity, we consider two fiducial stellar populations: $Z=\zsun$ and $Z=0.1 \zsun$. 
Our main results are presented in Fig.~\ref{02rates} and \ref{002rates}.
A broader metallicity study will be presented in paper II of this series, with
inclusion of the cosmic star formation history, and a calculation of the dependence
of rates on redshift. Paper III will examine the detection prognosis with 
advanced gravitational telescopes (e.g. Advanced LIGO/Virgo, Einstein 
Telescope), incorporating the latest gravitational waveforms and sensitivity curves 
for the various instruments. 

In order to make the results of our research easily accessible to the
community, we have provided an on-line {\tt Synthetic Universe} database,
available at {\tt www.syntheticuniverse.org}.
At present this contains detailed information on the population synthesis models
in this paper. In addition, a number of physical
parameters and their distributions that are not discussed here in detail
(e.g., initial binary parameters, component masses, mass ratios, evolutionary 
tracks) are also provided, for those who might want to include them in their own
research. We intend this database to act as an 
extension to our three papers, functioning as a repository of the full spectrum of
our double compact object populations. The ``Synthetic Universe" site will be 
updated in real time as new physical models are calculated, and will respond to 
requests from the scientific community. Eventually, the {\tt Synthetic Universe} 
will include various objects in addition to double compact object binaries, for
example stellar populations with white dwarfs 
(e.g., supernovae Ia progenitors; low frequency gravitational wave
sources; cataclysmic variables) and various types of X-ray binaries 
(e.g., persistent and transient sources; high- and low-mass X-ray binaries 
in Galactic and extra-galactic environments). 

In the next section we describe recent physical developments addressing the CE events, the wind mass loss rates from massive stars, supernovae explosions and 
compact object formation, incorporated into the {\tt StarTrack} code.
Section \ref{modeling}  
contains a general description of the most important physical parameters used in our
population synthesis, while in Section \ref{results} we present the models used
in this study, and 
our results. We summarize and conclude in Section \ref{dis}.

\section{{\tt StarTrack} improvements} \label{updates}
\subsection{The {\tt StarTrack} code}

We have used the population synthesis code {\tt StarTrack} to calculate 
the numbers and properties of DCOs. The full description of the code 
can be found in \cite{comprehensive,startrack}. The code utilizes a set of 
stellar models (\cite{hurley}; slightly modified from its original version) 
that allow for the evolution of stars at different metallicities.
The model for compact object formation adopted in the code has been significantly
revised; changes are described in Section \ref{cof}. During core collapse, 
fallback and direct BH formation is now accounted for \citep{bhmassdis} 
and the newly born objects receive natal kicks \citep{hobbs}. The formation of low mass
NSs through electron capture supernovae is also accounted for (e.g., \cite{pods}).
Binary interactions are treated in detail, and the various physical processes
have been calibrated
using either results of detailed evolutionary calculations (e.g., \cite{wellstein2} 
for mass transfer sequences), or specific sets of observations (e.g., \cite{levine}  
for tidal interactions). 

In the following sections we describe the major updates to the code that are 
relevant for the formation of BHs and NSs. These crucial 
improvements are based on advances in stellar evolution physics in the 
last decade. Included are revised rates for wind mass loss for massive 
stars that now allow for the formation of BHs with masses observed in the 
Milky Way and other galaxies (see Section \ref{winds}).
Another improvement relates to the CE coefficient,
$\lambda$, describing the binding energy of the envelope. It is now dependent
upon the parameters (mass, radius, etc.) and the evolutionary stage of the donor.
This is a major improvement when compared to most previous studies, where 
$\lambda$ was considered constant throughout the evolution of the donor 
(Section \ref{ce}). The third major update regards the core collapse/supernova 
explosions, and the resulting model for compact object formation. We introduce two 
convection-enhanced, neutrino driven supernova engines (based on 
Rayleigh-Taylor and Standing Accretion Shock instabilities) into 
our population synthesis code. This addition allows us to account for the dearth
of Galactic compact objects within the
mass range $2\mbox{--}5 \msun$ (the so called ``mass gap'' (Section \ref{cof})).

\subsection{Stellar winds} \label{winds}

The StarTrack code was recently updated with new wind mass loss rates (described 
in detail in \cite{onthemax}). One of the major
changes was the implementation of mass loss for O/B stars based on \cite{vink},
followed by new winds for Wolf-Rayet stars \citep{hamann,vink2005}. Furthermore, the Luminous Blue
Variable winds (based on \cite{lbv}) were calibrated so as to allow for the
formation of BHs with masses up to $15 \msun$ in a solar
metallicity environment. This was done in order to account for the observed
masses of known BHs in the Milky Way (e.g. \cite{oroszx1}). The most
important consequence of these updates is the
possibility of the formation of BHs with masses up to $30 \msun$ 
in sub-solar metallicity environments ($Z=0.3 Z_\odot$), which corresponds to
the observed mass of the most massive known stellar-mass BH in the IC10 X-1 
binary \citep{prestwich2007,silver2008}. 
These revisions also result in the formation of BHs with masses up to $80 \msun$ in a 
low metallicity ($Z=0.01 Z_\odot$) environment. 

\subsection{Common envelope updates} \label{ce}

\begin{figure}
\includegraphics[width=1.0\columnwidth,trim = 0 10.5cm 0 0,clip]{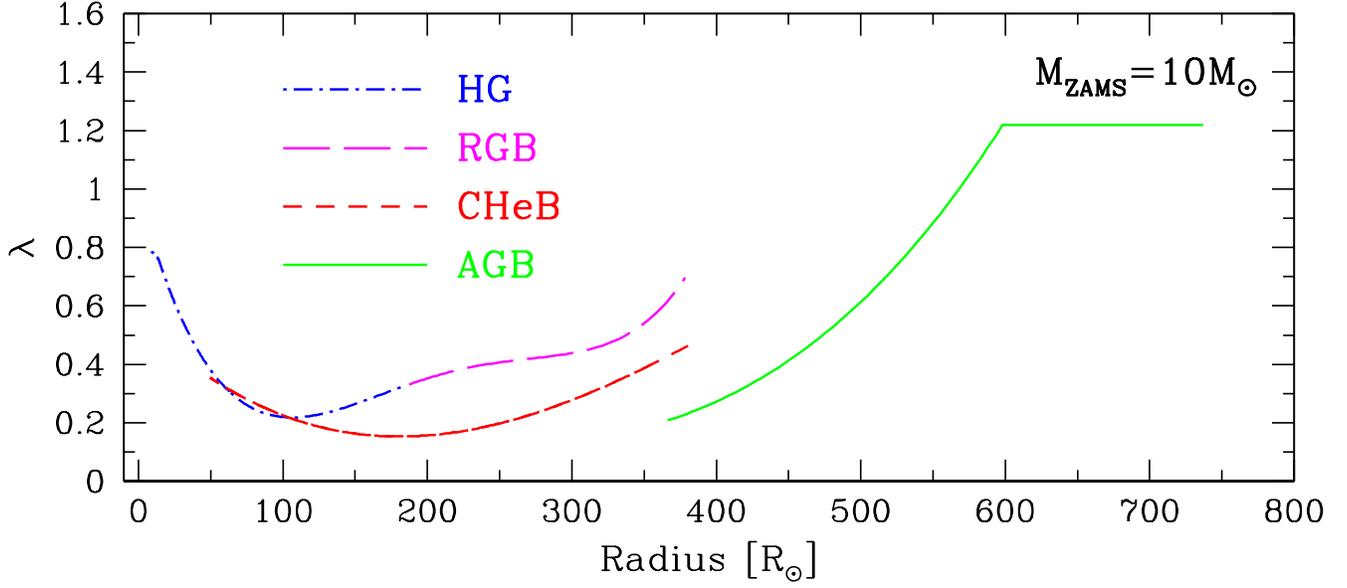}
\caption{
The average binding energy parameter, $\lambda$ (see Section \ref{newlam}), as a 
function of stellar radius, presented similarly as in~\cite{chlam} (\textit{Nanjing} $\lambda$), 
for a typical NS progenitor in a NS-NS system ($M_{\rm ZAMS}=10 \msun$) at$\zsun$. 
HG stands for Hertzsprung Gap, RGB for Red Giant Branch, CHeB for Core Helium Burning, 
and AGB for Asymptotic Giant Branch. The general behaviour of $\lambda$ is described in 
Section \ref{ce}.  
}
\label{lam10}
\end{figure}

\begin{figure}
\includegraphics[width=1.0\columnwidth,trim = 0 10.5cm 0 0,clip]{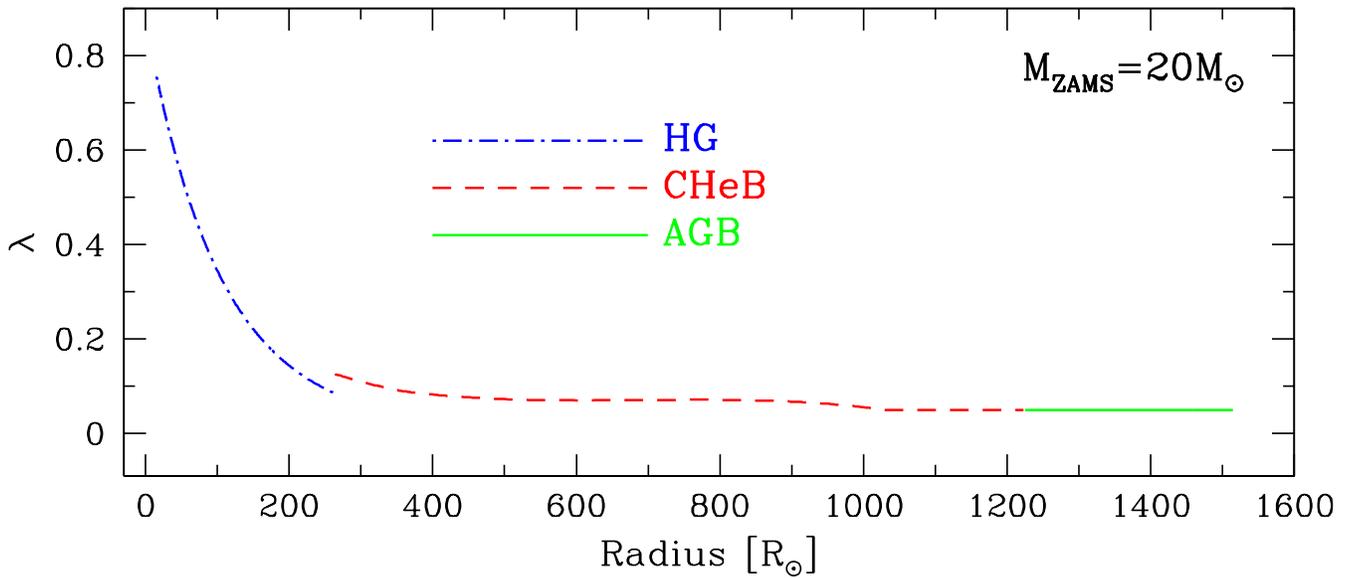}
\caption{
The same as Fig.~\ref{lam10} but for a BH-NS system ($M_{\rm ZAMS}=20 \msun$).
}
\label{lam20}
\end{figure}

\begin{figure}
\includegraphics[width=1.0\columnwidth,trim = 0 10.5cm 0 0,clip]{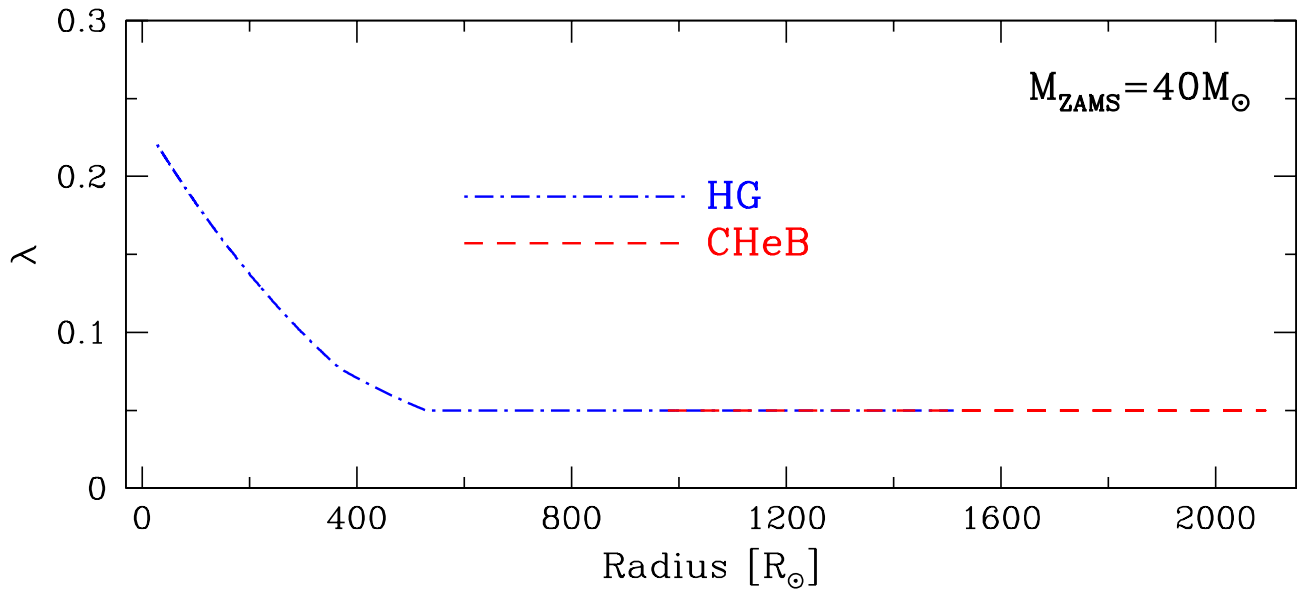}
\caption{\textit{Nanjing} $\lambda$ of a typical progenitor for the secondary component 
in a BH-BH system  ($M_{\rm ZAMS}=40 \msun$) at solar metallicity.
}
\label{lam40}
\end{figure}

\begin{figure}
\includegraphics[width=1.0\columnwidth,trim = 0 10.5cm 0 0,clip]{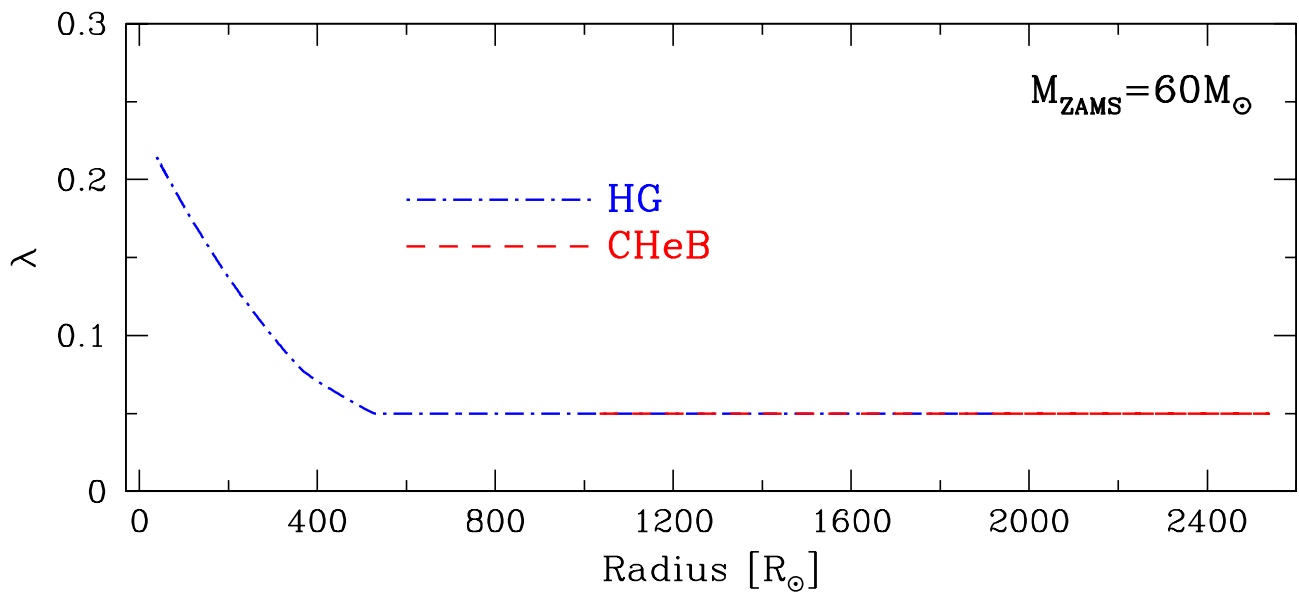}
\caption{Same as Fig.~\ref{lam40} but for the progenitor of the primary component.}
\label{lam60}
\end{figure}

We begin with an expression
describing the energy balance of the CE \citep{webbink84}:
\begin{equation} \label{ebaleq}
\alpha_{\rm CE} \left( \frac{GM_{\rm don,f}M_{\rm acc}}{2A_{\rm f}}-\frac{GM_{\rm don,i}M_{\rm acc}}{2A_{\rm i}} \right)=
\frac{GM_{\rm don,i}M_{\rm don,env}}{\lambda R_{\rm don,lob}}.
\end{equation}
Here, $R_{\rm don,lob}$ is the Roche lobe radius of the donor at the onset of
Roche lobe overflow (RLOF),
$A$ is the binary separation, $M_{\rm don,env}$ is the mass of the ejected envelope, and
$M_{\rm don}$ and $M_{\rm acc}$ are the masses of the donor and accretor, respectively. Indices
$i$ and $f$ stand for initial and final parameters, respectively. The
parameter $\alpha_{\rm CE}$ describes the efficiency of the transfer of orbital
energy into unbinding the envelope. If there is sufficient orbital energy to
unbind the envelope the binary may survive the CE phase,
otherwise the component merger and the formation of a peculiar single
object is assumed (an example: if the accretor is a compact object then in the outcome
of the CE a star harbouring a NS/BH may form). 
The value of $\alpha_{CE}$ has been set to $1$ throughout this
study. The range of values for the product $\alpha_{\rm CE}\lambda$ may therefore
be covered by $\lambda$ alone.
The parameter $\lambda$ describes the binding energy of the envelope,
and is defined as:
\begin{equation} \label{ebindeq}
E_{bind}=-\frac{GM_{\rm don}M_{\rm don,env}}{\lambda R},
\end{equation}
where  $R$ is the radius of the donor, which becomes equal to the Roche lobe radius of the donor
at the onset of the CE. 
The most significant changes of binding energy are usually associated with radial
expansion or contraction of a star. Additionally, the mass loss/gain may alter the binding 
energy. Generally, as the envelope of the star expands it becomes less dense and so its 
binding energy decreases, while a contraction of the envelope leads to an
increase in the binding energy. 
The parameter $\lambda$ is a function of binding energy, radius, and mass (Eq.~\ref{ebindeq}). 
For example,  $\lambda$ will decrease if the radius expands faster than the binding 
energy decreases , which is the case for stars on the
Hertzsprung gap (HG) stars (Fig.~\ref{lam10}--\ref{lam60}). 
If the radial expansion of a star remains moderate, $\lambda$ may increase
as it happens on the Red Giant Branch (RGB) and Asymptotic Giant Branch (AGB) 
(see Fig.~\ref{lam10}). 
For the most massive stars (initial mass above $\sim 20 \msun$) the mass loss
becomes a significant factor, and the changes in density profile, radius, and
binding energy result in an almost constant $\lambda$ during later evolutionary
stages (see Fig.~\ref{lam20}--\ref{lam60}).

Most importantly, Eq.~\ref{ebindeq} implies that given the mass, radius, and evolutionary
state (i.e. the envelope size) of a donor, a change of $\lambda$ will change the
binding energy of its envelope. For example, the binding energy of the envelope
of a star with $\lambda=0.1$ will be larger by an order of magnitude as compared
to an identical star with $\lambda=1$. During the CE 
phase, orbital energy is used to eject the envelope.
Therefore, a binary will have much more difficulty ejecting the tightly bound envelope
of a donor with $\lambda=0.1$, as compared with one with $\lambda=1$.

The consequences for the formation of double compact objects 
are twofold. If $\lambda$ has a low value, and the stars have insufficient separation
(orbital energy), they will merge during the CE phase. This terminates 
further binary evolution and prevents the formation of a DCO. If, on the other hand, 
$\lambda$ has a high value, it is most likely that a given system will eject the 
CE with a small resulting decrease in orbital separation. Binaries (later becoming DCOs) 
retaining a wide enough separation after the CE event may not be able to merge within a 
a Hubble time. Such DCOs become uninteresting from the gravitational wave detection perspective. 
Also, a supernova explosion occurring in a binary that maintains a wide orbital
separation tends to disrupt the system.

\subsubsection{Hertzsprung gap donors $--$ submodel A \& B} \label{hgtreatment}

The outcome of the CE phase depends strongly on the evolutionary 
stage of the donor, as first discussed in~\cite{rarity}. Even before considering
the energy balance, it is critical to incorporate an understanding of the core-envelope 
structure of the star. For example, Main Sequence (MS) stars do not have clear  
core-envelope division, as the helium core is still in the process of being
developed. Stars on the HG similarly lack a clear entropy
jump associated with the core-envelope structure \citep{cejump}, although when in
the evolution such a division appears (late HG or post HG?) remains unclear. 
In the case of a CE initiated by a MS or HG donor, therefore, the orbital 
energy is transferred into the entire star rather than just the envelope. This makes 
the ejection of the envelope difficult, and for MS stars we assume
that this will always result in a merger. However,
in the case of HG we extend the analysis by  
considering two possibilities. One is to ignore the core-envelope boundary
issue, and proceed with the calculation of the energy balance (submodel A). 
The second is to adopt a pessimistic approach (submodel B) in which each CE with
a HG donor leads (independent of energy balance) to a merger. 
Both submodels are calculated for each evolutionary model in this study.
Massive stars, beyond HG, that enter the core helium burning already possess
well-defined core-envelope structure and require only the energy balance calculation.
Exceptions are the Helium MS and Helium HG
stars, which have an analogous structure to their hydrogen counterparts, and are 
treated as such. 

\subsubsection{New calculation of the $\lambda$ parameter} \label{newlam}

The first attempts to use physical (calculated rather than assumed)
values of $\lambda$ coefficient were presented by \cite{dewi2000} and \cite{t2001}.
A study by \cite{pod2003} followed soon after, discussing in detail $\lambda$ values lower than $1$. 
The authors of the two former papers evaluated the binding energy for stars that are potential neutron star 
progenitors ($M_{\rm zams} \lesssim 20 \msun$). The actual $\lambda$ values were presented 
for about ten stellar models with different initial mass ($4<M_{\rm zams}<20 \msun$).
Later, \cite{vt2003} extended the initial work to calculate with physical $\lambda$ 
populations of Galactic double compact objects, including evolution of massive stars 
that are progenitors of black holes. Unfortunately, the values of $\lambda$ coefficients 
for massive stars were not tabulated and we were unable to base our calculations on this
study. However, it is worth noting that these initial studies, made it clear
that the value of physical $\lambda$ is rather sensitive to the definition of
the core-envelope boundary. In any realistic calculation of envelope binding
energy the exact location of this boundary in a given star needs to be
assumed and it is not entirely clear how to define this point \citep{t2001}.
Within the realistic limits on the exact location of core-envelope
boundary one may expect an uncertainty of a factor of $\sim 10$ on a double
compact object merger rates (2012, T.Tauris, private communication).

A physical estimate of $\lambda$ was recently presented by two groups:
\cite{chlam} and \cite{nwlambda}. We adopt the expressions from the former group
in the {\tt StarTrack} code, although both calculations yield similar results 
for NS/BH progenitors. The new $\lambda$ values cover CE events, 
and depend on the evolutionary stage of the donor, its mass at Zero Age Main Sequence (ZAMS) 
and the mass of its envelope, and its radius. In addition, all of these
quantities can depend on a wide range of metallicities 
($Z=10^{-4}$--$0.03$).
This represents a significant improvement in the physics
of the formation of DCOs, as most previous studies treated $\lambda$ as constant (usually $\lambda=1$) 
throughout the evolution of a star. In the new approach the authors 
calculate two values of $\lambda$ from detailed stellar models: in the first
they treat the binding energy of the donor's envelope as consisting only of its 
gravitational energy, while in the second the binding energy is decreased by the full 
internal energy of the envelope.
Since neither of these extremes is plausible \citep{chlam}, we
choose our $\lambda$ to be the average of the two. In the case of stars with masses leading to
the formation of BHs we find that $\lambda \lesssim 1$ 
(see Fig.~\ref{lam40},\ref{lam60}). We find higher values for typical NS progenitors, and especially 
in the late evolutionary stages we have $\lambda \gtrsim 1$  (see Fig.~\ref{lam10}). 
Since the group that developed this procedure is affiliated with Nanjing
University, we label this approach as the \textit{Nanjing} $\lambda$.
We have additionally requested that the authors compute a few extra models
for massive stars, as these are particularly relevant to our study. \cite{chlam} presents models up to $M_{\rm
zams}=20\msun$, and we have now obtained models up to $M_{\rm zams}=100
\msun$. We then
extrapolate these to our  entire mass range (up to $M_{\rm zams}=150 \msun$). 
For stars with $M_{\rm zams}>100 \msun$ we use the fitting function calculated for
$M_{\rm zams}=100 \msun$, but input the higher mass when necessary.

We acknowledge that recently similar population synthesis studies were performed.
Two most notable are: a study by \cite{seba} (performed with the SeBa code, more details in Section
\ref{compseba}) on the mergers of black hole systems and a study by \cite{church} (using the \cite{hurley} 
code fused with several {\tt StarTrack} procedures) on short gamma-ray bursts. However, in contrast 
to the input physics in our work, the mentioned codes utilize a constant $\lambda$ value. 

\subsection{Compact object formation} \label{cof}

To evolve a star from ZAMS 
toward its eventual supernova (SN) with the {\tt StarTrack}
code, we use the (slightly modified; see \cite{comprehensive}) procedure
presented in \cite{hurley}, with updated
wind mass loss rates as described in Section \ref{winds}.
The mass of the compact object is calculated from the properties of the 
pre-supernova star, and the type (whether it is a NS or a BH) is set solely
by its mass. 

We use a recent study by \cite{chrisija} to describe the SN explosion and the
resulting compact object formation. Our models allow for a successful
explosion without the need for the artificial injection of energy into the exploding star. 
This is a major update of the input physics used in the population synthesis 
of massive binaries. Previous studies at best used significantly outdated
supernova models, if they used any at all.  
We have introduced two alternative supernova models into our code: ``Delayed" and ``Rapid". 
Both are core-collapse scenarios, and they share the same convection-enhanced neutrino-driven 
explosion mechanism. The main factor that differentiates the two models is the type of 
instability which causes the macroscopic flows of matter (similar to convection) that 
eventually lead to the ejection of the infalling matter. The Delayed model is sourced
from the standing accretion shock instability (SASI), and can produce an
explosion as late as $1\,\mbox{s}$ after bounce, while the Rapid model starts from the Rayleigh-Taylor 
instability and occurs within the first $0.1$--$0.2\,\mbox{s}$. 
In the Rapid scenario we either end up with a very strong (high velocity kick) 
supernova in the case of a low mass star
($M_{\rm zams} \lesssim 25 \msun$), and produce a NS, or the supernova fails,
and there is direct collapse to a massive BH.
In the Delayed case the entire spectrum of explosion energies is allowed,
and this results in a wide range of compact object masses, from NSs to light
BHs to massive BHs. The formulae describing the two supernova engines adopted in {\tt StarTrack}
are provided in \cite{chrisija} (Eqs.~15--17 for the Rapid and 18--20 for the Delayed engine). 

We also allow for the formation of NSs through Electron Capture 
Supernovae (ECS, \cite{ecs}). These are weak supernovae (no natal kick assumed)
occurring for the lowest mass stars ($M_{\rm zams} \sim 7 \msun$), and they end
up forming NSs.

\section{Modeling} \label{modeling}

The physics underlying the formation of double compact objects is uncertain, notably
due to modelling challenges of the supernova and CE phases. Core collapse, 
usually (but not always) followed by a supernova explosion, forms a compact object. The 
mass and the type of compact object is determined by the details of the
event. Additionally, 
supernova asymmetry via mass loss and/or a natal kick may disrupt a binary,
depleting the population of double compact objects. At the same time, for
some binaries a natal kick of the right amplitude and direction may
produce a prematurely coalescing DCO.
The CE phase is present in all formation scenarios,
for all types of close double compact objects. It is the primary mechanism
for bringing initially widely-separated  binaries into close orbits, thereby
allowing them to coalesce within a Hubble time. 

To explore the uncertainties associated with our models we calculate a suite 
of population synthesis results, investigating a range of factors which have 
the largest impact on the rates and physical properties of DCOs.
As a reference we use the standard model introduced in Section~\ref{stan}, that resembles 
input physics described in detail in \cite{startrack}, 
with some important additions and modifications.
The subsequent models are variations on this standard model, each exploring a 
single parameter connected to either the core collapse/supernova explosion or
CE phases.  Despite the fact that we now have physically motivated values for
$\lambda$, we also explore models in which $\lambda$ is constant, with the
specific value set over a wide range. The range is chosen to be such as to encompass all plausible
values allowed by detailed physical calculation of the \textit{Nanjing} 
results \citep{chlam}. Variations 1--4 employ $\lambda=$0.01, 0.1, 1, and 10, respectively
(see Table \ref{pstudy}). 

To delineate between a NS and a BH we need to adopt a value for the maximum
NS mass. Theoretical studies of the equation of state allow values
in the range $1.5$--$3.0 \msun$ \citep{what}. Observations 
yield a narrower range, with the most massive NSs reaching $2 \msun$ \citep{2NS}. 
Statistical analyses of the measured BH masses indicate that the BH mass 
distribution is unlikely to extend below about $4.5 \msun$ \citep{mg1,mg2,ilya2011}, 
which leaves a significant range with no compact objects. This
might argue for a higher upper limit on the NS mass (or even potentially a lower limit on the BH
mass?). Utilizing the theoretical and  
observational estimates, we vary the maximum NS mass from $3.0 \msun$ (Variation 5), 
through $2.5 \msun$ (Standard), and as low as $2.0 \msun$ (Variation 6).

In the standard model for natal kicks in core collapse supernovae we employ a
Maxwellian kick distribution with $\sigma=265$ km s$^{-1}$, based on 
observed velocities of single Galactic pulsars \citep{hobbs}.
During the explosion, parts or all of the ejected mass may not reach escape velocity 
due to {\em (i)} the strong gravitational potential generated by the newly formed compact object 
and {\em (ii)} the decreasing explosion energy with the mass of exploding
star. In the asymmetric mass ejection kick mechanism
(adopted here) it makes BH natal kicks smaller as compared with NS kicks
(e.g., \cite{chrisija}).
To account for this effect we use a simple linear formula describing the reduction
of natal kick magnitude by the amount of fallback during a SN:
\begin{equation} \label{vkick}
V_k=V_{max}(1-f_{fb}),
\end{equation}
where $V_k$ is the final magnitude of the natal kick, $V_{max}$ is
the velocity drawn from a Maxwellian kick distribution, and $f_{fb}$
is the fallback factor. The values of $f_{fb}$ range between 0--1, with 0
indicating no fallback/full kick and 1 representing total fallback/no kick (in this
case a ``silent supernova"). This factor is calculated for both the Delayed
and Rapid SN engines according to \cite{chrisija}.
For the electron capture supernovae the explosions are found to be
symmetric \citep{dessart}, and we assume no natal kick. 
However, the orbit is still modified due to the mass loss in the explosion. 
BHs that form through partial fall back are assumed to receive
natal kicks that are drawn from the same distribution as for the NSs, but the
value is decreased in proportion to the amount of fall back. 
In particular, the most massive BHs form via full fall back,
which leads to no natal kick and ``silent'' BH formation. This may
be supported both theoretically \citep{bhmassdis} and from observations
\citep{mirabel}. 

There seems to be some evidence that NSs that are found in binaries
receive smaller natal kicks (e.g., \cite{pfahl2002,puzzle,wong,hmxb}), of the order 
of $100 $ km s$^{-1}$. Keeping the same Maxwellian distribution as in the standard model
, we modify the magnitude of natal kicks of NSs and BHs to 
$\sigma=132.5$ km s$^{-1}$ (Variation 7). 

Additionally, we calculate two models in which the BH kicks take on two extreme
values. In one, newly formed BHs receive full natal kicks as,
observed for single pulsars ($\sigma=265$ km s$^{-1}$) independent of the 
amount of fallback (Variation 8). In the other, no natal kicks are applied
to BHs, independent of their mass or the amount of fallback (Variation 9). 

As our standard model we select the Rapid supernova 
engine, since this supernova mechanism, combined with binary evolution, successfully 
reproduces the mass gap \citep{massgap} observed in Galactic X-ray binaries 
\citep{mg1,mg2}. On the other hand, the Delayed engine (Variation 10), as well as the routine
previously implemented within {\tt StarTrack} \citep{comprehensive},
generates a continuous spectrum of compact object masses. 

Observational and theoretical developments have led to the discovery of a number of
puzzling phenomena in the winds from massive stars. The two most prominent are the 
``weak wind problem" (e.g., \cite{chleb,kudr,herrero}) and ``wind clumping" (e.g., 
\cite{oster,markova,repo,lepine}). The former is related to the fact that wind mass 
loss rates from late O and early B type stars might be $\sim 1$--2 orders 
of magnitude lower than theoretically predicted. The latter suggests that winds 
might be forming dense clumps rather than being distributed uniformly, which 
may lead to an overestimate of mass loss rates by a factor of $\lesssim 2$.
Based on this we investigate the possibility that our wind mass loss rates are
too high, and so we calculate a model (Variation 11) in which the rates are reduced by 
a factor of 2. This is done for all stars at all points in their nuclear
evolution. 

The standard model utilizes detailed calculations of stable, nuclear and thermal timescale,
mass transfer episodes. The model distinguishes between nuclear and thermal timescale 
transfers, also accounting for contributions from magnetic braking, gravitational radiation
and tidal interactions (for details see \cite{startrack} -- Section 5). These calculations
have been calibrated on massive binaries that are relevant for DCO formation (e.g. \cite{well2001,
tau1999,dp2003}; for further details see \cite{startrack} -- Section 8). 
During mass transfer episodes it is not known how much of the transferred
mass is retained within the given binary and how much mass is lost entirely 
from the system. In our standard prescription we set the mass transfer to be 
half-conservative. This means that half of the mass that is transferred from
the donor is accreted onto the companion and the rest is expelled to infinity.
The specifics of the angular momentum loss associated with the expelled matter are
presented in Eq.~32 \& 33 in \cite{startrack}.
The accretion rate in an RLOF mass transfer ($\dot{M}_{\rm acc}$) 
is given by the formula:
\begin{equation} \label{mteq}
\dot{M}_{\rm acc}=f_{\rm a}\dot{M}_{\rm don},
\end{equation}
where $\dot{M}_{\rm don}$ is the donor's RLOF mass transfer rate and $f_{\rm a}$
sets the level of conservativeness ($f_{\rm a}=0.5$ for the standard model, half-conservative).
If the transfer rate exceeds the Eddington limit, then more accreted mass can be ejected. 
We additionally investigated two cases: a fully conservative (Variation 12, $f_{\rm a}=1.0$) 
and fully non-conservative (Variation 13, $f_{\rm a}=0$) mass transfer.  

We also include two additional models investigating the common envelope parameter 
$\lambda$. However, in contrast to Variations 1--4 we allow it to vary physically
according to the \textit{Nanjing} prescription while boosting it by a constant value.
In Variation 14 we multiply $\lambda$ by a factor of $5$, which causes the binding energy
of the envelope to decrease by the same factor. This is done to account for the possibility 
that not only internal energy but enthalpy may be the agent that causes the ejection of 
the envelope, making it easier in consequence \citep{entalpia}. To investigate the aforementioned
core-envelope boundary problem (see Section \ref{newlam}), which may cause $\lambda$ to vary 
significantly we present Variation 15. Here we divide $\lambda$ by a factor of $5$, which in 
turn increases the binding energy. The aforementioned boundary problem may also boost $\lambda$, 
however high values of this parameter are already covered in Variations 4 and 14.

Table~\ref{pstudy} lists all of our models, each with the relevant parameter to be
varied as compared to our standard model (listed in the first row).
Each model is calculated for two metallicity values: solar ($Z=\zsun=0.02$)
and $10\%$ solar ($Z=0.1\zsun=0.002$). 
Each model is further divided into submodel A (CE energy balance with HG donor) and 
submodel B (CE merger with HG donor) as discussed in Section~\ref{hgtreatment}.
We end up with $61$ distinctive
models (15 variations for 2 metallicities and 2 CE submodels, on top of our
standard model). 

For each model we evolve $2\times 10^6$ binaries (with one exception, see. Sec.~\ref{v4}), 
assuming that each component is created at the same time. Each binary system is initialized
by four parameters which are assumed to be independent. These are: primary
mass $M_1$ (initially more massive component), mass ratio $q=M_2/M_1$, where 
$M_2$ is the mass of the secondary component (initially less massive), the
semi-major axis $a$ of the orbit, and the eccentricity $e$. The mass of the primary component
is randomly chosen from the initial mass function adopted from \cite{kro1}, and
\cite{kro2},
\begin{equation} \label{imf}
\Psi (M_1) \propto \left\{
\begin{array}{l c}
M_1^{-1.3} & \quad 0.08 \ \msun \leq M_1 < 0.5 \ \msun \\
M_1^{-2.2} & \quad 0.5 \ \msun \leq M_1 < 1.0 \ \msun \\
M_1^{-\alpha} & \quad 1.0 \ \msun \leq M_1 < 150 \ \msun, \\
\end{array}
\right.
\end{equation}
where $\alpha=2.7$ is our standard choice for field populations. 
Stars are generated from within an initial mass range
$M_{min}$--$M_{max}$, with the limits  based on the targeted stellar population. For example, NS
studies require evolution of single stars within the range $8$--$25 \msun$, while
for BHs the lower limit is $25\msun$. Binary evolution
may broaden these ranges due to mass transfer episodes, and we therefore set the
minimum mass of the primary to $5\msun$. We assume a flat mass ratio distribution, $\Phi(q)=1$,
over the range $q=0$--1, in agreement with recent observations~\citep{kob}.
Given a value of the primary mass and the mass ratio, we
obtain the mass of the secondary from $M_2=qM_1$. However, for the same reasons
as for the primary, we don't consider binaries where the mass of
the secondary is below $3\msun$. 
The distribution of initial binary separations is assumed to be flat in 
$\log(a)$~\citep{abt}, and so $\propto \frac{1}{a}$, 
with $a$ ranging from values such that at ZAMS the primary fills no more than 50\%
of its Roche lobe to $10^5 \ \rsun $. For the initial eccentricity
we adopt a thermal equilibrium distribution (e.g., \cite{heggie,duq})
$\Xi(e)=2e$, with $e$ ranging from $0$ to $1$.

\section{Results} \label{results}

The double compact object merger rates for a synthetic galaxy resembling the Milky 
Way, but for two differing values of metallicity, are presented in Table~\ref{gmerger02} 
($Z=\zsun$) and Table~\ref{gmerger002} ($Z=0.1\zsun$). The corresponding plots of 
the merger rates are shown in Figures~\ref{02rates} and \ref{002rates}.
The physical properties of double compact objects for the standard model are 
presented in Figure~\ref{2011mch} (chirp mass), Figure~\ref{2011tdel} (delay 
time), and Figure~\ref{2011qdis} (mass ratio). 
The distribution of chirp masses (Figures~\ref{amch02var}--\ref{bmch002var}
(with corresponding Tables \ref{amch02char}--\ref{bmch002char}) and delay times 
(Figures \ref{atdel02var}--\ref{btdel002var}) for all the models are presented.
The delay time, $t_{del}$, is the sum of the time needed to form two compact objects
from a ZAMS binary and the time for the two compact
objects to coalesce due to the emission of gravitational radiation.  
For double compact objects the former evolutionary time interval ($\sim$ Myr) 
is usually much shorter than the latter merger time ($\sim$ Gyr), and the delay 
time is rather similar to the merger time.
Additional models and DCO population properties are available on-line at 
{\tt www.syntheticuniverse.com}. 

For each model we calculate the Galactic merger rates. These are
defined as the number of coalescences of DCOs per unit time occurring in a synthetic galaxy
similar to the Milky Way (with age of $10$ Gyr and a constant star formation
rate (SFR) of $3.5 \mpy$). In practice this is done by checking if the delay
time of a DCO (with a random starting point between $0$--$10$ Gyr) falls near
the current galaxy age ($10$ Gyr). However, the amount
of mass within the simulated binaries corresponds only to a part of the star
forming mass of the galaxy. In order to extrapolate the simulated mass to that
of the entire galaxy we employ the following procedure. First, the amount 
of mass contained within the simulated binary stellar population is estimated (the mass of 
the primary follows from Eq.~\ref{imf}, and the mass of the secondary follows from our assumed mass ratio
distribution).
Additionally, we assume a binary fraction of $50\%$, so that for each binary system
there is one additional individual star. The mass of each of the individual
stars is taken to be in the same range as for primary components in binaries.
The acquired mass, $M_{\rm acq}$,
is then divided by the age of the synthetic galaxy $t_{\rm gal}$ ($10$ Gyr) in order to get a constant 
star formation rate corresponding to the simulated stellar mass ($0.073 \mpy$).
To match this SFR to the one of the synthetic galaxy ${\rm SFR}_{\rm gal}$ ($3.5 \mpy$) one needs
a multiplication factor of $48$ ($f_{\rm SFR}$). The corresponding equation is:
\begin{equation}
f_{\rm SFR}={\rm SFR}_{\rm gal}\left(\frac{M_{\rm acq}}{t_{\rm gal}}\right)^{-1}
\end{equation}
Therefore, to extrapolate our results to the
entire mass in the synthetic Milky Way, we use each synthetic DCO binary
$48$ times. Each time the given binary is assigned a
new starting time (from a uniform distribution), and if its coalescence time falls
within $9\mbox{--}10$ Gyr it is included in our results.  

The typical range of the number of DCOs, $N$, generated in each simulation is 
$\sim 1000$--$10000$ for submodel A and $\sim 10$--100 for submodel B. Therefore
the relative statistical error ($\sqrt{N}/N$) is at most $\sim 10\%$--30\%
where the ranges are given by the values in submodels A--B). The errors arising
from uncertainties in various aspects of
the single and binary star evolution can change the Galactic merger 
rates of DCOs by $\sim 1$--2 orders of magnitude (as shown in the next sections), 
making the statistical errors irrelevant.

In addition to the population of DCOs with delay times below 10 Gyr (the 
merging population), in each model we acquire another population occupying 
the domain above this time limit (the non-merging population).
This population contains each type of DCO, and is available at {\tt www.syntheticuniverse.org}. 

\subsection{Standard Model} \label{stan}

At solar metallicity, the merger rates for a synthetic galaxy similar to the Milky Way are 
dominated by NS-NS systems (23.5--7.6 Myr$^{-1}$; submodel A--B),
with a smaller but still significant contribution from BH-BH systems ($8.2$--$1.9$ Myr$^{-1}$),
and with a minor contribution from BH-NS systems to the overall DCO merger rate
($1.6$-$-0.2$ Myr$^{-1}$; see Table \ref{gmerger02}). Qualitatively 
these findings are consistent with previous results~\citep{comprehensive}.
The quantitative results, however, are quite different, due to the many
improvements in the models over the intervening decade. \cite{comprehensive} found the following
mean rates:  NS-NS $53$ Myr$^{-1}$, BH-BH $26$ Myr$^{-1}$ and BH-NS $8.1$ Myr$^{-1}$.  
At sub-solar metallicity, the systems with BHs increase their relative
contribution to the overall rates, and the merger rate is dominated by BH-BH systems
(73.3--13.6 Myr$^{-1}$), with smaller contribution from NS-NS
(8.1--2.5 Myr$^{-1}$) and BH-NS systems (3.4--2.3 Myr$^{-1}$; see Table \ref{gmerger002}).
These results are qualitatively similar to our recent work on the dependence of
merger rates on metallicity \citep{nasza}. Again, quantitatively 
there are significant differences. \cite{nasza} found the following rates: $84$--$6.1$ Myr$^{-1}$ (BH-BH), 
$41$--$3.3$ Myr$^{-1}$ (NS-NS) and $12$--$7.0$ Myr$^{-1}$ (BH-NS) for $Z=0.1\zsun$. These changes 
reflect the fact that since the previous study we have introduced physical $\lambda$ values 
and observationally constrained SN models, which yields a new compact object mass spectrum.

A general division of these rates by DCO type may be understood in the following way.
The initial mass function (IMF) falls steeply with mass, and delivers more NS than  
BH progenitors (by a factor \footnote{This factor is the ratio of the 
number of stars between $8$--$20 \msun$ (NS progenitors) to the number of stars
between $20$--$150 \msun$ (BH progenitors) as calculated from the IMF (see Section 
\ref{modeling}).} of $\sim 4$). The supernova explosion is the major process that drastically 
affects the number of massive binaries , as the explosions tend to disrupt binaries. This is 
especially true in the case of NS progenitors, as these receive large natal kicks and as many 
as $\sim 90$--$95\%$ of potential binaries may end up disrupted after the first supernova 
explosion (e.g., \cite{lorimer2004,puzzle}). Supernovae do not affect binaries 
with BHs as much because it is believed that most of the massive stars producing BHs do not 
experience large kicks at core collapse. What follows is that the rates for NS-NS 
and BH-BH mergers are not as separated as would be simply deduced from the IMF. 
The rates are also very sensitive to the details of the CE, and
these are quite
different for NS and BH progenitors (as discussed below).
Additionally, the three types of double compact objects evolve along 
separate evolutionary channels (see Table \ref{channels}). This qualitative 
picture can explain the calculated ratio of NS-NS to BH-BH merger rates ($\sim 3$--$4$). 
The BH-NS merger rates are the smallest, as the majority of potential progenitor 
binaries have initially large mass ratios, and the first interaction of the two 
components leads to a CE phase and the inevitable merger. This happens because the 
massive envelope of the BH progenitor cannot be successfully dispersed by the much less massive NS 
progenitor.  Note that none
of the significant BH-NS formation channels starts with the CE phase, but
instead the progenitors of these systems originate from a narrow mass range
(mass ratio of the two components larger than $\sim \frac{1}{2}$--$\frac{1}{3}$ 
allowing for the first interaction to be a stable RLOF (see Table \ref{channels}).

At sub-solar metallicity other factors come into play and make BH-BH systems 
dominant in the overall merger rate. First, the smaller wind mass loss makes
pre-SN progenitors of NSs slightly more massive (by $\sim 10\%$). Heavier
NS progenitors tend to explode as core collapse supernovae (full kicks) rather 
than ECSe (no kicks, see Section~\ref{cof}).
This means that for sub-solar metallicity more NSs are formed with disruptive
natal kicks than for a solar environment. Additionally, smaller wind mass loss 
decreases the expansion of the separation between the components. This causes the 
progenitors of NS-NS systems, at sub-solar metallicities, to engage in a second CE
(see Table~\ref{channels2}) just after the first one. This increases the probability
of a merger during evolution when compared with$\zsun$, hence the drop of merger 
rates of NS-NS systems from$\zsun$ to $0.1\zsun$. The merger rates for BH-BH 
systems increase for sub-solar metallicity, as 
low wind mass loss rates allow the progenitors to remain more massive during 
evolution. This in turn makes the pre-SN stars more massive and allows for larger
amount of fallback. Increased fallback reduces the magnitude of the natal kicks
(see Eq.~\ref{vkick}) to almost none or none at all (direct collapse into a BH) 
and makes the SN significantly less disruptive, as explained in detail by \cite{nasza}.

The formation of BH-NS binaries is determined by the properties of the 
progenitors of both compact object types. Therefore, the behaviour of these systems
may be considered as a combination of effects noted in the formation of BH-BH and 
NS-NS systems. 

\begin{figure}
\includegraphics[width=1.0\columnwidth,trim = 0 0cm 0 0,clip]{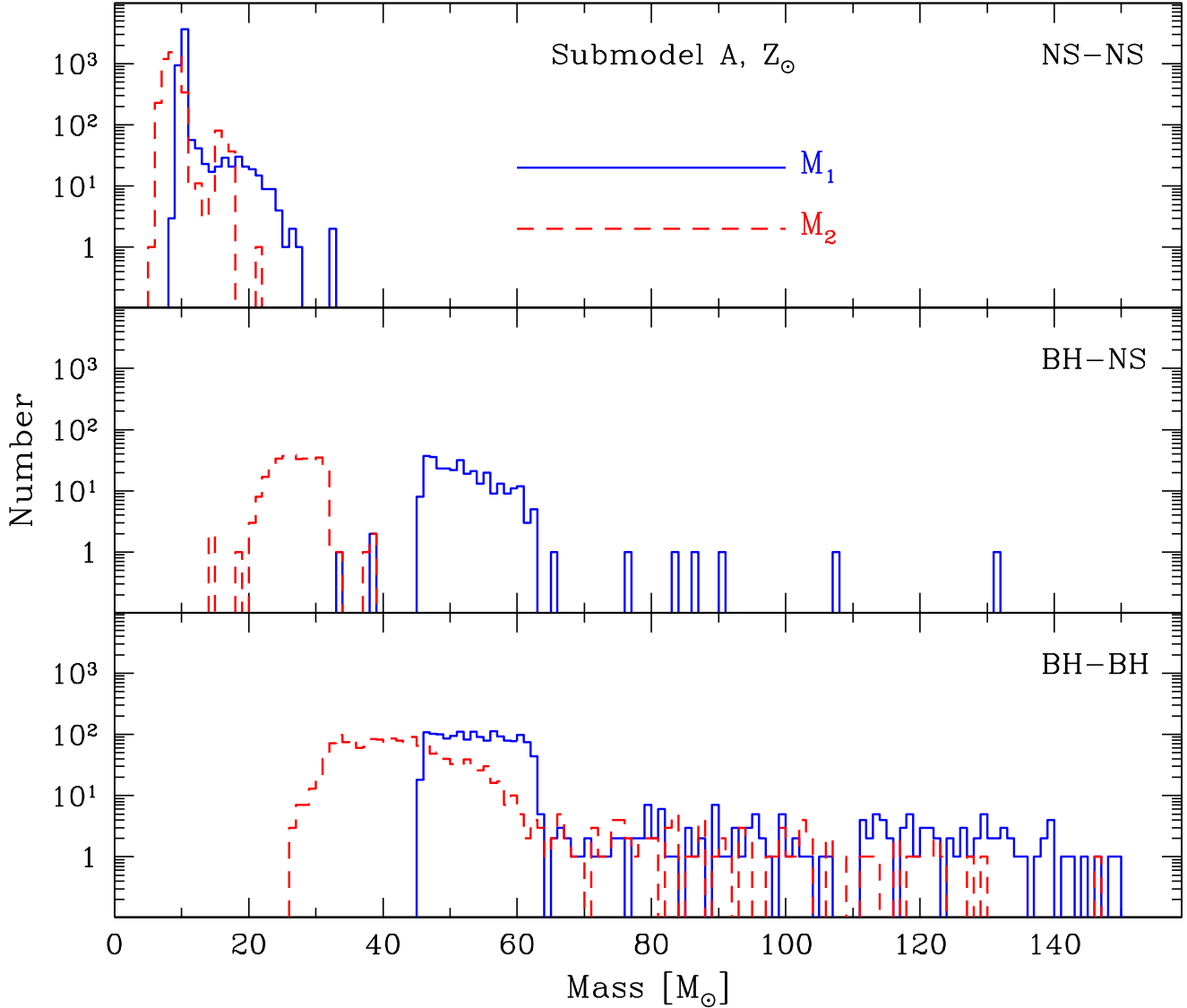}
\caption{The distribution of progenitor (ZAMS) masses of coalescing DCOs for the standard
model, submodel A (for submodel details see Section \ref{hgtreatment}),$\zsun$. 
The top panel presents the distribution for NS-NS, 
the middle panel for BH-NS, and the bottom panel for BH-BH progenitors. $M_1$ 
stands for the primary component (initially more massive, solid, blue line) and 
$M_2$ for the secondary (initially less massive, dashed, red line). The average mass
for NS-NS progenitors is $11$--$9 \msun$, for BH-NS progenitors $52$--$27 \msun$,
and for BH-BH progenitors $58$--$44 \msun$ ($M_{1}$--$M_{2}$). Note that binary
evolution blurs the ZAMS mass limits for NS/BH formation (see Section
\ref{stan}). 
}
\label{mpro}
\end{figure}

\begin{figure}
\includegraphics[width=1.0\columnwidth,trim = 0 0cm 0 0,clip]{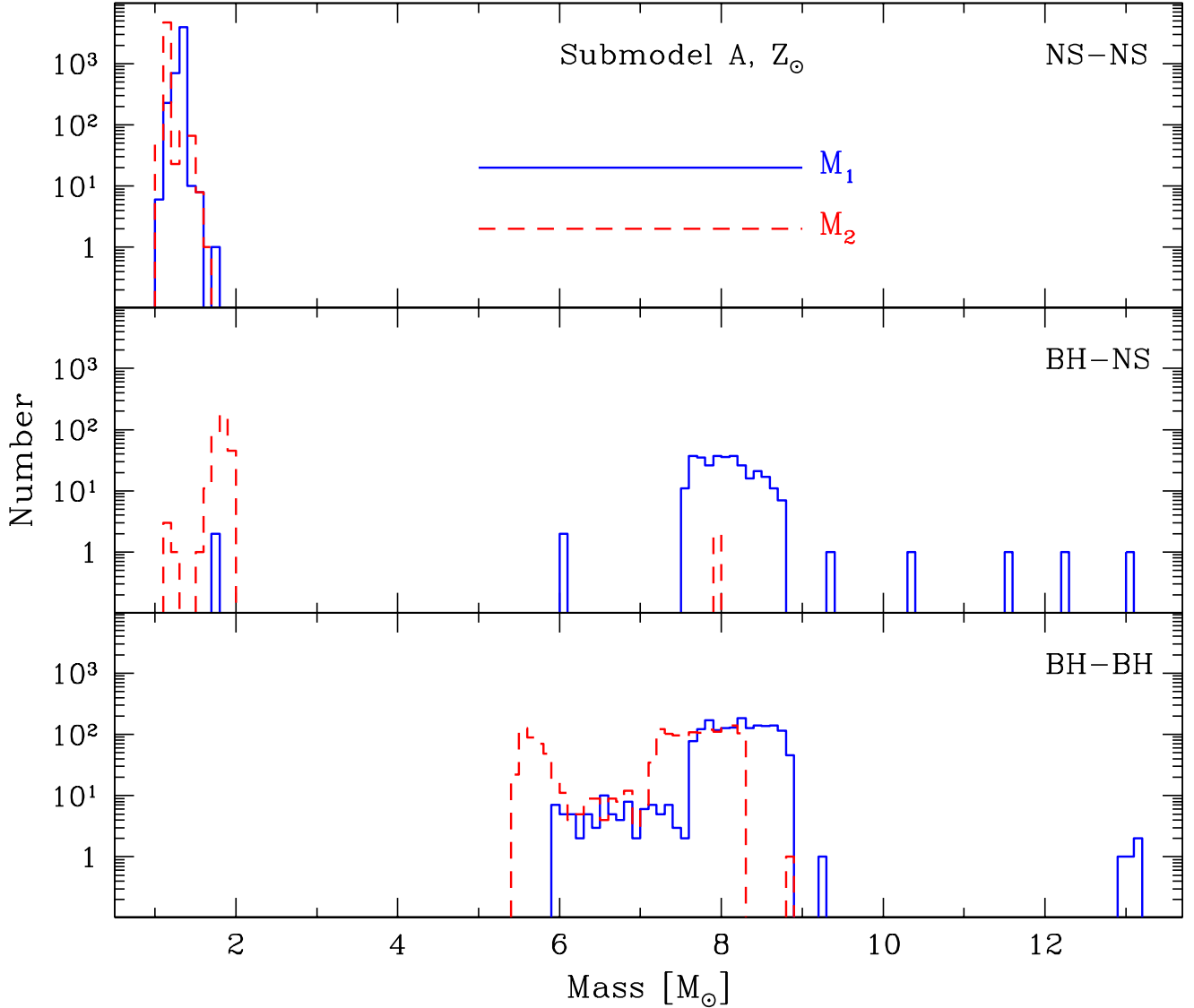}
\caption{The distribution of remnant masses of coalescing DCOs for the standard
model, submodel A,$\zsun$. The top panel presents the distribution for NS-NS, 
the middle panel for BH-NS, and the bottom panel for BH-BH systems. $M_1$
represents the primary remnant (corresponding to $M_{1}$ in Fig.~\ref{mpro}, solid, blue line) 
while $M_{2}$ is the secondary (corresponding to $M_{2}$ in Fig.~\ref{mpro}, dashed, 
red line). The average mass for NS-NS systems is $1.3$--$1.1 \msun$, for BH-NS systems $8.0$--$1.8 \msun$,
and for BH-BH systems $8.2$--$7.2 \msun$ ($M_{1}$--$M_{2}$). 
The gap between the upper mass of NSs ($2 \msun$) and the lowest
mass of BHs ($5\msun$) results from the use of the Rapid SN engine (see Section 
\ref{cof}). 
}
\label{mrem}
\end{figure}

\begin{figure}
\includegraphics[width=1.0\columnwidth,trim = 0 5cm 0 0,clip]{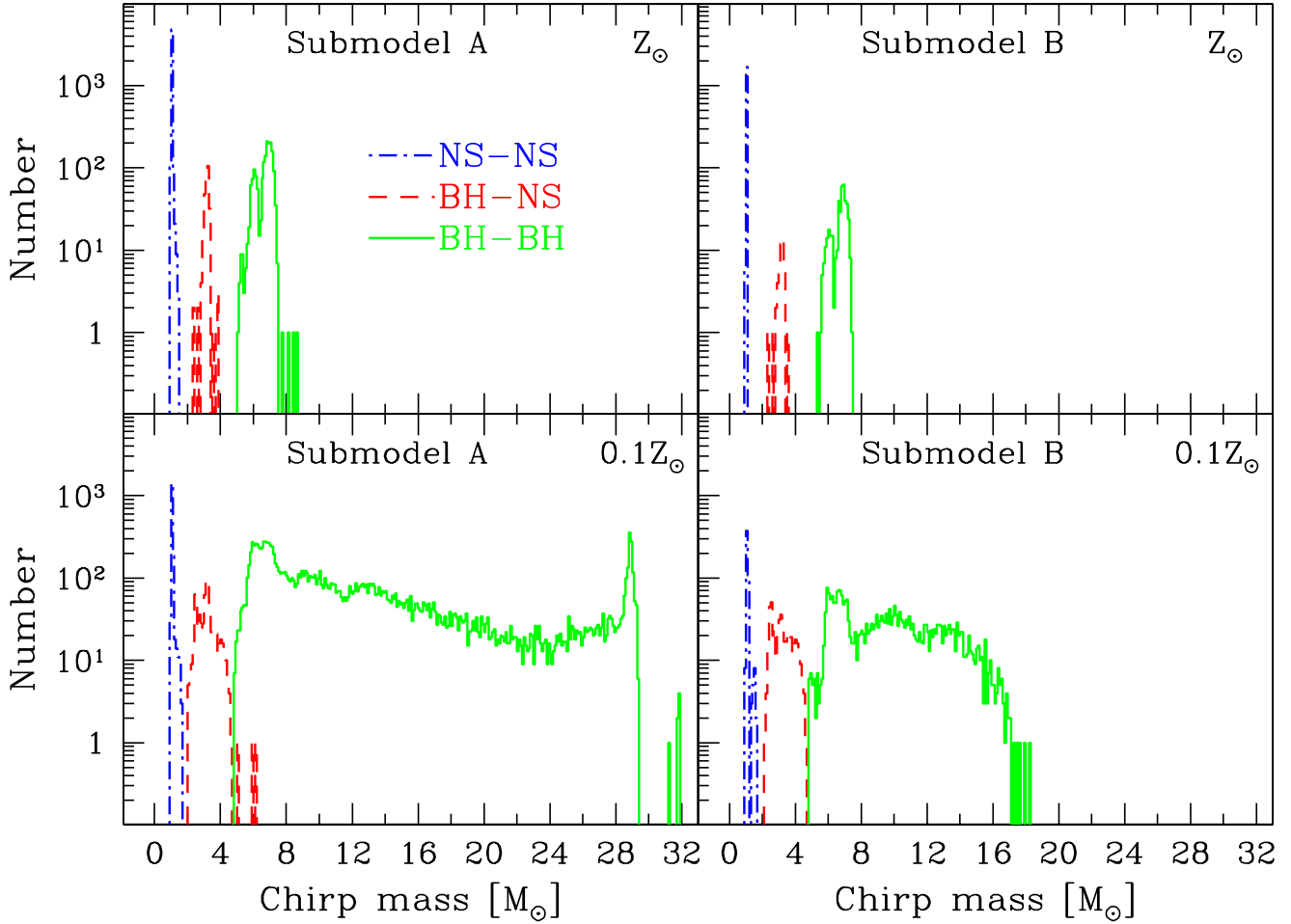}
\caption{The distribution of chirp masses of coalescing DCOs for the standard model. 
The average chirp masses for NS-NS and BH-NS systems are $\sim 1.1 \msun$ and
$3.2 \msun$, respectively, for both submodels and metallicities. The average chirp
mass for BH-BH systems, for$\zsun$, is $\sim 6.7\msun$ for both submodels. For
$0.1\zsun$ the masses are $13.2$--$9.7 \msun$ for submodel A and B,
respectively. 
The maximum chirp mass increases with metallicity as wind mass loss rates 
decrease, allowing for the formation of heavier BHs (see \cite{nasza} and
Section \ref{stan}).
}
\label{2011mch}
\end{figure}

\begin{figure}
\includegraphics[width=1.0\columnwidth,trim = 0 5cm 0 0,clip]{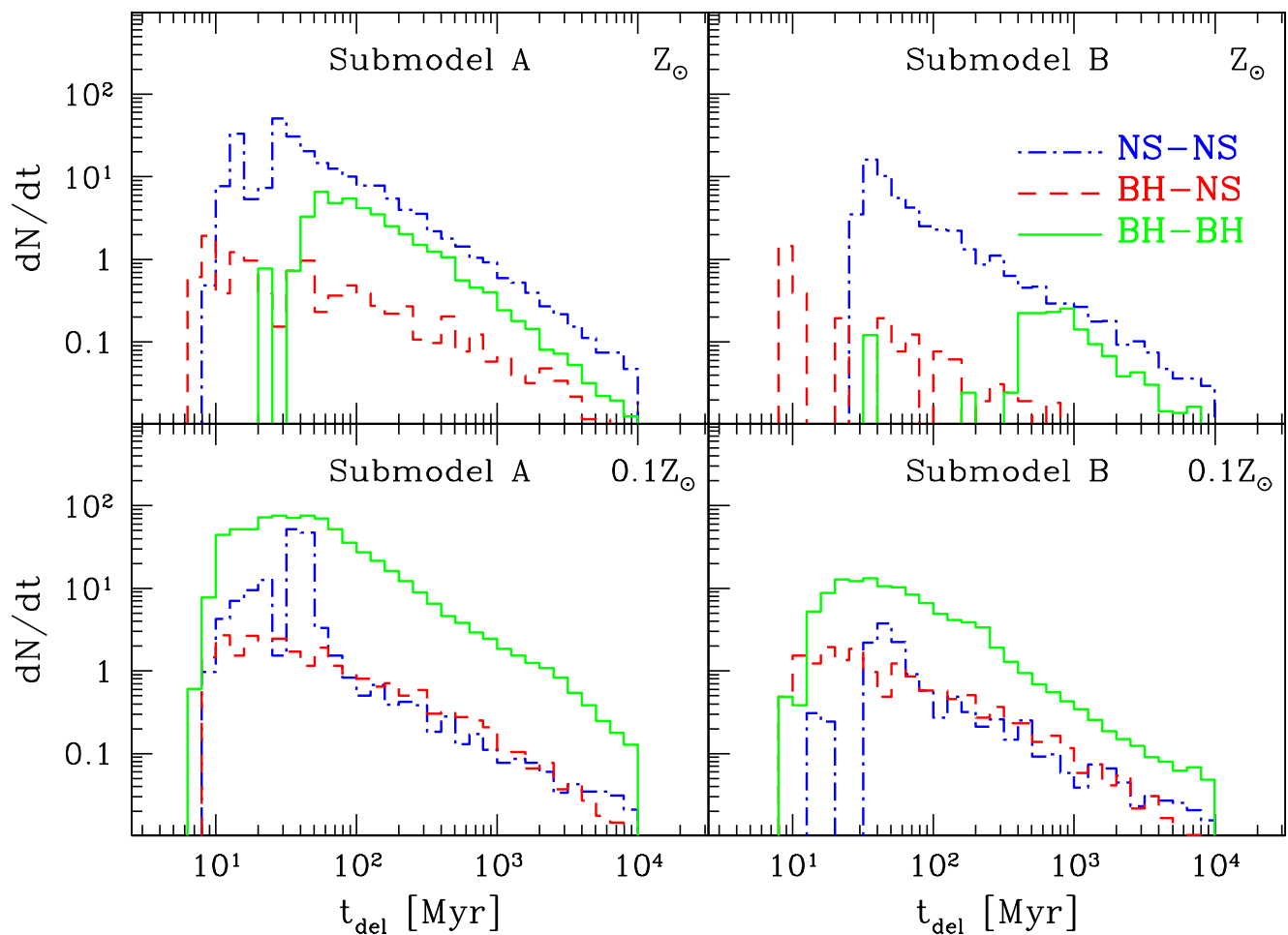}
\caption{The distribution of delay times for coalescing DCOs, for the 
standard model. The vertical axis present the number of DCOs per linear time. 
The average delay time for all binaries is $\sim 1$ Gyr. 
}
\label{2011tdel}
\end{figure}

\begin{figure}
\includegraphics[width=1.0\columnwidth,trim = 0 5cm 0 0,clip]{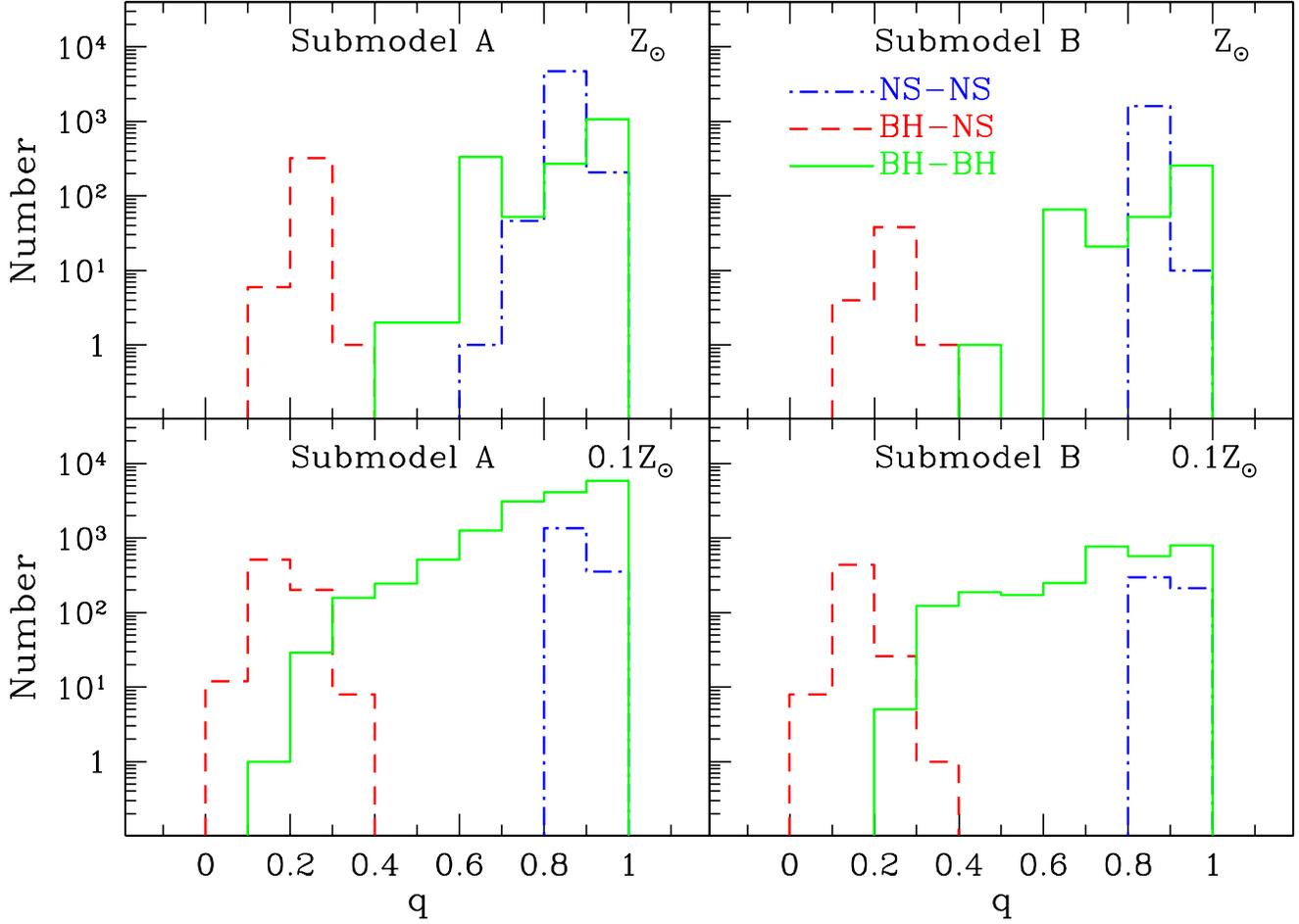}
\caption{The distribution of mass ratios of coalescing DCOs for the standard
  model. 
The mass ratio is defined as the ratio of the less massive to the more massive compact 
object in the binary. The average values for NS-NS systems are $\sim 0.85$ for both
submodels and metallicities. For BH-NS binaries the average is $\sim 0.22$ for$\zsun$, for both submodels and $\sim 0.15$ for $0.1\zsun$, also for both submodels.
For BH-BH systems the average value is $\sim 0.8$ for both submodels and metallicities.
}
\label{2011qdis}
\end{figure}

\begin{figure}
\includegraphics[width=1.0\columnwidth]{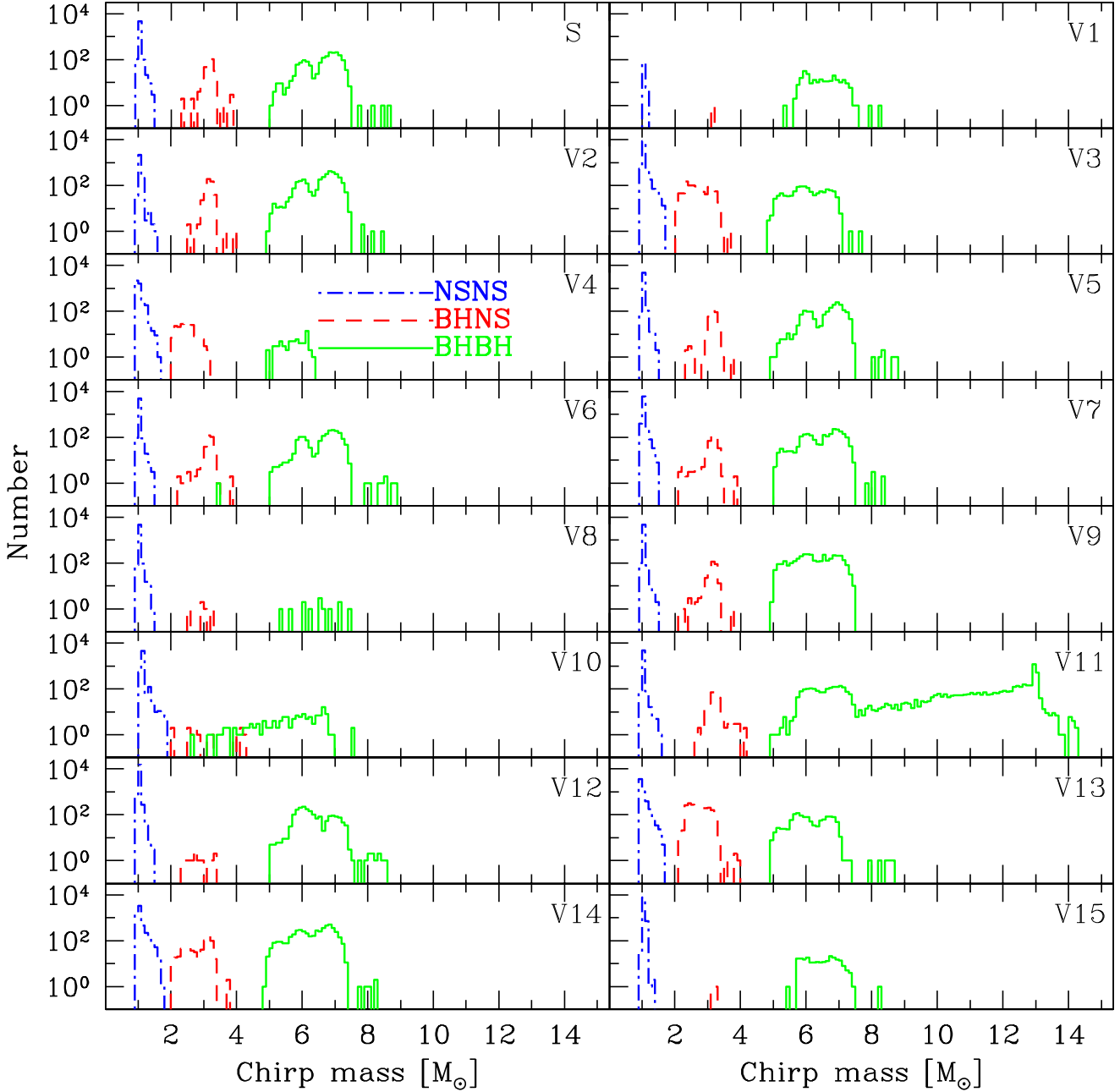}
\caption{
The chirp mass distribution of coalescing double compact objects for all 
variations for submodel A (for submodel details see Section \ref{hgtreatment}) and 
$Z=\zsun$. The maximum chirp mass is found for 
BH-BH systems, and may reach as high as $\sim 14\msun$. The typical chirp
mass for BH-NS systems is $\sim 2$--$3 \msun$, while the chirp mass for NS-NS
systems peaks around $\sim 1 \msun$ independent of the model. 
Note that the chirp masses of BH-NS systems are separated from the BH-BH
values. The only exception to this rule is the (most-likely unphysical)
V10 model, which employs the Delayed supernova engine (see Sec.~\ref{v10} for
details). 
}
\label{amch02var}
\end{figure}

\begin{figure}
\includegraphics[width=1.0\columnwidth]{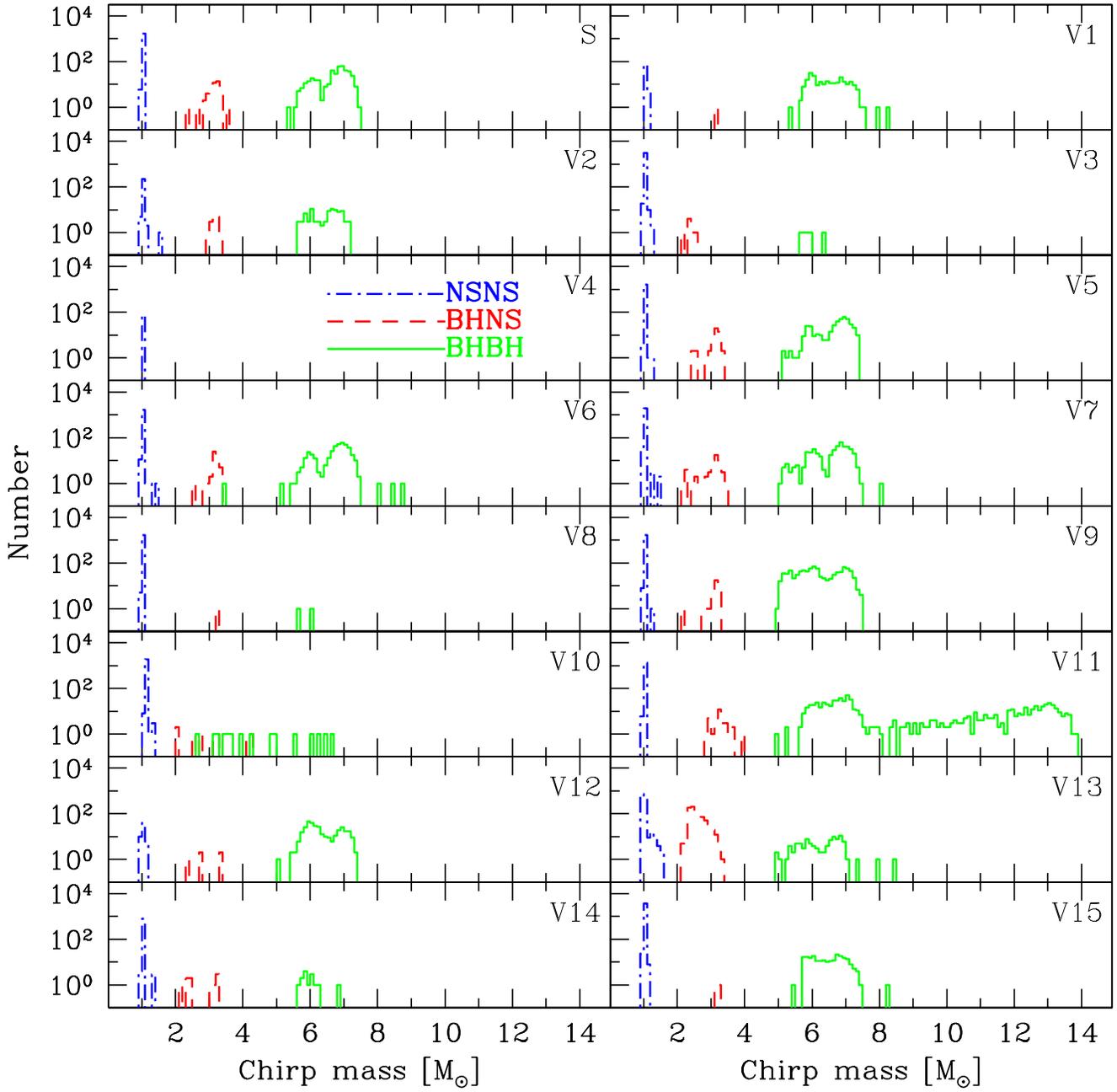}
\caption{
The chirp mass distribution of coalescing double compact objects for all 
variations for submodel B and $Z=\zsun$. 
}
\label{bmch02var}
\end{figure}

\begin{figure}
\includegraphics[width=1.0\columnwidth]{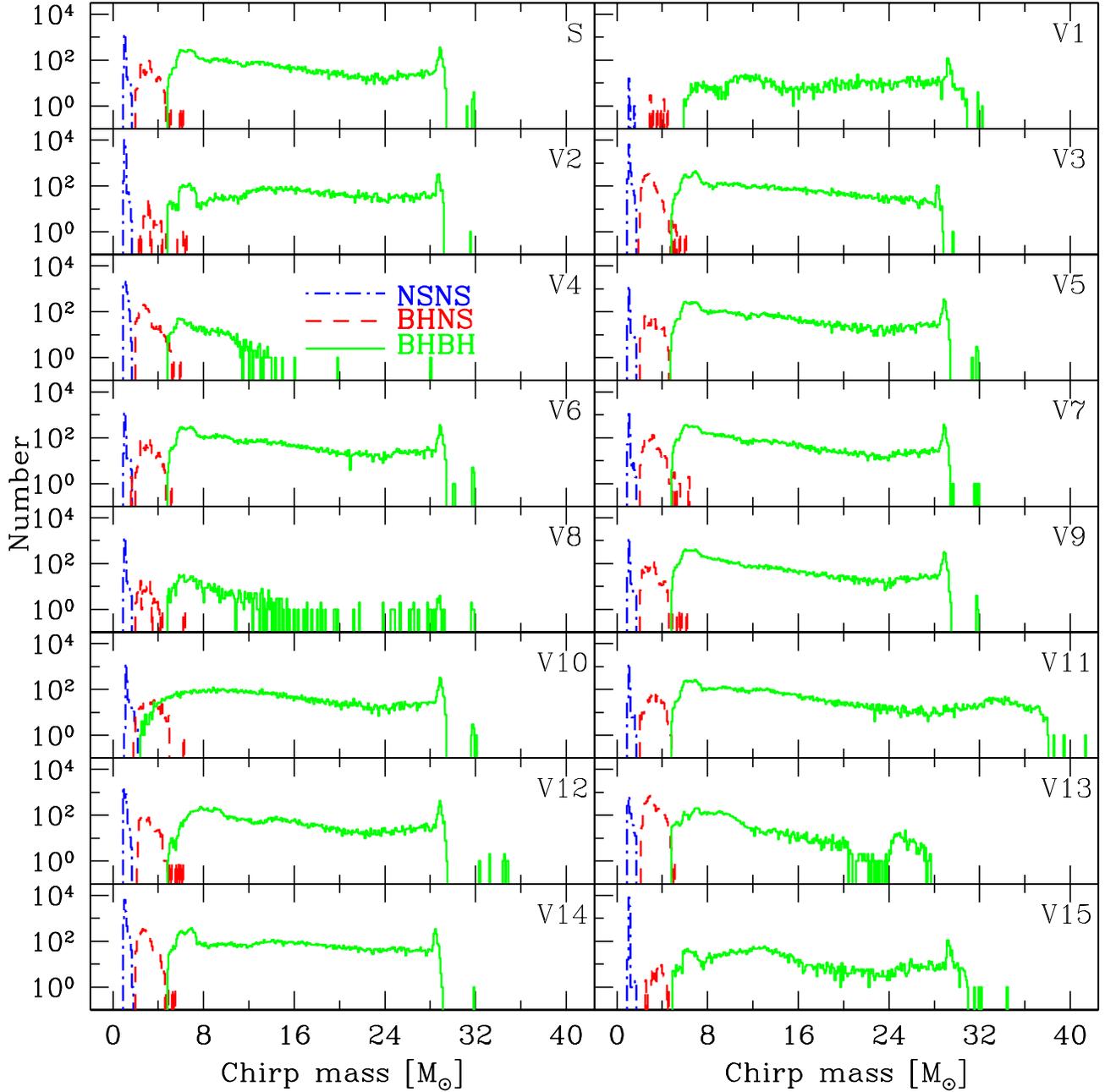}
\caption{
The chirp mass distribution of coalescing double compact objects for all 
variations for submodel A and $Z=0.1\zsun$. 
Note the dramatic increase of the maximum chirp mass with decreasing
metallicity. For solar metallicity, the chirp mass was always below 
$15\msun$ (Fig.~\ref{amch02var}), while for the majority of models shown
here, the chirp mass reaches $\sim 30 \msun$ for sub-solar metallicity. The 
lack of high chirp-mass systems in model V4 is explained in Sec.~\ref{v4}.
}
\label{amch002var}
\end{figure}

In Figures \ref{mpro}, \ref{mrem}, \ref{2011mch}, \ref{2011tdel} and \ref{2011qdis} 
we present the distributions of DCO progenitor ZAMS masses, DCO masses, chirp masses, 
delay times and DCO mass ratios for the standard model. For purposes of 
illustration, the distribution of progenitor and remnant masses are given for submodel 
A, $Z=\zsun$ only. Additional plots are available at {\tt www.syntheticuniverse.org}.

In the distribution of progenitor ZAMS masses (Fig.~\ref{mpro}) one 
can clearly see that binary evolution blurs the limits for ZAMS mass of the star for 
the formation of NS/BH. For NS-NS progenitors (top panel), the masses of the 
primary components (up to $\sim 30 \msun$) exceed the typical upper limit for 
the formation of a NS for single stars ($\sim 20 \msun$, also lower limit for the 
formation of a BH). Under favourable circumstances, binary evolution may push this limit
even further, up to $100 \msun$ (e.g., \cite{wellstein,cxo}). Note that for BH-NS systems 
(middle panel), a high progenitor mass does not necessarily imply that it will form a BH. 
During their evolution, progenitors of BH-NS systems may undergo a mass ratio reversal 
(due to mass transfer events) so the primary component (initially more massive) may become a NS
(as seen on the middle panel of Fig.~\ref{mrem}).  
For BH-BH progenitors the lowest mass of the primary component is $\sim
45\msun$, and the lowest mass of the secondary is $\sim 25\msun$.
Also, progenitor stars of BH-BH systems have a wide mass distribution (ranging up to $150 \msun$). 
However, the masses peak at $\sim 55 \msun$ for the primary
component and $\sim 40 \msun$ for the secondary.

The distribution of masses of DCO remnants (Fig.~\ref{mrem}) clearly shows the 
gap between the upper mass achievable by NSs ($2 \msun$, despite allowing for 
the formation of NSs with mass up to $2.5 \msun$) and the lower mass of BHs ($\sim 5 \msun$).
This is due to the implementation of the Rapid supernova engine, which is the 
first to successfully reproduce this ``mass gap" (for details see Section \ref{cof}).
Additionally, for BH-NS binaries some remnants formed out of the initially less massive
star in the binary fall within the BH mass regime while some primary
components (initially more massive) may end up as NSs. This is due to the
aforementioned mass ratio reversal. 
Despite a wide range of BH-BH progenitor masses, the component masses of remnant systems are
mostly clustering around $5$--$9 \msun$. Such a drastic reduction in mass range
comes from the significant wind mass loss for massive BH progenitors and mass ejection in
CE events. Both factors usually reduce the masses of the whole progenitor 
stars to the masses of their cores, for which the mass range is narrow.   

To calculate the chirp mass $M_{\rm chirp}$ of a DCO we use the following formula:
\begin{equation}
M_{\rm chirp}=(M_{\rm 1}M_{\rm 2})^{\frac{3}{5}}(M_{\rm 1}+M_{\rm 2})^{-\frac{1}{5}},
\end{equation}
where $M_{\rm i}$ are the masses of the components.
The most notable aspect of the chirp mass distributions (Fig.~\ref{2011mch}) is the 
maximum value for BH-BH systems ($\sim 30 \msun$), for submodel A, $0.1\zsun$. As metallicity 
decreases from solar to sub-solar, so do the wind mass loss rates. This allows for 
the formation of more massive BHs. For submodel B, $Z=0.1\zsun$ the maximum value is 
lower ($18 \msun$) than for submodel A for the same metallicity. In submodel A the 
most massive progenitors of BHs experience significant expansion during evolution 
leading to the CE events with a HG donor, rather than a donor beyond the HG which is 
more likely for lower mass progenitors. Submodel B does not allow
for the survival of HG donors during a CE (see Section \ref{hgtreatment})
and the most massive BHs disappear.
The average chirp masses for $Z=\zsun$ are the same for both submodels: $1.05$ for NS-NS, 
$3.2 \msun$ for BH-NS, and $6.7 \msun$ for BH-BH systems. For $0.1\zsun$ the
average chirp masses are: $1.1$--$1.1 \msun$ (submodel A--B) for NS-NS, $3.2$--$3.1 \msun$ 
for BH-NS, and $13.2$--$9.7 \msun$ for BH-BH systems.

The delay time distributions (Fig.~\ref{2011tdel}) are proportional to
$t_{\rm del}^{-1}$. 
The average delay time for systems merging within 10 Gyr at$\zsun$ is: $1.1$--$1.5$ Gyr (submodel A--B) for
NS-NS, $1.7$--$1.7$ Gyr for BH-NS, and $1.0$--$2.5$ Gyr for BH-BH systems. For sub-solar 
metallicity the average is: $1.0$--$2.3$ Gyr for NS-NS, $1.5$--$1.5$ Gyr for BH-NS, and
$1.0$--$1.4$ Gyr for BH-BH systems. 

The delay times for submodel B are in general higher than for submodel A.
This is a direct result of our assumption on the outcome of the CE phase with HG
donor. Binary populations are identical in submodel A and B until a CE
event. If the separation is relatively small, the CE may be initiated early 
on in the evolution, specifically by a HG donor. The survival of such an event
may lead to the formation of a close binary in submodel A. In submodel B CEs with
this type of donor are always considered a merger (see Section \ref{hgtreatment}).
Binaries with separation large enough to prevent the rapidly expanding HG star from
overfilling its Roche lobe will initiate the CE in later stages of evolution. This
scenario meets the criteria of A and B submodels and allows the binaries to be
accounted in both (in terms of forming merging DCOs).  Submodel B, therefore,  
favours binaries with larger separations, which translate into larger merger times. 

Additionally, the delay times decrease with metallicity. This comes from the fact
that for sub-solar environments stars lose less mass in winds and therefore, form 
more massive remnants. The more massive the components of a DCO, the less time it 
takes for a system to merge. 
A secondary effect of the reduced wind mass loss rates for $Z<\zsun$ on the delay times
is the smaller expansion of the separation during evolution as a smaller amount of mass
is removed from the system. 

The distribution of mass ratios $q$ is similar for both submodels and both metallicities.
NS-NS systems group around $q=0.9$ as the NS masses found in our models are
on average similar in each formed DCO system ($\sim 1.1$--$1.4 \msun$; see Fig.~\ref{mrem}). 
BH-NS systems group around small mass ratio values ($q=0.2$), which means
a large difference in mass between both remnant compact objects. This follows
from the fact that the typical NS mass is $1.3 \msun$ and the typical BH mass
is found at $5$--$9 \msun$ (see Fig.~\ref{mrem}).
BH-BH systems have the widest range of $q$, and typically our simulations
show mass ratios in the range $0.4$--$1.0$ with increasing probability toward $q=1$.
This simply reflects the fact that BH progenitors come from a wide range
of masses, and that binaries with similar masses more readily survive binary
interactions (mass transfers and supernovae explosions). 

\subsection{Variation 1} \label{v1}

This is the first of four variations addressing the CE binding 
energy parameter, $\lambda$. In this model we fix $\lambda=0.01$. 
This It is found that in this evolutionary variation all merger rates significantly
decrease as compared to the standard model (see Table \ref{gmerger02} and \ref{gmerger002}).
For solar metallicity the Galactic
merger rates are dominated by BH-BH systems ($1.1$ Myr$^{-1}$ for both
submodels), followed by NS-NS ($0.4$ \pmyr for both submodels), and BH-NS systems ($0.002$ \pmyr,
also for submodels A and B). For $0.1\zsun$ the rates for BH-BH systems are $12.5$--$8.1$ \pmyr
(submodel A--B), $0.06$ \pmyr for NS-NS systems (for both submodels), and $0.03$ \pmyr for
BH-NS systems (also for both submodels). The behaviour of the population of
double compact objects in this model can be understood as follows. We
have chosen $\lambda$ to have a very low value, which translates into a very
high binding energy of the CE  
(see Eq.~\ref{ebindeq}). 
This binding is so strong that most of the binaries experience a merger during
the first CE phase in their evolutionary history. This terminates the evolution of the 
binary and prevents the formation of a DCO. The least affected are
BH-BH systems, since the fixed $\lambda$ value used in this variation is
comparable to (although lower than) the  \textit{Nanjing} $\lambda$ 
(see Fig.~\ref{lam40},\ref{lam60}) of the typical BH progenitor.

In earlier stages, such as on the Hertzsprung gap, the values are higher (\textit{Nanjing} $\lambda
\approx 0.1$--$0.2$; see Fig.~\ref{lam40}). This
means that in this fixed $\lambda$ scenario the value of the binding energy of the CE of
a HG donor is an order of magnitude larger than for the standard (more physical) model. 
Therefore, many CE events with a HG donor end in a merger, as the reservoir 
of the orbital energy is insufficient to eject the envelope. The surviving binaries are the 
ones in which the CE was initiated by a Core Helium Burning (CHeB) donor and/or that have very 
massive accretors and therefore a large orbital energy reservoir. Submodel B does not allow for 
CE events with a HG donor. In effect, the very small $\lambda$ (very large binding energy)
adopted in submodel A also prevents binaries from surviving a CE event with a HG donor, just
as in submodel B. As a result, the binary populations formed in the two submodels are very
similar.

For $Z=\zsun$ the average chirp mass for NS-NS systems is
$1.1 \msun$, $3.2 \msun$ for BH-NS, and $6.5 \msun$ for BH-BH systems (for both 
submodels). In both submodels the distributions of chirp masses in the standard model 
and V1 are similar. The slight differences come from the fact that the population
of merging DCOs in V1 is composed of binaries that would form the non-merging 
population in the standard model. 
For $Z=0.1\zsun$ the corresponding values are $1.1 \msun$ for NS-NS and $3.6 \msun$ for BH-NS systems (for
both submodels). In the case of BH-BH systems the average chirp mass is $20.0 \msun$
for submodel A and $16.1 \msun$ for submodel B.
For $Z=0.1\zsun$, submodel A, the distribution of BH-BH chirp masses is flatter
in V1 than for the standard model. Specifically, there are fewer low chirp mass
systems in V1 than for the standard model. There are two mechanisms determining this
outcome. The first is the increased number of mergers during the CE in this variation, which
reduces the overall number of DCOs. The second is the fact that less massive BH-BH 
systems have relatively light progenitors. These binaries have smaller chances of 
ejecting such a strongly bound CE due to a smaller orbital energy 
reservoir, and merge in the process.

Since the populations in submodels A and B are similar, we find similar distributions 
of chirp masses and delay times. For $Z=\zsun$ the average chirp mass for NS-NS systems is 
$1.09 \msun$, $3.2 \msun$ for BH-NS, and $6.5 \msun$ for BH-BH systems. 
For $Z=0.1\zsun$ the values are $1.11 \msun$ for NS-NS, $3.6 \msun$ for BH-NS systems
for both submodels, while for the BH-BH systems the average chirp mass is $20.0 \msun$ for 
submodel A and $16.1 \msun$ for submodel B. The components of these 
BH-BH systems belong to the most massive compact objects, and this results in the 
high cutoff point of the chirp mass distribution in submodel B for $0.1\zsun$ ($\sim 
30 \msun$), when compared to other variations. Due to the very low $\lambda$ in V1, 
only very wide progenitor binaries can survive a CE phase. These binaries would form 
non-merging ($t_{\rm del}>10$ Gyr) BH-BH systems in the standard model. Since the 
progenitor binaries are very wide, the CE is typically initiated when the donor expands 
significantly. This happens at late evolutionary stages and/or for very massive 
donors; therefore, the donor in most cases has already evolved past the HG and
is in the CHeB phase. 
As most donors are CHeB stars, both submodels produce very similar
populations. Furthermore, since many progenitors are very massive stars, they produce very massive BH-BH 
systems in both submodels.

The average delay times for$\zsun$ are $\sim 52$\myr for NS-NS, $\sim 1.7$ Gyr
for BH-NS, and $\sim 1.5$ Gyr for BH-BH systems. For $0.1\zsun$ these are $\sim 82$\myr 
for NS-NS, $\sim 0.8$ Gyr for BH-NS for both submodels. For BH-BH systems
we find average delay $\sim 480$ Myr for submodel A and $\sim 680$ Myr for submodel B. 
The short delay times for NS-NS systems is a direct
consequence of the very low value of $\lambda$. The
high binding energy of the envelope following from this causes significant orbital energy 
dissipation. This prevents the formation of NS-NS systems with delay times
over $300$ Myr (see Fig.~\ref{atdel02var}--\ref{btdel002var}).
 
Due to the severe reduction in the total number of systems in this model, the
least populated group, BH-NS systems, is subject to larger statistics errors.
For example, for solar metallicity we find only $1$ merging BH-NS system in the V1
model. However, as we will see below, this model can be excluded based
on observations of known NS-NS Galactic systems, and therefore we do not
follow up with additional computations. 

\subsection{Variation 2} \label{var2}

In this model we fix $\lambda=0.1$. The rates for solar metallicity, for the DCOs
are: $11.8$--$1.1$ \pmyr for NS-NS, $2.4$--$0.08$ \pmyr for BH-NS, and $15.3$--$0.4$ \pmyr for 
BH-BH systems (submodel A--B). For sub-solar metallicity NS-NS ($65.9$--$6.9$ Myr$^{-1}$) and 
BH-BH systems ($56.7$--$16.1$ Myr$^{-1}$) strongly dominate the DCO merger rate, with relatively 
insignificant rates for BH-NS systems ($0.5$--$0.4$ Myr$^{-1}$). 

The most notable aspect of this variation is the high Galactic NS-NS merger
rate: $65.9 \myr^{-1}$ for submodel A and sub-solar metallicity. When compared to the standard 
model, these systems merge $\sim 8$ times more often (from $8.1 \myr^{-1}$ to $65.9 \myr^{-1}$). 
In this variation the $\lambda$ is smaller (by at least factor of $2$) than the one used in the 
standard model for a NS progenitor (typically a $10 \msun$ star; see Fig.~\ref{lam10}).
It means that the envelope is strongly bound in V2. As a consequence the binaries will end the CE 
with smaller orbital separations than they would in the standard model. This effect is 
additionally enhanced since NS-NS progenitors, for $Z=0.1\zsun$, often experience two
CE events (see NSNS01 in Tab.~\ref{channels2} in standard model; a similar
channel is also found in V2). A decreased semi-major axis increases the chances of survival 
through a SN.  The vast majority of disruptions occur corresponding to the first SN, when the systems are
still wide. For example, in the standard model $\sim 94\%$ of disruptions are encountered at 
the first SN, while $\sim 6\%$ at the second SN. In V2, there are virtually no disruptions at 
the second SN leading to the $8$-fold increase of merger rates. 
A secondary effect of the strongly bound envelope is the migration of the progenitors toward 
low merger times. This has two competing consequences. One is the increased number of mergers 
during CE and a moderate reduction of merger rates. The second is the efficient dissipation of 
orbital energy that allows wide progenitors (that in the standard model would produce non-merging
NS-NS) to form merging NS-NS systems moderately increasing the merger rates. 

For solar metallicity the effect of SN survival does not manifest itself because progenitors 
of NSs often experience ECSe (high wind mass loss reduces pre-SN progenitor mass), which 
generally do not disrupt binaries (no kick). However, for$\zsun$ there is a significant NS-NS 
merger rate drop in submodel B (from $7.6 \myr^{-1}$ in the standard model to $1.1 \myr^{-1}$ in this 
variation). 
The main formation channel in the standard model (see NSNS01 in Table \ref{channels})
involves a CE and non-conservative mass transfer after the first SN. In V2 after the
CE the orbital separation is smaller than in the standard model. Therefore the secondary, 
a low mass helium star that is a progenitor of the second NS, experiences RLOF much earlier 
in its evolution in V2. A combination of shorter orbital period and the earlier evolutionary 
stage (typically helium HG) leads to the development of the second CE in V2 (as opposed to 
stable mass transfer in the standard model). Since the survival through CE with HG donors 
is not allowed in model B, the merger rates significantly decrease. There is also a small 
decrease of merger rates in submodel A. Although HG CE is allowed in submodel A, small 
orbital separation in V2 sometimes leads to a merger in the second CE event. 

The most notable change in the merger rates of BH-NS systems is the drop for $Z=0.1\zsun$:
$3.4$--$0.5$ \pmyr (standard--this variation) for submodel A and $2.3$--$0.4$ \pmyr for
submodel B. This comes from the fact that, for low metallicity environments, the progenitors
of BH-NS systems quite often engage in a CE at the beginning of their evolution (see BHNS01 in
Tab.~\ref{channels2}). In this early CE event, the donor is the more massive component and a progenitor of
the BH. The $\lambda$ value used in this variation ($0.1$) is higher than in the standard
model for a BH progenitor (see Fig.~\ref{lam40},\ref{lam60}).
This means that the envelope is easier to eject and after the first CE the binaries will
not lose as much orbital energy remaining above the $10$ Gyr merger time. Additionally a
wider separation after the first CE increases the chances of disruption by the supernovae.
A combination of both effects accounts for the drop in BH-NS merger rates for both submodels
for $Z=0.1\zsun$. This does not manifest itself for the $Z=\zsun$ model. High wind mass loss rates
present at solar metallicity often reduce the initial mass ratio of the progenitors ($\gtrsim 
\frac{1}{2}$--$\frac{1}{3}$)
and the first interaction instead of being a CE is a non-conservative mass transfer (see BHNS01 in
Tab.~\ref{channels}). Since $\lambda$ plays role on in CE events, the BH-NS rates are comparable in
V2 and the standard model. The factor of $\sim 2$ difference arises from the later CE events initiated
by the NS progenitor.

For solar metallicity the BH-BH merger rates increase by a factor of $\sim 2$
(from $8.2 \myr^{-1}$ to $15.3 \myr^{-1}$) for submodel A, and decrease by factor of $\sim 5$ 
(from $1.9 \myr^{-1}$ to $0.4 \myr^{-1}$) for submodel B, in contrast with the standard model. 
The $\lambda$ value fixed in this variation (at $0.1$) is larger than the average 
\textit{Nanjing} values for typical BH progenitors (see Figs.~\ref{lam40} and~\ref{lam60}). 
High $\lambda$ and correspondingly low binding energy leads to small orbital contraction
during the CE phase. This can (counter-intuitively) both increase and decrease merger rates of 
various BH-BH populations. For example, consider BH-BH populations in submodel A and B. From the relative 
merger rates (A much higher than B) it is clear that submodel A is dominated by BH-BH 
binaries that evolved through the CE phase with a HG donor. In contrast, submodel B 
includes only systems that have formed through CE with a CHeB donor (by assumption). Since HG stars are 
smaller (for the same mass and metallicity) than CHeB stars, progenitors in 
submodel A have relatively small orbital separations, while in submodel B they have relatively wide 
separations. 
On one hand, short period progenitor binaries do not merge in the CE phase as often as in the standard 
model, since the $\lambda$ chosen in V2 is large, and that leads to an increased BH-BH 
merger rate in submodel A. 
On the other hand, long period progenitor binaries are not sufficiently contracted (in term
of separation) due to 
the high $\lambda$ in V2. They form long-period BH-BH systems with merger times exceeding 
$10$ Gyr, and thus the merger rate decreases in submodel B. 
For sub-solar metallicity the effects of an increased $\lambda$ are compensated
for by the smaller 
radial expansion of stars as compared to solar metallicity, and the V2 merger rates for both 
submodels are similar to the standard model rates. 

The distribution of chirp masses for both submodels and metallicities closely
resembles that of the standard model. The average chirp masses for $Z=\zsun$ are $1.08$--$1.06 \msun$ 
(submodel A--B) for NS-NS, $3.2$--$3.2 \msun$ for BH-NS, and $6.5$--$6.5 \msun$ for BH-BH systems.
For $Z=0.1\zsun$ the corresponding values are $1.09$--$1.07 \msun$ (submodel A--B) for NS-NS, $3.5$--$3.3 \msun$ 
for BH-NS, and $17.2$--$9.3 \msun$ for BH-BH systems.

The average values of delay times for$\zsun$ are: $90$--$618$ \myr (submodel A--B) for
NS-NS, $2.2$--$2.6$ Gyr for BH-NS, and $1.8$--$6.5$ Gyr for BH-BH systems. For $0.1\zsun$
the values are: $0.2$--$1.6$ Gyr (NS-NS), $1.6$--$1.7$ Gyr (BH-NS), $1.2$--$1.9$ Gyr (BH-BH 
systems).  

\subsection{Variation 3}

This variation fixes the binding to $\lambda=1$. The Galactic merger rates for$\zsun$ are 
dominated by NS-NS systems ($48.8$--$14.3$ Myr$^{-1}$, submodel A--B) followed by BH-BH 
($5.0$--$0.03$ \pmyr) and BH-NS systems ($4.6$--$0.03$ Myr$^{-1}$). For sub-solar metallicity, 
the BH-BH systems merge most often ($90.2$--$7.9$ Myr$^{-1}$, submodel A--B), then NS-NS
systems ($44.1$--$4.2$ Myr$^{-1}$) and BH-NS systems ($15.8$--$8.4$ \pmyr).

For solar metallicity there are no significant changes in merger rates for submodel A.
However, the rise in rates for NS-NS systems by a factor of $\sim 2$ when compared 
to the standard model (from $23.5$ \pmyr to $48.8$ Myr$^{-1}$) makes it the most frequently
merging population of double NSs, for$\zsun$, in this study. The increase of merger
rates for NS-NS and BH-NS systems (from $1.6$ \pmyr to $4.6$ Myr$^{-1}$) is associated 
with fewer binaries merging during the CE due to the high $\lambda$ value used in this 
variation. However, the high $\lambda$ values have the opposite effect for BH-BH systems.
The resulting low binding energy of the CE is responsible for the slight 
drop in merger rates for BH-BH systems (from $8.2$ \pmyr to $5.0$ Myr$^{-1}$), as fewer 
binaries cross the $10$ Gyr point (toward the merging population) due to insufficient 
orbital energy loss when ejecting their envelopes. 

The population of NS-NS systems in submodel B follows the behaviour described in submodel A.
What is noticeable for BH-NS and BH-BH systems in submodel B, at$\zsun$, is a
drop in the merger 
rates (by an order of magnitude with respect to the standard model). By assumption, submodel B 
favours DCOs which initiated a CE with a donor beyond the HG, in this case during CHeB. Donors of 
this type that are progenitors of components in BH-NS or BH-BH systems have
lower (see Figs.~\ref{lam20}--\ref{lam60}) $\lambda$ values in the standard
model (more physical) than in this variation 
($\lambda=1$). A direct consequence is the reduced orbital energy loss during a CE event.
This allows binaries with CHeB CE to retain a large separation at the end of the
CE phase, 
preventing a DCO merger within $10$ Gyr. Therefore, the population of merging DCOs containing 
BHs for submodel B is reduced.

For sub-solar metallicities the high $\lambda$ effect is still relevant. However,
in these chemical environments donors initiate CE more often as CHeB stars (instead 
of HG) due to smaller radial expansion during the evolution. This increases the number and
merger rates of NS-NS and BH-NS systems in submodel B. This also allows for more
BH-BH systems to be formed in this submodel but for this type of DCO the high $\lambda$
causes a counter-effect, and efficiently reduces the merger rates of double BHs. 

The average chirp masses for$\zsun$ are: $1.06$--$1.05 \msun$ (submodel A--B) for NS-NS, 
$2.7$--$2.4 \msun$ for BH-NS, and $6.0$--$5.9 \msun$ for BH-BH systems. For $0.1\zsun$ the
values are: $1.09$--$1.12 \msun$ for NS-NS, $2.9$--$2.9 \msun$ for BH-NS, and $12.5$--$6.8$ for 
BH-BH systems.

The average delay times for$\zsun$ are: $1.2$--$2.2$ Gyr (submodel A--B) for NS-NS, $2.0$--$2.7$
Gyr for BH-NS, and $4.2$--$2.2$ Gyr for BH-BH systems. The values for $0.1\zsun$ are: $0.9$--$2.4$
Gyr for NS-NS, $1.8$--$2.3$ Gyr for BH-NS, and $1.6$--$3.3$ Gyr for BH-BH systems. These average
values are greater than for the standard model, as the high $\lambda$ value used in this
variation (low binding energy) allows the binaries to retain a larger orbital separation 
after the CE phase.

\subsection{Variation 4} \label{v4}

In this model we fix the $\lambda$ value at $10$. The Galactic merger rates for$\zsun$ 
are: $20.8$--$0.3$ \pmyr (submodel A--B) for NS-NS, $0.9$--$0.0$ \pmyr for BH-NS, and $0.3$--$0.0$ \pmyr 
for BH-BH systems. The rates for $0.1\zsun$ are: $29.5$--$1.4$ \pmyr for NS-NS, $8.8$--$1.6$ \pmyr
for BH-NS, and $5.9$--$0.3$ \pmyr for BH-BH systems.  

For$\zsun$ the populations of DCOs of all types experience a reduction in the merger rate for 
both submodels (see Table \ref{gmerger02}) when compared with the standard model. The mechanism 
responsible for this result is straightforward. The very high $\lambda$ value means a low binding energy 
of a star's envelope, and so it is easily ejected during a CE event. As a consequence, little orbital energy 
is lost, and the separation of the components is reduced insignificantly. The weak
orbital contraction causes the DCOs to retain merger times larger than $10$ Gyr, thus creating a 
large non-merging population at the cost of the merging one. The mechanism 
described above holds true for all types of DCOs, but manifests most clearly for BH-BH systems. For 
example, for submodel A, for each BH-BH system formed below the $10$ Gyr merger time limit, $\sim 240$ 
are formed above it. For submodel B the rates are null for BH-NS and BH-BH
systems. In other words, despite high statistics---6 million binary stars simulated---
no DCO systems containing BHs were produced. In
order to understand this result one needs to recall  
that the CE phase is usually initiated by HG and CHeB donors, due to significant radial expansion during 
these stages of evolution. Submodel B does not allow for a HG donor so the only systems left to populate 
this submodel are those with a CHeB donor during the CE, in their evolutionary history. This evolved CE
donor indicates that these binaries had initial separations large enough to be able to bypass the CE 
on the HG, and this initial orbital condition places them in the long merger time regime. 
For progenitors of BH-NS and BH-BH systems the $\lambda$ value in the standard model during CHeB is 
much lower (see Figs.~\ref{lam20}--\ref{lam60}) when compared to the one used in this variation ($10$). 
So, due to insufficient orbital energy loss during CE, no binaries are drawn toward the merging 
population, making it void of system containing BHs. This migration of DCOs between the merging and 
non-merging populations is illustrated is Figure~\ref{mig}. 

For $0.1\zsun$ more merging BH-NS/BH systems are formed in general due to weaker SNe (see 
Sec.~\ref{stan}). The probability of producing such a system for the merging population increases, 
yielding non-zero merger rates for submodel B. Additionally, the merger rates for BH-NS system for 
submodel A increase by a factor of $\sim 3$ (from $3.4$ to $8.8$ Myr$^{-1}$) when contrasted with 
the standard model. This result is opposite to the one for$\zsun$, where the merger rate 
of BH-NS systems for submodel A drops slightly (from $1.6$ to $0.9$ \pmyr). A weakly bound CE (in this 
variation) causes insignificant orbital reduction and increases the chances of disruption by the 
following SNe (after the first or second CE, see Table \ref{channels2}). However the stalled SN kicks 
associated with low metallicity environments reduce the disruption probability, accounting for the 
increase in $0.1\zsun$ in V4.

For NS-NS systems the Galactic merger rates increase from $8.1 \myr^{-1}$ (standard model)
to $29.5 \myr^{-1}$ (this variation) at $0.1\zsun$. A characteristic of these DCOs for sub-solar
metallicities is the second CE initiated just after the first one (see Table~\ref{channels2}).
Two weakly bound CE phases allow the binary to avoid a merger
in either of these events, effectively increasing the 
number and merger rates of NS-NS systems. 

The average chirp masses for$\zsun$ are $\sim 1.03 \msun$ for both submodels for NS-NS systems.
For BH-NS and BH-BH systems, submodel A, these are: $2.5 \msun$ and $5.8 \msun$, respectively.
For $0.1\zsun$ these values are: $1.1$--$1.2 \msun$ (submodel A--B) for NS-NS, $2.9$--$2.7 \msun$ for
BH-NS, and $7.6$--$6.7 \msun$ for BH-BH systems.

The average delay time values for$\zsun$, for NS-NS systems are $0.5$ Gyr and $2.0$ Gyr, for 
submodel A and B, respectively. For BH-NS and BH-BH systems these are $2.5$ Gyr and $3.2$ Gyr, 
respectively, for submodel A. For $0.1\zsun$ these values are: $0.7$--$2.0$ Gyr (submodel A--B) for 
NS-NS, $2.4$--$2.2$ Gyr for BH-NS, and $3.8$--$2.7$ for BH-BH systems.

\subsection{Variation 5 \& 6}

In variations 5 and 6 we set the maximum mass achievable for a NS to $3.0 \msun$
and $2.0 \msun$, respectively (this is $2.5 \msun$ in the standard model). These models produce no 
statistically significant changes, when compared to the standard model. The reason for
the population of DCOs being insensitive to the maximum mass of a NS is the usage of the 
Rapid supernova mechanism. The main characteristic of this engine is the lack of 
compact objects formed in the range $\sim 2$--$5 \msun$. 

The distribution of delay times and chirp masses, as well as their average values, are 
nearly identical (within statistical accuracy) to the standard model.

\subsection{Variation 7}

In this variation we set the $\sigma$ of the Maxwellian distribution of the 
natal kicks to $132.5$ km/s (it is $265$ km/s in the standard model). The Galactic
merger rates for$\zsun$ are: $32.4$--$9.5$ \pmyr (submodel A--B) for NS-NS, $1.9$--$0.3$
\pmyr for BH-NS, and $10.4$--$2.1$ \pmyr for BH-BH systems. For sub-solar metallicity 
the rates are: $8.3$--$2.2$ \pmyr for NS-NS, $6.1$--$4.3$ \pmyr for BH-NS, and $83.7$--$15.1$
\pmyr. 

For solar metallicity we find slightly increased (by less than a factor of $\sim 2$) 
Galactic rates for all DCO types, for both submodels. In practice, reducing the $\sigma$
value means that natal kicks of high magnitude are chosen less frequently from the 
Maxwellian distribution. As a consequence fewer binaries are disrupted during the SN
explosions. This increases the pool of potential merging DCOs, and accounts for higher
merger rates in this variation. The least affected, when compared to the standard model,
are BH-BH systems, since the progenitors of these systems are massive stars,
and thus produce weak SNe (stalled or no kicks, direct collapse). Therefore, statistically, these
systems are minimally impacted by decreasing the occurrence of high magnitude natal kicks in
the binary population. 

For sub-solar metallicity this effect is even less significant, due to naturally
weaker explosions. Low wind mass loss rates associated with low metallicity 
environments produce more massive pre-SN progenitors. These stars cause more fallback of 
ejecta during the explosion, due to their strong gravitational potential. Therefore, the 
magnitude of the velocity of the kicks is in general smaller than for$\zsun$. Further
reduction of the occurrence of high magnitude kicks yields no statistically significant
effects. 

In the case of NS-NS systems for $0.1\zsun$ the rates are almost unchanged when compared
with the standard model. Note that most of these DCOs have two CE events (exactly like in 
the standard model, see Table \ref{channels2}), which makes them susceptible to
merging during one of these episodes. Reducing the probability of high velocity kicks 
results in smaller separations after the first SN. A consequence is an
increased number of CE mergers that counters increased SN survivability. 

The average chirp masses for$\zsun$ are: $1.05 \msun$ for NS-NS,
$3.1 \msun$ for BH-NS, and $6.6$ for BH-BH systems (all for both submodels). 
For $0.1\zsun$ the values are: $1.08 \msun$ for NS-NS and $3.1 \msun$ for BH-NS
systems (for both submodels). For BH-BH systems these are: $12.4 \msun$ for 
submodel A and $9.7 \msun$ for submodel B.

The average values of delay time distribution are: $1.3$--$2.2$ Gyr (submodel A--B) 
for NS-NS, $2.7$--$4.3$ Gyr for BH-NS, and $1.2$--$2.9$ Gyr for BH-BH systems. For
$0.1\zsun$ the values are: $1.0$--$2.7$ Gyr for NS-NS, $1.9$--$1.9$ Gyr for BH-NS
and $1.1$--$1.3$ Gyr for BH-BH systems. 

\subsection{Variation 8}

In this model we only enforce full natal kicks for the BHs. The
results therefore only change for BH-BH/NS systems. The Galactic merger
rates for$\zsun$ for BH-NS systems are $0.03$--$0.004$ \pmyr (submodel A--B).
For BH-BH systems: $0.05$--$0.005$ Myr$^{-1}$. For $0.1\zsun$ the rates are: $0.7$--$0.2$ \pmyr
for BH-NS and $4.2$--$0.8$ \pmyr for BH-BH systems.   

The disruption rate of binary progenitors of BH-BH/NS systems during the SN, due to the 
strong velocity kicks, is significant. When contrasted with the standard model, the Galactic 
merger rates for$\zsun$ drop by $2$--$3$ orders of magnitude (see Table \ref{gmerger02}). For 
example, for BH-BH, submodel B at$\zsun$, the merger rate drops from $1.9 \myr^{-1}$ (standard) 
to $0.005 \myr^{-1}$. For $0.1\zsun$ the effect is smaller but still noticeable, as the 
rates drop by a factor of $\sim 10$ (see Table \ref{gmerger002}). The pre-SN progenitors of 
BHs in this environment are, on average, more massive when compared to$\zsun$. This 
results in strongly bound binaries, which makes them less susceptible
  to being disrupted from natal kicks.

The average chirp masses for$\zsun$ are: $3.0$--$3.2 \msun$ (submodel A--B) for BH-NS systems. 
For BH-BH systems these are $6.5 \msun$ and
$5.9 \msun$ for submodels A and B, respectively. For $0.1\zsun$ these values are $1.09 \msun$
(both submodels) for NS-NS systems and $3.0 \msun$ (also both submodels) for BH-NS systems.
For BH-BH systems these are $9.0 \msun$ and $7.2 \msun$ for submodels A and B, respectively.

The mean delay times for$\zsun$ are: $1.2$--$1.7$ Gyr (submodel A--B) for NS-NS systems and
$2.0$--$7.0$ Gyr for BH-BH systems. For BH-NS, submodel A, the mean value is $0.4$ Gyr.
For $0.1\zsun$ the mean values are: $0.9$--$2.2$ Gyr (submodel A--B) for NS-NS, $1.2$--$1.3$ Gyr 
for BH-NS, and $0.6$--$1.2$ Gyr for BH-BH systems.

\subsection{Variation 9}

In this variation BHs do not receive natal kicks. We
therefore describe the results for BH-NS and BH-BH systems  
(NS-NS systems behave as in the standard model). The Galactic merger rates for BH-NS 
systems are $1.4$--$0.2$ \pmyr (submodel A--B) for$\zsun$ and $5.2$--$3.7$ \pmyr ($0.1\zsun$).
For BH-BH systems the rates are: $16.9$--$4.2$ \pmyr ($\zsun$) and $92.3$--$19.3$ \pmyr 
($0.1\zsun$).

For$\zsun$ the merger rates for BH-BH systems increase by a factor of $\sim 2$ for both
submodels, when compared to the standard model. This is a straightforward consequence of
the assumption of no natal kicks used in this variation. The survival rate of BH-BH/NS
systems increases due to fewer disruptions occurring during the SN. This allows more 
DCOs with BHs, which in turn increases the merger rates. 
For $0.1\zsun$ the gain is smaller. Pre-SN progenitors of BHs are heavier in this
chemical environment due to low wind mass loss. This causes a lot of ejecta fallback 
(almost all matter is pulled back onto the newly formed BH). Setting no natal kicks for 
these systems changes little as these BHs ``naturally" suppress the SN explosion.

The behaviour of BH-NS systems is explained analogously to BH-BH DCOs.
However the effects are smaller since the survival rates due to disruption are
dominated by SNe resulting in the formation of NSs. Pre-SN
progenitors of NSs are light and experience low fallback (stronger kicks). This
allows for more powerful explosions, which disrupt the binaries and prevent
the creation of DCOs.

The average chirp masses for BH-NS systems are: $3.1 \msun$ for $\zsun$ and 
$3.0 \msun$ for $0.1\zsun$ (all for both submodels). For BH-BH systems the
values are: $6.3$--$6.2 \msun$ (submodel A--B) for$\zsun$ and $12.1$--$9.3 \msun$ for 
$0.1\zsun$.

The mean values of delay times for BH-NS systems are: $1.7$--$3.0$ Gyr (submodel A--B)
for$\zsun$ and $1.7$--$1.7$ Gyr ($0.1\zsun$). For BH-BH systems these are: $1.5$--$3.4$ Gyr 
for$\zsun$ and $1.2$--$1.4$ Gyr for $0.1\zsun$
  
\subsection{Variation 10} \label{v10}

In this variation we use the Delayed supernova engine (as compared with the Rapid engine
used in the standard model). The Galactic merger rates for$\zsun$ are:
$25.6$--$8.9$ \pmyr (submodel A--B) for NS-NS, $0.07$--$0.03$ \pmyr for BH-NS, and
$0.6$--$0.08$ \pmyr for BH-BH systems. For $0.1\zsun$ the rates are: $8.6$--$2.6$
\pmyr for NS-NS, $2.3$--$2.0$ \pmyr for BH-NS, and $62.0$--$11.5$ \pmyr for BH-BH
systems.

For all metallicities and submodels we note a drop in the merger rates for systems 
containing BHs (see Table \ref{gmerger02} 
and \ref{gmerger002}). The reason for this is the difference in the amount of fallback 
matter acquired in SN explosions resulting from the two engines. To illustrate,
we present the results of an example calculation using Eq.$16$, $18$, and $19$ from 
\cite{chrisija}, estimating the fallback factor $f_{\rm fb}$. For a star prior to the SN explosion
with a total mass $M=8.7 \msun$ and carbon-oxygen core $M_{\rm CO}=6.5 \msun$ (typical core mass
collapsing to a BH), we get the following values of $f_{\rm fb}$: $1.0$ (full fallback) for the Rapid 
and $0.46$ (partial fallback) for the Delayed engine.
For progenitors of BHs the amount of fallback is lower (more mass is ejected)
in the case of the Delayed engine. This results in natal kicks of
a higher magnitude (see Eq.~\ref{vkick}), which are more capable of disrupting 
the binary and preventing the creation of a DCO.
For progenitors of NSs the fallback is similar in both V10 and the standard
model, and therefore we find similar
magnitudes of natal kicks, and a similar chance for disruption by SN. 

Another significant result produced by this model is the lower limit of the chirp
mass distribution for BH-BH systems: $\sim 2$ (for an example see Figure \ref{amch02var}).
The Delayed supernova engine creates remnant objects (NSs and BHs) with a continuous spectrum of 
masses, contrary to the Rapid engine (for 
details see \cite{massgap}). This suggests observational tests to distinguish
the models~\citep{chrisija,massgap}. 

The average chirp mass for$\zsun$ is: $1.14$--$1.13 \msun$ (submodel A--B) for NS-NS,
$3.1$--$3.0 \msun$ for BH-NS, and $5.7$--$4.6 \msun$ for BH-BH systems. For $0.1\zsun$
these are: $1.2$--$1.2 \msun$ for NS-NS, $3.4$--$3.4 \msun$ for BH-NS, and $14.4$--$10.2 \msun$
for BH-BH systems.

The mean delay time values for$\zsun$ are: $1.2$--$1.8$ Gyr (submodel A--B) for NS-NS,
$2.5$--$1.4$ Gyr for BH-NS, and $1.7$--$2.6$ Gyr for BH-BH systems. For $0.1\zsun$ these are:
$0.9$--$2.1$ Gyr for NS-NS, $1.8$--$1.7$ Gyr for BH-NS, and $1.2$--$1.5$ Gyr for BH-BH systems.    

\subsection{Variation 11}

In this variation we reduce the stellar wind mass loss rates by a factor of $2$.
The Galactic merger rates for$\zsun$ are: $24.2$--$6.5$ \pmyr (submodel A--B) for NS-NS,
$1.2$--$0.2$ \pmyr for BH-NS, and $29.7$--$3.6$ \pmyr for BH-BH systems. For $0.1\zsun$ the
rates are: $7.7$--$2.3$ \pmyr for NS-NS, $3.8$--$2.4$ \pmyr for BH-NS, and $79.2$--$17.1$ for 
BH-BH systems.

The most significant result due to this variation is the increase in the Galactic merger
rate for BH-BH systems for$\zsun$. The rates increase by a factor of $\sim 3$ and 
$\sim 2$ for submodels A and B, respectively. The effect of the lowered wind mass 
loss rates on double BHs is manifold. First, the decreased reduction of mass of BH
progenitors causes them to retain more mass just prior to the SN. This results in a 
larger amount of fallback matter during the explosion, which stalls the natal kicks.
This in turn increases the survivability rate of binaries and produces more merging
systems. Second, the increased pre-SN mass also results in more massive remnants, which is
clearly visible in Figure~\ref{amch02var} and \ref{bmch02var}. 
Third, the minimum mass of a BH progenitor is reduced to lower values (from the usual 
$\sim 20 \msun$ to $\sim 18 \msun$). For sub-solar environments
these effects are weak as the wind mass loss rate for $0.1\zsun$ is already low. Decreasing
this even further yields no new qualitative results.

For NS-NS systems these effects are also insignificant. The progenitors of NSs 
in these systems are lighter than those of BHs, and for these stars the wind mass loss rates 
are smaller than for their heavier counterparts. Therefore, decreasing wind mass loss even further plays
no significant role in the survivability or the properties of these systems. 

The merger rates for BH-NS systems do not increase in the way they do for BH-BH systems, despite 
their harboring a BH and a massive NS progenitor. Binary stars that would have formed BH-NS systems
in the case of ``standard" winds, now retain more mass during evolution (due to
the reduced winds),
and form BH-BH systems instead. This reduces the BH-NS population and adds
merging DCOs to the BH-BH systems.

The average chirp masses for $\zsun$ are: $1.05$--$1.05 \msun$ (submodel A--B) for NS-NS, 
$3.2$--$3.3 \msun$ for BH-NS, and $10.5$--$9.1 \msun$ for BH-BH systems. Note that for the latter
DCOs the range of chirp mass is the largest ($\sim 14 \msun$) for this metallicity of all
variations. For $0.1\zsun$ the average chirp masses are: $1.08$--$1.08 \msun$ for NS-NS, 
$3.3$--$3.4 \msun$ for BH-NS, and $14.3$--$10.6 \msun$ for BH-BH systems.

The mean delay times for $\zsun$ are: $1.0$--$1.6$ Gyr (submodel A--B) for NS-NS, $1.3$--$1.9$ Gyr
for BH-NS, and $0.6$--$2.8$ Gyr for BH-BH systems. For $0.1\zsun$ these values are: $1.0$--$2.3$ Gyr
for NS-NS, $1.5$--$1.5$ Gyr for BH-NS, and $1.0$--$1.2$ Gyr for BH-BH systems.

\subsection{Variation 12}

In this section we investigate the model with a fully conservative mass transfer
through RLOF (see Eq.~\ref{mteq}). In this model we assume that the entire transferred mass is
attached to the accretor. 
The Galactic merger rates for $\zsun$ are: $77.4$--$0.3$ (submodel A--B) \pmyr for NS-NS,
$0.06$--$0.02$ \pmyr for BH-NS and $8.9$--$1.6$ \pmyr for BH-BH systems. For $0.1\zsun$
the rates are: $17.1$--$4.4$ \pmyr for NS-NS, $4.1$--$3.0$ \pmyr for BH-NS and $68.8$--$6.6$
\pmyr for BH-BH systems.

For $\zsun$ this model yields the most numerous and therefore, the most frequent merging
population of NS-NS systems for submodel A in this study. The increase in merger rates by
a factor of $\sim 3-4$, when contrasted with the standard model, is primarily associated with the
first occurrence of mass transfer through RLOF. In this variation, in the outcome of RLOF,
the secondary component (initially less massive) gains more mass from the primary component 
(initially more massive) as compared with the same event in the standard model. In the latter
only half of the donated mass is accreted. This additional gain of mass for the secondaries
can alter their evolution in a relevant way. Now, (in this variation) many secondary
components which normally underwent the RGB phase (in the standard model), skip this evolutionary
period. The additional mass increases the core temperature of the star and allows it to begin
Helium burning in the core (CHeB star) immediately after the HG. Note that RGB and CHeB phases
are where the secondary components often initiate the common envelope, however
with different
$\lambda$ values ($\sim 0.5$ for RGB and $\sim 0.3$ for CHeB, see Fig.~\ref{lam10}). In 
consequence, the outcome of the CE for a binary with a CHeB donor will result in a tighter
separation than with an RGB donor. In the latter case, which often occurs in the standard model,
the binaries undergo an additional mass transfer from the secondary (see channel NSNS01 in 
Table \ref{channels}) after the CE. The resulting separation is often insufficient to prevent
the disruption by the following second SN. However, if the CHeB star is the donor, which 
is often the case in this variation, NS-NS progenitors will engage in a second CE immediately 
after the first one, due to the smaller separation. The outcome of two consecutive CEs is an
orbit tight enough (small separation) to increase the chances of survival of the 
subsequent second SN. In other words the NS-NS systems that form the merging population 
in this variation originate
from the population of disrupted systems in the standard model. 
A secondary effect of 
fully conservative mass transfer is that stars with mass that is close to
borderline between white dwarf and neutron star often become the latter. 
Overall more NS-NS are formed relative to white dwarf-neutron star systems
in this model. 
For submodel A the NS-NS rates increase due to the aforementioned evolutionary scenarios 
with two consecutive CEs (e.g.,NSNS03). It is noted that the second CE is initiated by an 
evolved Helium star. 
Submodel B does not allow for the survival of the CE with evolved Helium donors. 
Therefore, the merger rates drop drastically (from $7.6$ to $0.3$ \pmyr) for 
submodel B in this variation, when contrasted with the standard model. 

For $0.1\zsun$, in this variation the NS-NS systems arrive from a different population of 
progenitors than the ones in the standard model. 
In the standard model many progenitors that engage in a stable mass 
transfer at first contact do not survive the following first supernova explosion. However, 
in this variation the fully conservative mass transfer forces the stars to engage in CE upon first
contact. The full mass transfer causes the Roche lobe around the donor to shrink rapidly
due to the companion becoming more massive and the transfer starts to be unstable.
The reduction of the separation between the components in the CE outcome allows
for the survival of the first SN and eventually the formation of a merging DCO.

The population of $\zsun$ BH-NS systems is severely reduced in this variation when compared
to the standard model. In the standard, BH-NS progenitors evolved mostly through a stable mass
transfer upon first contact (see BHNS01, Table \ref{channels}). However, in this variation
the fully conservative mass transfer almost immediately forces the binary into a CE. The outcome
is usually a coalescence of the components and further binary evolution is terminated. 
Additionally to the above, many binaries that formed BH-NS systems in the standard model now 
form BH-BH systems. The fully conservative mass transfer allows the secondary component to 
accumulate more mass during the first 
contact. This gain allows the component to cross the threshold for neutron star/black hole formation
and become the latter. For $0.1\zsun$ the populations of BH-NS in the standard model and this 
variation are very similar and so are the merger rates in both models.

The populations of BH-BH systems in the standard model and this variation are also very similar
for both solar and sub-solar metallicities. What stands out is the lower merger rate for submodel B
for $0.1\zsun$ in this variation (by a factor of $\sim 2$). Again the first contact in the binary
is crucial at this point. The primary component initiates a stable mass transfer and its fully 
conservative character (in this variation) allows the secondary to gain more mass than in the standard model.
Later, the secondary component will expand more significantly due to its higher mass. Therefore, it
will be more likely to initiate CE while on the Hertzsprung gap rather than at later stages of 
evolution. Submodel B does not allow for the survival of CEs with HG donors, thus the merger rates
are lower.

The average chirp masses for $\zsun$ are: $1.07$--$1.04 \msun$ (submodel A--B) for NS-NS,
$2.9$--$2.9 \msun$ for BH-NS, and $6.3$--$6.3 \msun$ for BH-BH systems. 
For $0.1\zsun$ the average chirp masses are: $1.06$--$1.18 \msun$ for NS-NS,
$3.1$--$3.2 \msun$ for BH-NS, and $14.4$--$10.1 \msun$ for BH-BH systems.

The mean delay times for $\zsun$ are: $0.05$--$1.8$ Gyr (submodel A--B) for NS-NS, $2.4$--$4.1$ Gyr
for BH-NS, and $1.6$--$3.3$ Gyr for BH-BH systems. For $0.1\zsun$ these values are: $0.5$--$2.1$ Gyr
for NS-NS, $1.9$--$1.6$ Gyr for BH-NS, and $1.1$--$1.7$ Gyr for BH-BH systems.

\subsection{Variation 13}
In this section we investigate the model with a fully non-conservative mass transfer
through RLOF (see Eq.~\ref{mteq}).
The Galactic merger rates for $\zsun$ are: $26.1$--$6.2$ \pmyr 
(submodel A--B) for NS-NS, $10.6$--$3.9$ \pmyr for BH-NS and $5.8$--$0.5$ \pmyr for 
BH-BH systems. For $0.1\zsun$ the Galactic merger rates are $5.9$--$1.4$ \pmyr for NS-NS,
$33.0$--$30.1$ \pmyr for BH-NS and $39.0$-$28.9$ \pmyr for BH-BH systems.

For $\zsun$ the most significant difference occurs for BH-NS systems. When compared
with the standard model the rates increase by an order of magnitude for both submodels.
The reason for this is the character of the mass transfer at first contact in the binary.
The first mass transfer through RLOF occurs when the primary component (initially more massive)
evolves off the Main Sequence, expands rapidly on the Hertzsprung gap and overfills its 
Roche lobe. In the standard model (half-conservative character) the resulting mass transfer 
is often unstable, leading to a CE event. As the outcome the components are so close 
to each other that a merger occurs at the first passage of the periastron. 
However, in this 
variation, this first mass transfer is often stable due to its non-conservative character.
The reason is that the Roche lobe around the donor will not decrease as rapidly as in the standard
model (where the accretor gains more mass). So in this case the lack of a CE allows for an 
undisturbed evolution of the binary at this stage, thus yielding more DCOs in the end.
The same mechanism acts for $0.1\zsun$ and increases the Galactic merger rates by a factor
of $\sim 10$ in this variation when compared with the standard model.

The NS-NS systems experience a drop in Galactic merger rates by a factor of $\sim 2$ for 
both submodels. This is caused by a simple fact that in a non-conservative mass transfer
the accretor does not gain any mass from the donor. The most significant stable RLOF event
(in terms of the amount of mass flow) is the one at first contact, where the primary component
is the donor. If the secondary component is deprived of mass at this point, then it cannot
grow enough to become a NS progenitor and produces a white dwarf remnant instead. In other words
this variation yields NS-WD more often when contrasted with the standard model.
For $0.1\zsun$ this mechanism is still present however, it is somewhat suppressed by the low wind mass
loss caused by the low metallicity composition of the stars. In effect more stars are
able to retain enough mass to create an NS remnant rather than a WD. The result is a slightly
lower Galactic merger rate for both submodels when compared to the standard model.

For $\zsun$ BH-BH systems experience a similar effect  as the aforementioned NS-NS systems. 
At first contact the secondary component does not gain any mass (non-conservative mass transfer)
and is more likely to become a NS instead of a BH. Therefore there are more BH-NS systems 
created at the cost of BH-BH systems. 
Additionally, secondary stars that make BHs are lighter than in the standard model. 
Light BHs receive high natal kicks and are subject to disruption. 
Both effects cause a slight decrease in Galactic merger rates for $\zsun$ in both submodels. 

For $0.1\zsun$ BH-BH systems experience the same effects resulting from the secondary 
component gaining no mass at first contact. This produces BH-NS binaries
(instead of BH-BH binaries in standard model) and makes stars smaller.
Additionally, as compared to models with solar metallicity, expansion of stars is slower 
for $0.1\zsun$. Therefore, secondaries typically initiate CE late on the CHeB. Typical 
values of $\lambda$ for BH progenitors at CHeB are lower (higher binding energy) than 
for HG stages (see Fig.~\ref{lam60}). It means that the envelope will be more difficult to 
expel in this variation for the low metallicity and a premature CE merger is more likely to 
occur. The formation of BH-NS systems (instead of BH-BH) and additional
merging are two effects that reduce the overall number of BH-BH mergers for
$0.1\zsun$, submodel A.
However, in submodel B the rates are high when compared with the standard model ($28.9$ \pmyr
in this variation and $13.6$ \pmyr in the standard, see Table \ref{gmerger002}). This comes
from the fact that secondary components bypass the CE event while on the HG
and initiate it later (while on CHeB). Despite the aforementioned fact that often occurring
CE mergers reduce the overall number of BH-BH systems in this variation, the Galactic merger
rates in submodel B increase. Submodel B does not allow for a CE with a HG donor
so the increased presence of CHeB donors fuels the growth of the population of BH-BH systems
in this submodel.

The average chirp masses for $\zsun$ are: $1.02$--$1.0 \msun$ (submodel A--B) for NS-NS,
$2.7$--$2.6 \msun$ for BH-NS, and $6.1$--$6.3 \msun$ for BH-BH systems.
For $0.1\zsun$ the average chirp masses are: $1.1$--$1.1 \msun$ for NS-NS,
$3.1$--$3.1 \msun$ for BH-NS, and $9.7$--$9.4 \msun$ for BH-BH systems.

The mean delay times for $\zsun$ are: $0.6$--$2.0$ Gyr (submodel A--B) for NS-NS, $1.9$--$2.2$ Gyr
for BH-NS, and $2.4$--$4.2$ Gyr for BH-BH systems. For $0.1\zsun$ these values are: $0.5$--$1.9$ Gyr
for NS-NS, $1.4$--$1.4$ Gyr for BH-NS, and $0.9$--$1.0$ Gyr for BH-BH systems.

\subsection{Variation 14}
In this section we investigate the common envelope parameter $\lambda$, however we 
approach it differently as in Variations 1--4. Here we allow $\lambda$ to vary 
physically, according to the \textit{Nanjing} prescription but we multiply its value
by a factor of $5$ (following \cite{entalpia}).

The Galactic merger rates for $\zsun$ are: $28.2$--$3.7$ \pmyr (submodel A--B) for NS-NS, 
$3.4$--$0.05$ \pmyr for BH-NS, and $23.0$--$0.07$ for BH-BH systems. For $0.1\zsun$ the rates
are $47.0$--$1.0$ \pmyr for NS-NS, $15.5$--$5.7$ \pmyr for BH-NS and $90.5$--$14.9$ \pmyr for 
BH-BH systems.

For $\zsun$ we can see an increase in Galactic merger rates for submodel A and a decrease
for submodel B for all DCO types, when contrasted with the standard model. This result is 
strongly correlated to the initial parameters of binaries that form the DCO populations of both
submodels. For example binaries in submodel B are those that initiated CE when the donor was
at later stages of evolution (past the Hertzsprung gap). They were allowed to bypass the CE on
the HG because of sufficiently large initial separation. In general these are wide binaries
at the risk of not losing enough orbital energy to achieve merger times lower than $10$ Gyr. 
On the other hand stars that populate submodel A have on average smaller initial separations.
This means they can enter CE on the Hertzsprung gap, which is often the case. However, these
binaries are close (in terms of separation) and are at risk of merging during CE. Increasing
$\lambda$ by a factor of $5$ means that the binding energy of the envelope is this many times
weaker and easier to expel. 

For $0.1\zsun$ there is an overall increase in Galactic merger rates for all types of DCOs
in both submodels, with the exception of NS-NS systems for submodel B. For sub-solar metallicities
the aforementioned initial separation conditions do not determine as strongly the CE onset time.
As described in \cite{nasza} low metallicity HG stars do not expand to the extent of their high 
metallicity counterparts. This means that even relatively close binaries may initiate CE
at later stages of evolution. Boosting $\lambda$ is this case will primarily prevent mergers
during CEs and therefore increase Galactic DCO merger rates. 
This explanation in not applicable do NS-NS systems as the progenitors of these DCOs are 
relatively low mass stars and are not as prone to metallicity related expansion effects. In
other words they behave similarly to their $\zsun$ counterparts. 

The average chirp masses for $\zsun$ are: $1.07$--$1.05 \msun$ (submodel A--B) for NS-NS,
$2.9$--$2.8 \msun$ for BH-NS, and $6.4$--$6.0 \msun$ for BH-BH systems.
For $0.1\zsun$ the average chirp masses are: $1.10$--$1.26 \msun$ for NS-NS,
$2.9$--$2.8 \msun$ for BH-NS, and $14.6$--$6.7 \msun$ for BH-BH systems.

The mean delay times for $\zsun$ are: $0.8$--$1.8$ Gyr (submodel A--B) for NS-NS, $2.4$--$3.0$ Gyr
for BH-NS, and $3.0$--$2.0$ Gyr for BH-BH systems. For $0.1\zsun$ these values are: $0.4$--$2.2$ Gyr
for NS-NS, $2.0$--$2.4$ Gyr for BH-NS, and $1.1$--$2.1$ Gyr for BH-BH systems.

\subsection{Variation 15}
In this section, similarly to the previous one, we allow the $\lambda$ parameter to vary
physically, accordingly to the \textit{Nanjing} prescription (see Section \ref{newlam}) but
we decrease its value by a factor of $5$. This is done to account for the problem with defining
the core-envelope boundary, that is an extra source of uncertainty in the
DCO formation (Thomas Tauris, private communication).

The Galactic merger rates for $\zsun$ are: $39.8$--$17.8$ \pmyr (submodel A--B) for NS-NS, 
$0.01$-$0.007$ \pmyr for BH-NS, and $1.1$-$1.0$ \pmyr for BH-BH systems. The rates for 
$0.1\zsun$ are: $54.4$--$7.8$ \pmyr for NS-NS, $0.4$--$0.3$ \pmyr for BH-NS and $21.7$--$10.2$ 
\pmyr for BH-BH systems.

The value of the common envelope parameter $\lambda$ used in this variation is very low.
This means that the binding energy of the donor's envelope is very high. In consequence,
binaries will lose large amounts of orbital energy in order to expel the envelope. This causes
two major effects. The first is that binaries that would normally (in the standard model) form close 
DCOs will often merge during the CE. The second is that wide binaries (that would form 
non-coalescing DCOs in the standard model) now lose enough orbital energy to be able to
merge as DCOs within Hubble time. For solar and sub-solar metallicities the impact of the very low
$\lambda$ value is similar. For the population of NS-NS systems the second effect dominates, 
which causes an increase in merger rates for both submodels when compared to the standard model. 
However, for systems containing BHs the premature CE merger occurs very often. This causes a 
drastic decrease in merger rates of BH-NS systems for both submodels when contrasted with the 
standard model. This effect for BH-BH systems is still significant but to a smaller extent. 
Due to the  overall high mass, black hole progenitors may overcome the high binding energy of
the envelope during CE. This allows the population to preserve some of its members.

The average chirp masses for $\zsun$ are: $1.07$--$1.05 \msun$ (submodel A--B) for NS-NS,
$3.2$--$3.2 \msun$ for BH-NS, and $6.5$--$6.5 \msun$ for BH-BH systems.
For $0.1\zsun$ the average chirp masses are: $1.09$--$1.06 \msun$ for NS-NS,
$3.7$--$3.7 \msun$ for BH-NS, and $15.0$--$11.8 \msun$ for BH-BH systems.

The mean delay times for $\zsun$ are: $0.8$--$1.6$ Gyr (submodel A--B) for NS-NS, $0.04$--$0.04$ Gyr
for BH-NS, and $1.5$--$1.5$ Gyr for BH-BH systems. For $0.1\zsun$ these values are: $0.4$--$2.1$ Gyr
for NS-NS, $2.0$--$2.0$ Gyr for BH-NS, and $0.5$--$0.8$ Gyr for BH-BH systems.

\section{Comparison with the SeBa code} \label{compseba}
In this section we compare our findings with those acquired with another population 
synthesis code -- SeBa (e.g. \cite{seba2006,seba}).

{\tt Startrack} and SeBa population synthesis codes have significantly
different input physics. {\tt Startrack} uses metallicity dependent wind mass 
loss rates, while SeBa utilizes only solar metallicity models. To account for CE events 
{\tt Startrack} uses physical values of the $\lambda$ parameter (depending on the donor's 
mass, radius, evolutionary stage, metallicity, etc., see Section \ref{newlam}); SeBa uses 
fixed values of $\lambda$, which are at some level unphysical. Additionally {\tt Startrack} 
accounts for the possibility that CE events with HG donors may always lead to mergers due 
to the undeveloped core of the donor, while SeBa does not. Another major difference is in 
the physics describing SN explosion and the mass of the remnant: {\tt Startrack} utilizes 
the most recent supernovae calculations \citep{chrisija}, while SeBa utilizes the theoretical 
predictions by \cite{bhmassdis}. Both codes use similar natal kicks acquired by compact objects:
for NSs {\tt Startrack} uses a Maxwellian distribution of kicks with the $\sigma=265$ km/s, while
in SeBa the kick velocity ($v$) is drawn from a Paczynski distribution \citep{paczkick} divided 
by a $\sigma$ factor equal $300$ km/s, the final velocity being $u=v/ \sigma$.
For BHs {\tt Startrack} uses the same kick distribution as for NSs however, scaled down linearly
by the amount of ejecta falling back on the BH (see Eq.~\ref{vkick}); in SeBa the
kick is similar as for NSs however, scaled down by a factor dependent on the
mass of the BH.

In the most recent paper \citep{seba} utilizing the SeBa code, the authors
calculate the gravitational wave background generated by BH-BH systems. However, only
solar metallicity results are presented. The model used in this study that is the 
closest in terms of input physics to the results present in the aforementioned paper 
is the unphysical Variation 3, submodel A. In the paper, the authors present Galactic 
merger rates (Table III, therein) for all types of DCOs (NS-NS, BH-NS, BH-BH), which 
allows for an immediate comparison. When contrasted with our results the rates for NS-NS 
systems are: $20$--SeBa, vs. $48.8$--{\tt StarTrack} Myr$^{-1}$. Rates for BH-NS systems 
are slightly higher for SeBa ($6.2$ Myr$^{-1}$) than the ones obtained with
our code ($4.6$ Myr$^{-1}$). The rates for BH-BH systems are: $1.8$--SeBa,
vs. $8.2$--{\tt StarTrack} Myr$^{-1}$.

The masses of coalescing BH-BH systems are also apart in both models (SeBa model and Variation 3).
The mass range of the primary (initially more massive component) is $6\msun$--$10\msun$ in 
V3 (not listed in this paper) and $10\msun$--$18\msun$ for SeBa (Fig. 2 in \cite{seba}).
The mass range of the secondary (initially less massive): $5\msun$--$10\msun$ in V3 and
$10\msun$-$12\msun$ for SeBa. It seems then, that although SeBa predicts smaller BH-BH
rates, the employed in SeBa input physics makes BHs more massive then in {\tt StarTrack}.
It seems quite possible that these both effects cancel out and the resulting
LIGO/VIRGO detection rates will come out very similar at least if compared
with our unphysical models.

\section{Summary}
\label{dis}

We have investigated the major parameters and input physics involved in binary evolution 
leading to the formation of double compact objects: the CE coefficient 
$\lambda$, the supernova mechanisms (remnant formation and natal kick
magnitude), the maximum mass 
of NSs, and the wind mass loss rates. The study was performed by calculating a suite 
of population synthesis models, allowing us to estimate the associated uncertainties 
in the formation of DCOs. 

The calculations were done using the {\tt StarTrack} population synthesis code, recently 
updated with wind mass loss rates for massive stars, 
a realistic CE treatment, and convection-driven neutrino-enhanced supernovae.
These updates are incorporated in our revised Standard Model (see Section \ref{stan})
of binary evolution and DCO formation. Our newer version  yields lower merger rates than the equivalent
Standard Model from~\citet{comprehensive}. This difference arises mainly from the \textit{Nanjing} 
treatment of the CE parameter, $\lambda$ (see Section
\ref{newlam}; \citet{chlam}),
which we have now incorporated.
Significant changes in the merger rates  are also caused by varying the 
value of $\lambda$ (Variations 1--4, 14 and 15, see Table \ref{gmerger02} and \ref{gmerger002}). 
We therefore identify a strong dependence of the merger rates on  
the binding energy of the CE. 

Emerging from the simulations are two populations of DCOs.
The first is that of double compact objects with merger times less than $10$ Gyr (merging
in a Milky Way-like galaxy, the \textit{merging population}). The second is of objects 
having merger times higher than this limit (distant future mergers, the \textit{non-merging
population}). The $\lambda$ parameter shifts DCOs between these two populations. 
On the one hand, lowering $\lambda$ increases the CE binding energy, and thus tends to draw
the binaries from the non-merging population toward the merging one, as
progenitor binaries must lose
more orbital energy to unbind the envelope. This results in a larger
number of system in the merging population, and increases the Galactic merger rates. 
On the other hand, this also causes progenitors of DCOs from the merging population
to merge during the CE more often, terminating binary evolution and therefore
reducing the number of binary systems, and thus merger rates. The final outcome is set by
the specific value of $\lambda$, and varies from population to population. For
example, a given change in $\lambda$ may cause an
increase in the merger rate of BH-BH systems while decreasing the rate of
NS-NS systems, due to the differing orbital separations of BH-BH and NS-NS
progenitors.

The types of DCOs most affected by this mechanism are NS-NS and BH-NS systems. The range 
of merger rates for these systems spans three orders of magnitude between solar and sub-solar 
metallicities (see Table \ref{gmerger02} and Table \ref{gmerger002}).
The large range of merger rates obtained in this and previous studies using
calculations with a fixed value of $\lambda$ is an artifact of the unphysical
treatment of the CE phase. When a more physically motivated  treatment is applied, in which the
value of $\lambda$ is chosen to correspond to the given star and its evolutionary
stage, it is found that the merger rates are on the high side of the ranges
computed with variations that assume non-physical, fixed values of $\lambda$.
However, despite having a physical description of this parameter, 
the details of the common envelope event are still elusive. First, it is not
known how and with what efficiency the orbital energy is transferred into the
envelope. Second, we lack the full description  of the core-envelope boundary. 
We have attempted to manipulate the value of $\lambda$ to estimate a potential effect
of these unknowns on the DCO merger rates.
The most significant manifestation of this exercises is a factor of $\sim 10$--$100$ 
reduction in Galactic merger rates for systems containing a BH (see Tables \ref{gmerger02} 
and \ref{gmerger002}).

In particular, for the standard model, for either choice of metallicity of submodel (A or B), 
the BH-BH Galactic merger rate is at least $1.9$ per Myr.  This is a factor of $\sim 5$ higher 
than the ``realistic'' estimate used in \cite{abadie2010}, which translates into a detection rate 
of $20$ BH-BH mergers per year with Advanced LIGO and Virgo detectors in that paper.  Therefore, 
if the standard model is correct, we may expect that advanced gravitational-wave detectors 
will be able to capture above a hundred BH-BH coalescences per year.  This is a clearly positive 
prediction from the perspective of the ongoing searches for gravitational-wave signals. However, 
we are unable to provide error estimate on this improvement, since there are no 
alternative physical estimates for $\lambda$ readily available to us. Therefore,
one should treat this estimate with some caution. 

\begin{figure}
\includegraphics[width=1.0\columnwidth]{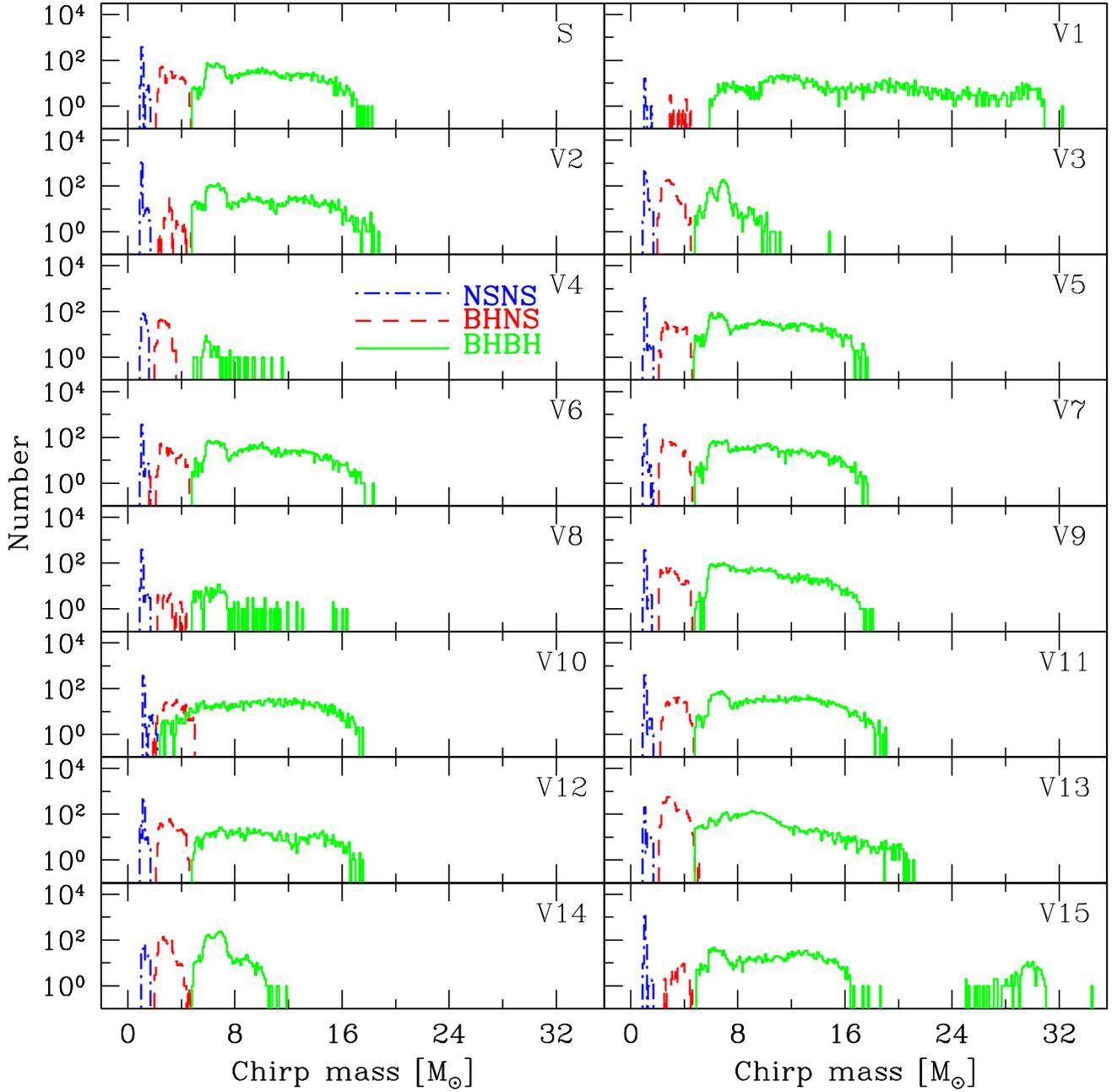}
\caption{
The chirp mass distribution of coalescing double compact objects for all 
variations for submodel B and $Z=0.1\zsun$. 
The maximum chirp mass for submodel B 
typically reaches only $\sim 15 \msun$, as contrasted with $\sim 30 \msun$ for submodel
A (see Fig.~\ref{amch002var}).
The reason why the V1 model allows for chirp mass as high as $\sim 30 \msun$, even for
submodel B, is explained in Sec.~\ref{v1}. For the same reason V15 harbours alarge chirp 
mass range.
}
\label{bmch002var}
\end{figure}

\begin{figure}
\includegraphics[width=1.0\columnwidth]{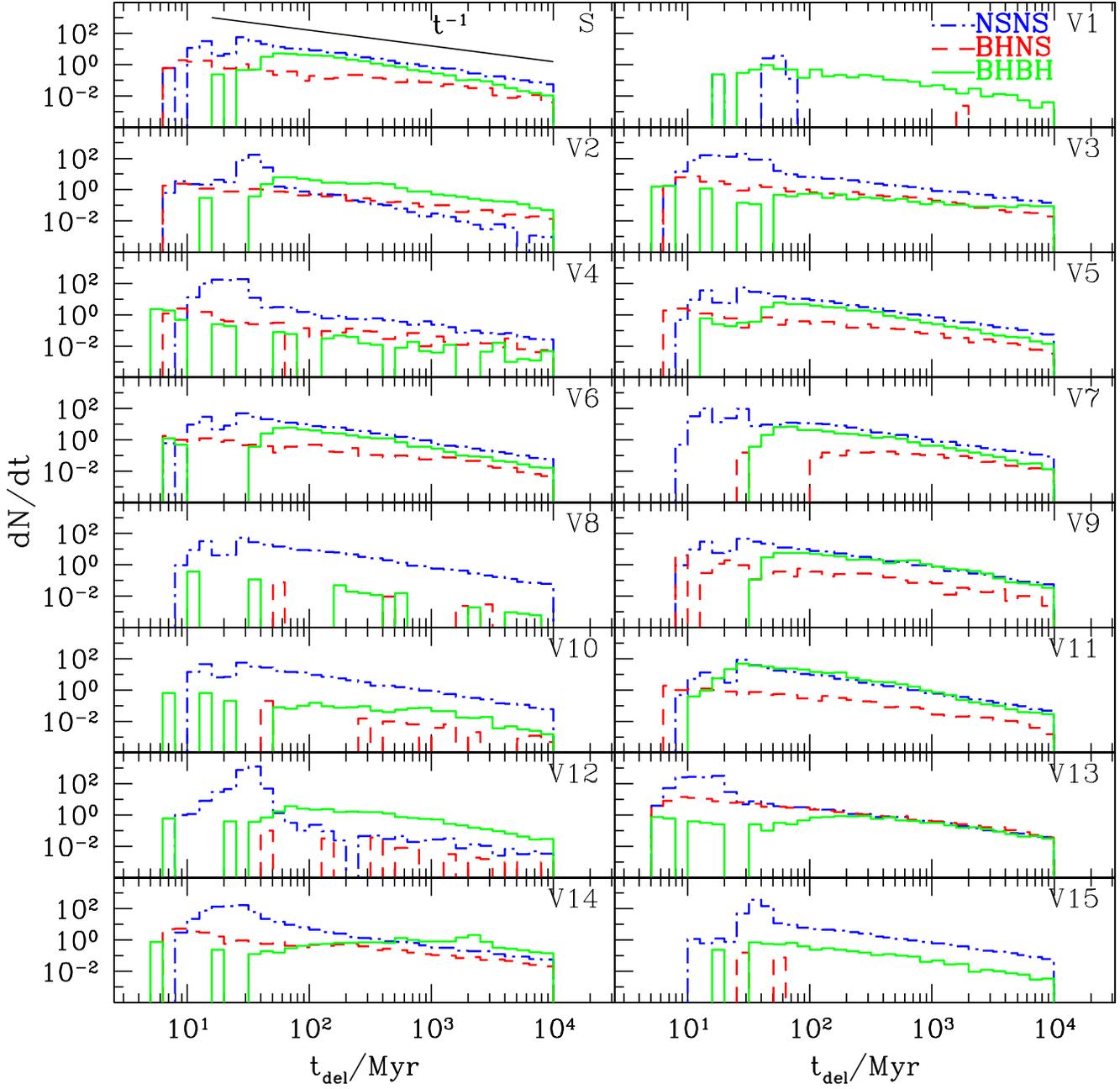}
\caption{
The delay time distribution of coalescing double compact objects for all 
variations for submodel A and $Z=\zsun$. Note that especially at
later times the number of sources typically falls off as $\propto t_{\rm del}^{-1}$. 
}
\label{atdel02var}
\end{figure}

\begin{figure}
\includegraphics[width=1.0\columnwidth]{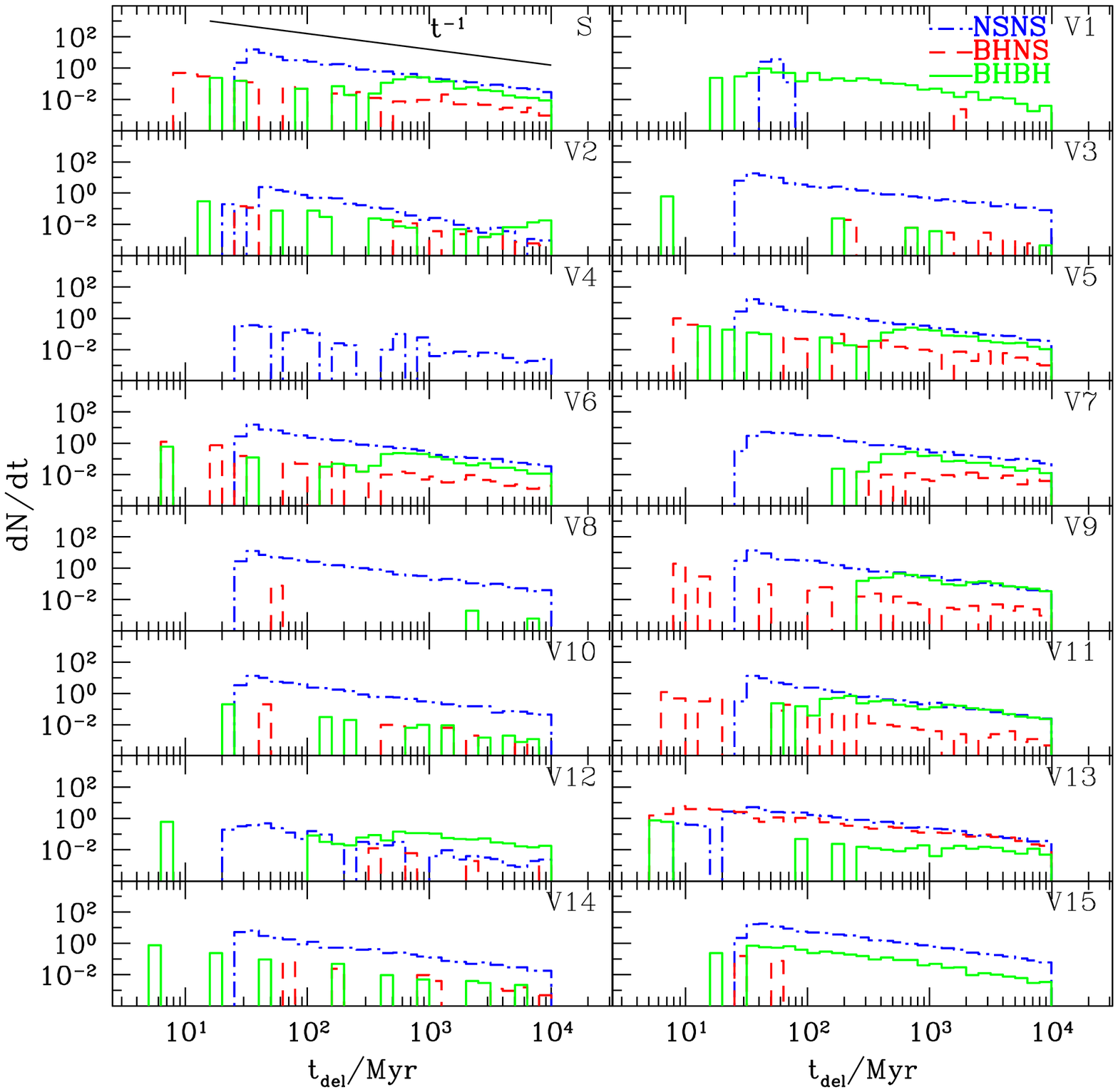}
\caption{Same as Fig.~\ref{atdel02var} but for submodel B.}
\label{btdel02var}
\end{figure}

\begin{figure}
\includegraphics[width=1.0\columnwidth]{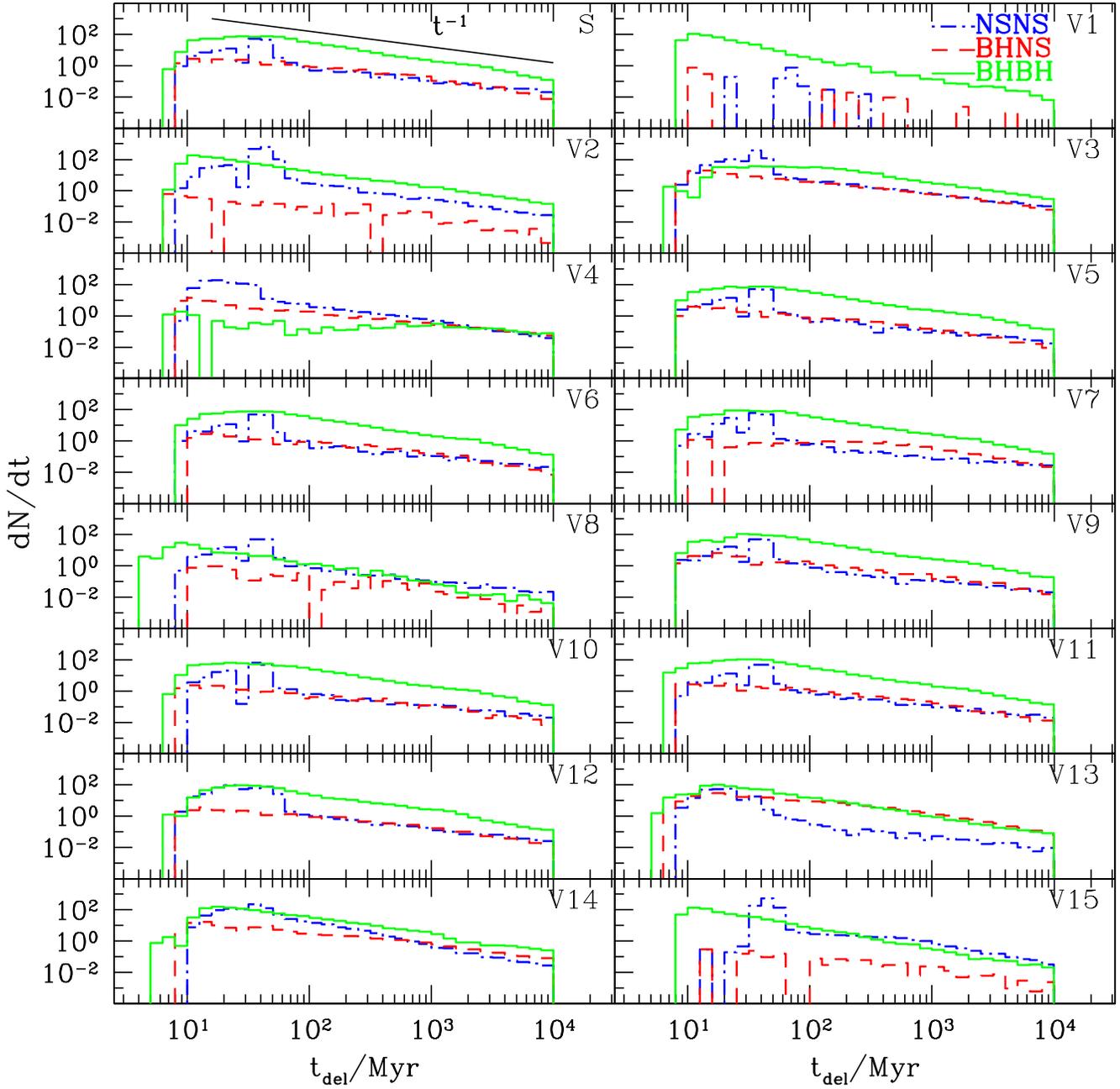}
\caption{Same as Fig.~\ref{atdel02var} but for $0.1\zsun$. The average delay 
time for systems containing BHs decreases with decreasing metallicity as 
the components become more massive.
}
\label{atdel002var}
\end{figure}

\begin{figure}
\includegraphics[width=1.0\columnwidth]{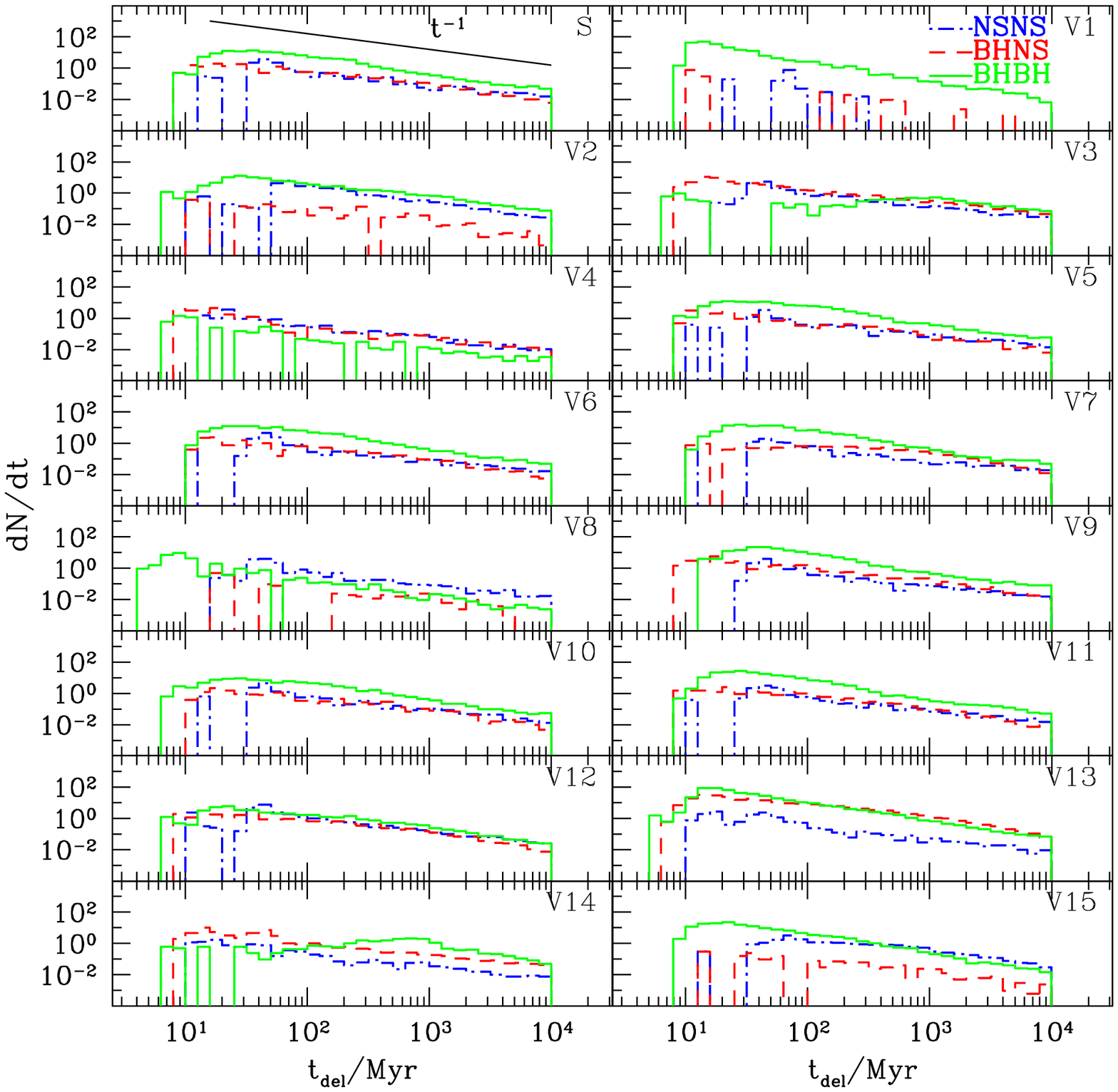}
\caption{Same as Fig.~\ref{atdel002var} but for submodel B.
}
\label{btdel002var}
\end{figure}

\begin{figure}
\includegraphics[width=1.0\columnwidth]{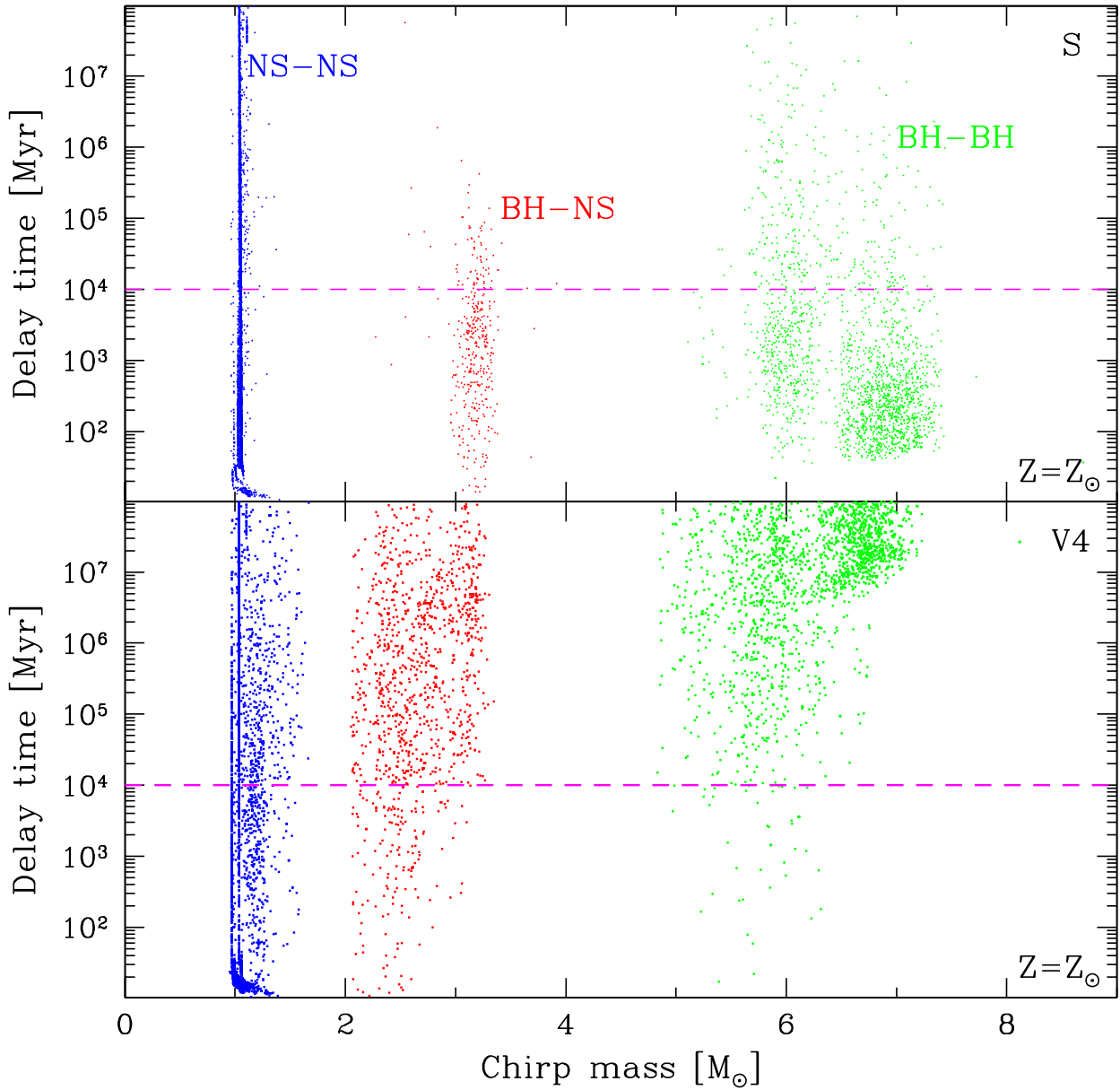}
\caption{
Chirp mass---delay time diagrams for merging DCOs for submodel A, $Z=\zsun$.
\textit{Top panel}. Results for the standard model. Significant populations
of DCOs containing BHs are present below the $10$ Gyr delay time limit (the
horizontal dashed magenta line).
\textit{Bottom panel}. Results for V4. Weakly bound envelopes,
represented by a very high $\lambda$ value, cause insignificant separation
reduction during the CE phase. This results in only a few systems crossing the $10$ Gyr
time limit and into the merging population.
Both of these panels illustrate the migration of DCOs between
the merging and non-merging populations with varying $\lambda$ values. The
clustering of chirp masses around higher values for merging DCOs with BHs in the 
standard model comes from the fact that more massive progenitors are more likely 
to survive a strongly bound CE phase due to a larger orbital energy reservoir. Note that 
these diagrams do not show the full range of the non-merging populations; systems 
with delay times much larger than $10^8$ Myr and masses larger than $8 \msun$ may 
also be formed. 
}
\label{mig}
\end{figure}

\begin{figure}
\includegraphics[width=1.0\columnwidth]{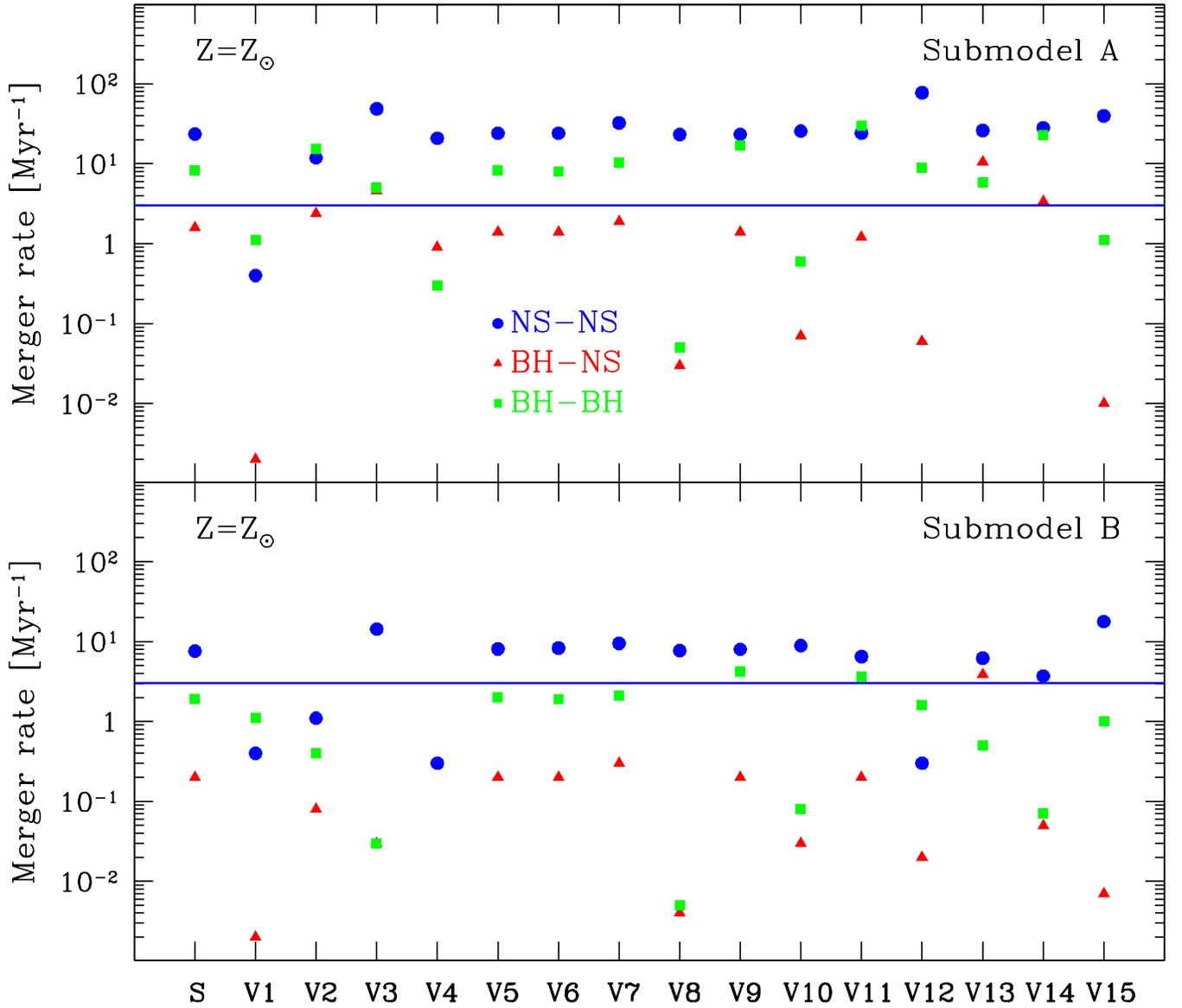}
\caption{
Galactic merger rates from all of our models, with submodel A in the top panel
and B in the bottom, for $Z=\zsun$. The blue solid line represents the lower limit
for predicted merger rates of NS-NS systems observed in our Galaxy (at $3$ \pmyr) 
as shown in \cite{kimkal}. Models yielding merger rates of NS-NS systems lower than this 
value are disfavored; these are V1-submodel A, V1-submodel B, 
V2-submodel B, V4-submodel B and V12-submodel B. \textit{Reminder:} the described models are: 
V1--V4, changing $\lambda$ from $0.01$ to $10$; V5--V6, changing $M_{\rm NS,max}$
from $3.0 \msun$ to $2.5 \msun$;
V7 -- reducing natal kicks for all DCOs, V8--V9, full and no natal kicks for BHs,
respectively; V10 -- investigating the Delayed SN engine; V11 -- reducing wind mass
loss rates by half; V12--V13, investigating fully conservative and non-conservative mass
transfer episodes, respectively; V15-V16, boosting and reducing the physical \textit{Nanjing} 
$\lambda$ value by a factor of $5$, respectively.  
}
\label{02rates}
\end{figure}

\begin{figure}
\includegraphics[width=1.0\columnwidth]{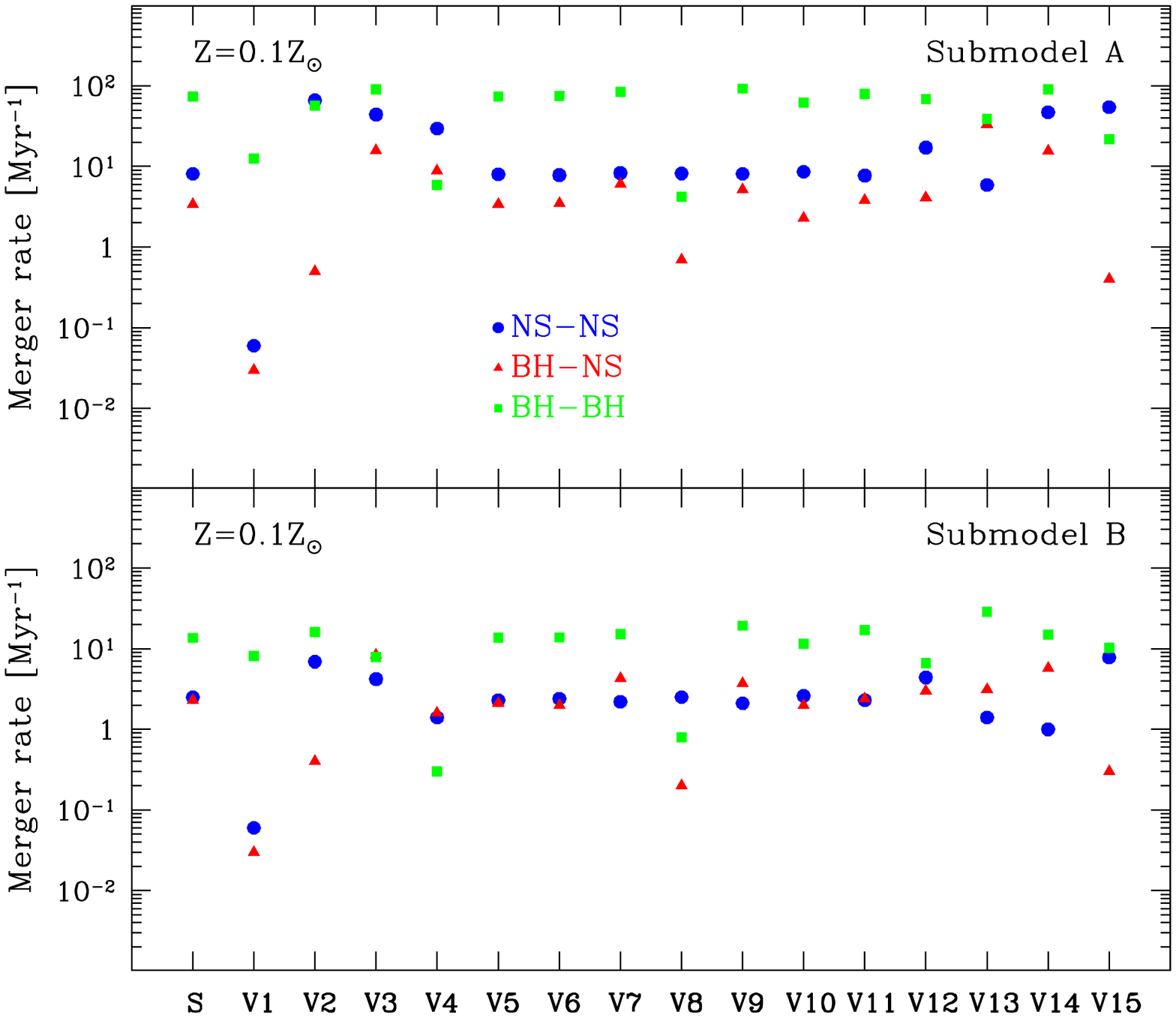}
\caption{Same as Fig.~\ref{02rates} but for $Z=0.1\zsun$.
}
\label{002rates}
\end{figure}

We can test (similarly to \cite{rich3}) the validity of certain models using the observed limits 
on the Galactic NS-NS merger rate~\citep{kimkal}. Their upper limit ($190$ Myr$^{-1}$) is not 
violated by any of our models. However, the lower limit ($3$ Myr$^{-1}$) is not met by several
of our models, and they would therefore be ruled out. 
Models (for$\zsun$) that do not meet this condition are: Variation 1-submodel A 
(NS-NS Galactic merger rate $0.4$ Myr$^{-1}$), Variation 1-submodel B ($0.4$ Myr$^{-1}$), 
Variation 2-submodel B ($1.1$ Myr$^{-1}$), Variation 4-submodel B ($0.3$ Myr$^{-1}$ and 
Variation 12-submodel B ($0.e$ Myr$^{-1}$);
this is illustrated in Figure~\ref{02rates}. The last variation utilizes the fully 
conservative mass transfer scenario. This points to a conclusion that a RLOF event, in which
no mass is lost form the system seems unlikely. All other employ a fixed
value of $\lambda$, and the observations of Galactic NS-NS systems
lends further support to our claim that such a treatment is unphysical. 
Additionally we can use the estimates on the predicted chirp mass of 
IC10 X--1 ($0.3 \zsun$ environment) to further validate our results. 
We test our chirp mass distribution for $0.1\zsun$ against the predicted minimal
predicted chirp mass for IC10 X--1 equal to $15 \msun$ \citep{bulikic10}. From
Fig.~\ref{bmch002var} and Table \ref{bmch002char} one can see that models V3--submodel B 
(max. chirp mass $14.9 \msun$), V4--submodel B (max. chirp mass $11.5 \msun$) and 
V14--submodel B (max. chirp mass $11.8 \msun$) do 
not reach this limit. All of these models use high $\lambda$ values (the former two 
a constant value). This is further evidence to support the claim that treatment of $\lambda$
as a constant value is unphysical. 

Another influential factor in DCO formation is the treatment of SN
explosions. Specifically, the magnitude of the natal kick plays a significant 
role, as the high kicks (assumed in our standard model) tend to disrupt DCO 
progenitors, reducing the merger rates. 
The rates for reduced (by half) kicks are given in model V7. 
The corresponding increase in the rates is rather modest, due to the fact that for
NS-NS systems some NSs explode as electron capture supernovae and do not receive kicks 
at all. BH kicks are already reduced by the amount of fall
back and further kick decrease (coming from reducing $\sigma$ in V7) does not significantly increase the
rates. However, the effect of kicks becomes significant for BH-BH systems once the
assumption on the reduction of kicks by fallback is relaxed. 
We have allowed for the very wide range of possible BH kicks in models V8
(high kicks) and V9 (no kicks). It is found that the Galactic merger rates of BH-BH
systems change by $\sim 2$--$3$ orders of magnitude (see Table \ref{gmerger02},
\ref{gmerger002}). Therefore, the mostly unknown magnitude of BH natal kick
limits our predictive power for the BH-BH rate estimates.

Despite the fact that we lack strong observational or theoretical constraints on BH 
natal kicks, in our standard model we have adopted the most likely model of
natal kicks (ones that decrease with increasing BH mass). This model is supported by existing 
observations (most massive BHs seem to form without a kick) and can be intuitively understood 
in the framework of natal kicks arising from asymmetric supernova mass ejection.  
The standard model BH-BH merger rates are close to upper limit set by the
full allowed range of possibilities (V8--V9), and again it appears that
our more realistic treatment of DCO formation favors higher rates, and sets an optimistic 
tone for near-future gravitational wave inspiral detection. 

We find that the fully conservative mass transfer scenario (Variation 12) for the pessimistic 
submodel B for $\zsun$ generates Galactic NS-NS merger rates below the
empirically predicted $3$ Myr$^{-1}$ for the Milky Way \citep{kimkal}.
This indicates that the assumption of fully conservative mass transfer is
most likely unphysical. 

This is the first in a series of three papers. Results presented here, and their extensions
in the {\tt Synthetic Universe} on-line database ({\tt www.syntheticuniverse.org}), will 
be used in the second study, where we will investigate the NS-NS/BH-NS/BH-BH merger rates 
as a function of cosmological redshift, star formation rate, and metallicity.
The third paper will focus on gravitational wave detection rates for upcoming
observatories (advanced LIGO/Virgo and Einstein Telescope). 

\acknowledgements

We would like to thank Francesca Valsecchi for providing detailed calculations of specific
mass transfer scenarios. 
We would also like to thank the N. Copernicus Astronomical Centre in Warsaw, Poland,
and the University Of Brownsville, Texas, USA,  
for their courtesy in allowing us to use their computational resources.
Work at LANL was done under the auspices of the National Nuclear Security 
Administration of the U.S. Department of Energy at Los Alamos National Laboratory 
under Contract No. DE-AC52-06NA25396. 
EB is supported by NSF Grant PHY-0900735 and by NSF CAREER Grant PHY-1055103.
ROS is currently supported by NSF award PHY-0970074, the Bradley Program Fellowship, and the UWM Research
Growth Initiative.
TB is supported by the Polish grants N N203 511238 and  DPN/N176/VIRGO/2009
KB is supported by the Polish grant N N203 302835. 
MD is supported by the Polish grant GR--4071.


\bibliographystyle{apj}

\clearpage

\begin{deluxetable}{c l l}
\tablewidth{330pt}
\tablecaption{Summary of Models\tablenotemark{a}}
\tablehead{Model & Parameter & Description}
\startdata
S  & Standard            & $\lambda=$\textit{Nanjing}, $M_{\rm NS,max}=2.5\msun$, $\sigma=265$\\
&&                         km s$^{-1}$ BH kicks: variable, SN: Rapid \\ 
&&                         half-cons mass transfer \\
&&\\
V1 & $\lambda=0.01$      & very low $\lambda$: fixed \\
&& \\
V2 & $\lambda=0.1$       & low $\lambda$: fixed \\
&& \\
V3 & $\lambda=1$         & high $\lambda$: fixed \\
&& \\
V4 & $\lambda=10$        & very high $\lambda$: fixed \\
&& \\
V5 & $M_{\rm NS,max}=3.0 \msun$    & high maximum NS mass \\
&& \\ 
V6 & $M_{\rm NS,max}=2.0 \msun$    & low maximum NS mass \\
&& \\
V7 & $\sigma=132.5$ km s$^{-1}$     & low kicks: NS/BH\\
&& \\
V8 & full BH kicks       & high natal kicks: BH \\
&& \\ 
V9 & no BH kicks         & no natal kicks: BH \\
&& \\
V10& Delayed SN          & NS/BH formation: Delayed SN engine \\
&& \\
V11& weak winds        & Wind mass loss rates reduced to $50\%$ \\
&& \\
V12& cons MT    & Fully  conservative mass transfer \\
&& \\
V13& non-cons MT & Fully non-conservative mass transfer \\
&& \\
V14& $\lambda \times 5$ & \textit{Nanjing} $\lambda$ increased by 5 \\
&& \\
V15& $\lambda \times 0.2$ & \textit{Nanjing} $\lambda$ decreased by 5 \\
\enddata
\label{pstudy}
\tablenotetext{a}{
All parameters, except for the one listed under ``Description'', retain their Standard
model (``S'') values.
}
\end{deluxetable}
\clearpage

\begin{deluxetable}{c c c c}
\tablewidth{340pt}
\tablecaption{Galactic Merger Rates,$\zsun$ [$\myr^{-1}$] \tablenotemark{a}}
\tablehead{Model    & NS-NS      & BH-NS         & BH-BH }
\startdata
S  & 23.5 (7.6)  & 1.6   (0.2)    & 8.2  (1.9) \\
V1 & 0.4  (0.4)  & 0.002 (0.002)  & 1.1  (1.1) \\
V2 & 11.8 (1.1)  & 2.4   (0.08)   & 15.3 (0.4) \\
V3 & 48.8 (14.3) & 4.6   (0.03)   & 5.0  (0.03)\\
V4 & 20.8 (0.3)  & 0.9   (0.0)    & 0.3  (0.0) \\
V5 & 24.1 (8.1)  & 1.4   (0.2)    & 8.3  (2.0) \\
V6 & 24.1 (8.3)  & 1.4   (0.2)    & 8.0  (1.9) \\
V7 & 32.4 (9.5)  & 1.9   (0.3)    & 10.4 (2.1) \\
V8 & 23.3 (7.7)  & 0.03  (0.004)  & 0.05 (0.005)\\
V9 & 23.4 (8.0)  & 1.4   (0.2)    & 16.9 (4.2) \\
V10& 25.6 (8.9)  & 0.07  (0.03)   & 0.6  (0.08)\\
V11& 24.2 (6.5)  & 1.2   (0.2)    & 29.7 (3.6)\\
V12& 77.4 (0.3)  & 0.06  (0.02)   & 8.9  (1.6)\\
V13& 26.1 (6.2)  & 10.6  (3.9)    & 5.8  (0.5)\\
V14& 28.2 (3.7)  & 3.4   (0.05)   & 23.0 (0.07)\\
V15& 39.8 (17.8) & 0.01  (0.007)  & 1.1  (1.0)\\
Range & 0.4--77.4 (0.3--17.8) & 0.002--10.6 (0.0--3.9) & 0.05--29.7 (0.0--4.2)\\
\enddata
\label{gmerger02}
\tablenotetext{a}{
Rates are calculated for a synthetic galaxy similar to the Milky Way (solar
metallicity, and $10$ Gyr of continuous star formation at the level of $3.5
\mpy$). Rates are presented for submodel A (CE HG donor allowed), with the rates
for submodel B listed in parentheses (CE HG donor forbidden); see Section \ref{hgtreatment} for
details. The range presents the minimum and maximum value for each DCO
type. 
}
\end{deluxetable}

\begin{deluxetable}{c c c c}
\tablewidth{340pt}
\tablecaption{Galactic Merger Rates, $0.1 \zsun$ [$\myr^{-1}$] \tablenotemark{a}}
\tablehead{    & NS-NS      & BH-NS         & BH-BH }
\startdata
S   & 8.1  (2.5)  & 3.4  (2.3)  & 73.3 (13.6) \\
V1  & 0.06 (0.06) & 0.03 (0.03) & 12.5 (8.1)  \\
V2  & 65.9 (6.9)  & 0.5  (0.4)  & 56.7 (16.1) \\
V3  & 44.1 (4.2)  & 15.8 (8.4)  & 90.2 (7.9)  \\
V4  & 29.5 (1.4)  & 8.8  (1.6)  & 5.9  (0.3)   \\
V5  & 8.0  (2.3)  & 3.4  (2.1)  & 73.4 (13.7) \\
V6  & 7.8  (2.4)  & 3.5  (2.0)  & 74.5 (13.8) \\
V7  & 8.3  (2.2)  & 6.1  (4.3)  & 83.7 (15.1) \\
V8  & 8.2  (2.5)  & 0.7  (0.2)  & 4.2  (0.8)  \\
V9  & 8.1  (2.1)  & 5.2  (3.7)  & 92.3 (19.3) \\
V10 & 8.6  (2.6)  & 2.3  (2.0)  & 62.0 (11.5) \\
V11 & 7.7  (2.3)  & 3.8  (2.4)  & 79.2 (17.1) \\
V12 & 17.1 (4.4)  & 4.1  (3.0)  & 68.8 (6.6)  \\
V13 & 5.9  (1.4)  & 33.0 (30.1) & 39.0 (28.9) \\
V14 & 47.0 (1.0)  & 15.5 (5.7)  & 90.5 (14.9)\\
V15 & 54.4 (7.8)  & 0.4  (0.3)  & 21.7 (10.2)\\
Range & 0.06--65.9 (0.06--7.8) & 0.03--33.0 (0.03--30.1) & 4.2--92.3 (0.3--28.9)\\
\enddata
\label{gmerger002}
\tablenotetext{a}{Same as Table~\ref{gmerger002} but for sub-solar
metallicity.} 
\end{deluxetable}

\begin{deluxetable}{c c l}
\tablewidth{410pt}
\tablecaption{Formation channels of DCOs for$\zsun$ \tablenotemark{a}}
\tablehead{    & Channel & Fraction}
\startdata 
NSNS01 & NC:a$\rightarrow$b, SN:a, CE:b$\rightarrow$a, NC:b$\rightarrow$a, SN:b & $79.3\%$ \\
NSNS02 & NC:a$\rightarrow$b, CE:b$\rightarrow$a, NC:b$\rightarrow$a,
AIC(WD$\rightarrow$NS):a, NC:b$\rightarrow$a, SN:b & $8.0\%$\\

NSNS03 & NC:a$\rightarrow$b, SN:a, CE:b$\rightarrow$a,
CE:b$\rightarrow$a\tablenotemark{b}, SN:b & $6.9\%$\\
NSNS04 & Other & $5.8\%$\\
&& \\
BHNS01 & NC:a$\rightarrow$b, SN:a, CE:b$\rightarrow$a, SN:b & $95.4\%$ \\
BHNS02 & NC:a$\rightarrow$b, SN:a, CE:b$\rightarrow$a, NC:b$\rightarrow$a, SN:b & $1.8\%$ \\
BHNS03 & Other & $2.8\%$\\
&& \\
BHBH01 & NC:a$\rightarrow$b, SN:a, CE:b$\rightarrow$a, SN:b & $98.9\%$ \\
BHBH02 & Other & $1.1\%$\\

\enddata
\label{channels}
\tablenotetext{a}{
Coalescing DCOs' formation channels for the standard model, submodel A at
solar metallicity. NC: non-conservative mass transfer, SN: supernova, 
CE: common envelope, AIC: accretion induced collapse of oxygen/neon white
dwarf into NS. The arrows show the direction of transfer and ``a" 
stands for the primary (initially more massive) component, ``b" for the 
secondary.} 
\tablenotetext{b}{The first CE is initiated by the H-rich 
Hertzsprung gap donor (allowed in model A). The second starts when the 
exposed core of the donor becomes an evolved helium star.
}  
\end{deluxetable}

\begin{deluxetable}{c c l}
\tablewidth{410pt}
\tablecaption{Formation channels of DCOs for $0.1\zsun$ \tablenotemark{a}}
\tablehead{    & Channel & Fraction}
\startdata
NSNS01 & NC:a$\rightarrow$b, SN:a, CE:b$\rightarrow$a, CE:b$\rightarrow$a, SN:b & $49.1\%$ \\
NSNS02 & NC:a$\rightarrow$b, SN:a, CE:b$\rightarrow$a, NC:b$\rightarrow$a, SN:b & $21.2\%$ \\
NSNS03 & NC:a$\rightarrow$b, SN:a, CE:b$\rightarrow$a, SN:b                     & $18.2\%$ \\
NSNS04 & NC:a$\rightarrow$b, CE:b$\rightarrow$a, SN:a, CE:b$\rightarrow$a, SN:b & $3.3\%$\\
NSNS05 & Other & 8.2\%\\
&& \\
BHNS01 & CE:a$\rightarrow$b, SN:a, CE:b$\rightarrow$a, NC:b$\rightarrow$a, SN:b & $40.8\%$ \\
BHNS02 & CE:a$\rightarrow$b, SN:a, CE:b$\rightarrow$a, SN:b                     & $17.4\%$ \\
BHNS03 & NC:a$\rightarrow$b, SN:a, CE:b$\rightarrow$a, SN:b                     & $13.4\%$\\
BHNS04 & NC:a$\rightarrow$b, SN:a, CE:b$\rightarrow$a, NC:b$\rightarrow$a, SN:b & $12.2\%$ \\
BHNS05 & NC:a$\rightarrow$b, CE:b$\rightarrow$a, NC:a$\rightarrow$b, SN:a, SN:b & $8.8\%$ \\
BHNS06 & Other & $6.4\%$ \\
&& \\
BHBH01 & NC:a$\rightarrow$b, SN:a, CE:b$\rightarrow$a, SN:b                     & $90.6\%$ \\
BHBH02 & CE:a$\rightarrow$b, SN:a, CE:b$\rightarrow$a, SN:b                     & $4.0\%$ \\
BHBH03 & NC:a$\rightarrow$b, SN:a, NC:b$\rightarrow$a, CE:b$\rightarrow$a, SN:b & $1.4\%$ \\
BHBH04 & Other & $4.0\%$\\
\enddata
\label{channels2}
\tablenotetext{a}{
Same as Table \ref{channels} but for sub-solar metallicity.
}
\end{deluxetable}
\clearpage

\begin{deluxetable}{c c c c c c c c c c}
\tablewidth{390pt}
\tablecaption{Chirp mass characteristics for$\zsun$, submodel A\tablenotemark{a}}
\tablehead{& \multicolumn{3}{c}{NS-NS}    & \multicolumn{3}{c}{BH-NS}   & \multicolumn{3}{c}{BH-BH} \\
             & min & avg & max  & min & avg & max & min & avg & max}
\startdata
S   &        0.96 & 1.05 & 1.40 & 2.3 & 3.2 & 3.7 & 5.1 & 6.7 & 8.7\\
V1  &        1.08 & 1.09 & 1.14 & 3.2 & 3.2 & 3.2 & 5.3 & 6.5 & 8.3\\ 
V2  &        0.96 & 1.08 & 1.53 & 2.5 & 3.2 & 4.0 & 5.0 & 6.6 & 8.4\\
V3  &        0.94 & 1.06 & 1.69 & 2.1 & 2.7 & 3.6 & 4.9 & 6.0 & 7.7\\
V4  &        0.95 & 1.03 & 1.64 & 2.1 & 2.5 & 3.1 & 5.0 & 5.8 & 6.3\\
V5  &        0.96 & 1.05 & 1.42 & 2.4 & 3.2 & 3.8 & 5.0 & 6.7 & 8.7\\
V6  &        0.96 & 1.05 & 1.44 & 2.2 & 3.2 & 3.9 & 3.5 & 6.7 & 8.8\\
V7  &        0.96 & 1.05 & 1.45 & 2.1 & 3.1 & 3.9 & 5.0 & 6.5 & 8.7\\
V8  &        0.96 & 1.05 & 1.42 & 2.6 & 3.0 & 3.2 & 5.4 & 6.5 & 7.4\\
V9  &        0.96 & 1.05 & 1.44 & 2.2 & 3.1 & 3.7 & 5.0 & 6.3 & 7.5\\
V10 &        1.01 & 1.14 & 1.86 & 2.0 & 3.1 & 4.2 & 2.7 & 5.7 & 7.6\\
V11 &        0.96 & 1.05 & 1.50 & 2.7 & 3.2 & 4.2 & 4.9 & 10.5 & 14.3\\
V12 &        0.96 & 1.07 & 1.44 & 2.4 & 2.9 & 3.6 & 5.0 & 6.3  & 8.6 \\
V13 &        0.94 & 1.02 & 1.63 & 2.1 & 2.7 & 4.0 & 4.9 & 6.1  & 8.6 \\
V14 &        0.96 & 1.07 & 1.70 & 2.1 & 2.9 & 3.8 & 4.9 & 6.4  & 8.3 \\
V15 &        0.95 & 1.07 & 1.40 & 3.1 & 3.2 & 3.2 & 5.5 & 6.5  & 8.2 \\
Range&       & 0.94--1.86 &   &   & 2.0--4.2 & &   & 2.7--14.3 &\\
\enddata
\label{amch02char}
\tablenotetext{a}{
The chirp mass distribution for merging DCOs, for $\zsun$ and submodel A. 
The values of chirp mass presented are: minimum, average, and maximum
in units of $\msun$. The range represents the minimum--maximum value of the
chirp mass from the entire
suite of models for each DCO type. This table corresponds to Fig.~\ref{amch02var}.
}
\end{deluxetable}

\begin{deluxetable}{c c c c c c c c c c}
\tablewidth{390pt}
\tablecaption{Chirp mass characteristics for$\zsun$, submodel B\tablenotemark{a}}
\tablehead{& \multicolumn{3}{c}{NS-NS}   & \multicolumn{3}{c}{BH-NS} & \multicolumn{3}{c}{BH-BH}\\
     & min & avg & max & min & avg & max & min & avg & max }
\startdata

S  & 0.96 & 1.05 & 1.17 & 2.3 & 3.2 & 3.3 & 5.2 & 6.7 & 7.4 \\
V1 & 1.08 & 1.09 & 1.14 & 3.2 & 3.2 & 3.2 & 5.3 & 6.5 & 8.3 \\
V2 & 0.96 & 1.06 & 1.53 & 3.0 & 3.2 & 3.3 & 5.6 & 6.5 & 7.2 \\
V3 & 0.96 & 1.05 & 1.22 & 2.2 & 2.4 & 2.6 & 5.7 & 5.9 & 6.4 \\
V4 & 1.03 & 1.03 & 1.04 & &no systems&  & &no systems&     \\
V5 & 0.96 & 1.05 & 1.28 & 2.4 & 3.1 & 3.3 & 5.2 & 6.7 & 7.4 \\
V6 & 0.96 & 1.05 & 1.43 & 2.6 & 3.2 & 3.4 & 3.5 & 6.7 & 8.7 \\
V7 & 0.96 & 1.05 & 1.45 & 2.2 & 3.1 & 3.5 & 5.1 & 6.5 & 8.0 \\
V8 & 0.96 & 1.05 & 1.08 & 3.2 & 3.2 & 3.2 & 5.6 & 5.9 & 6.1 \\
V9 & 0.96 & 1.05 & 1.21 & 2.2 & 3.1 & 3.3 & 5.0 & 6.2 & 7.4 \\
V10& 1.04 & 1.13 & 1.34 & 2.0 & 3.0 & 4.2 & 2.7 & 4.6 & 6.6 \\
V11& 0.96 & 1.05 & 1.09 & 2.9 & 3.3 & 4.0 & 4.9 & 9.1 & 13.8 \\
V12& 0.96 & 1.04 & 1.17 & 2.4 & 2.9 & 3.4 & 5.0 & 6.3 & 7.4 \\
V13& 0.95 & 1.00 & 1.59 & 2.1 & 2.6 & 3.3 & 4.9 & 6.3 & 8.4 \\       
V14& 0.96 & 1.05 & 1.34 & 2.2 & 2.8 & 3.3 & 5.7 & 6.0 & 6.9 \\
V15& 0.95 & 1.05 & 1.13 & 3.1 & 3.2 & 3.2 & 5.5 & 6.5 & 8.2 \\
Range&  &0.95--1.59& &  & 2.0--4.2 &  &  & 2.7--13.8 &\\
\enddata
\label{bmch02char}
\tablenotetext{a}{
Same as Table \ref{amch02char} but for submodel B. This table corresponds to Fig.~\ref{bmch02var}.
}
\end{deluxetable}

\clearpage

\begin{deluxetable}{c c c c c c c c c c}
\tablewidth{390pt}
\tablecaption{Chirp mass characteristics for $0.1\zsun$, submodel A\tablenotemark{a}}
\tablehead{& \multicolumn{3}{c}{NS-NS}  & \multicolumn{3}{c}{BH-NS} & \multicolumn{3}{c}{BH-BH}\\
& min & avg & max & min & avg & max & min & avg & max }
\startdata
S  & 0.96 & 1.09 & 1.67 & 2.1 & 3.2 & 6.1 & 4.8 & 13.2& 31.8 \\ 
V1 & 1.08 & 1.11 & 1.56 & 2.9 & 3.6 & 4.4 & 5.9 & 20.0& 32.3 \\ 
V2 & 0.96 & 1.09 & 1.66 & 2.3 & 3.5 & 6.5 & 4.8 & 17.2& 31.6 \\ 
V3 & 0.96 & 1.09 & 1.68 & 2.0 & 2.9 & 6.1 & 4.8 & 12.5& 29.7 \\ 
V4 & 0.95 & 1.10 & 1.64 & 2.1 & 2.9 & 5.9 & 4.8 & 7.6 & 28.1 \\ 
V5 & 0.96 & 1.09 & 1.66 & 2.0 & 3.1 & 4.6 & 4.8 & 13.3& 31.8 \\ 
V6 & 0.97 & 1.09 & 1.65 & 1.7 & 3.1 & 5.2 & 4.8 & 13.1& 31.8 \\ 
V7 & 0.96 & 1.08 & 1.70 & 2.0 & 3.1 & 6.4 & 4.9 & 12.4& 32.0 \\ 
V8 & 0.96 & 1.09 & 1.64 & 2.0 & 3.0 & 6.4 & 4.8 & 9.0 & 31.9 \\ 
V9 & 0.97 & 1.08 & 1.66 & 2.0 & 3.0 & 6.1 & 4.9 & 12.1& 31.9 \\ 
V10& 1.10 & 1.20 & 2.16 & 1.8 & 3.4 & 6.3 & 2.4 & 14.4& 32.0 \\ 
V11& 0.97 & 1.08 & 1.61 & 2.0 & 3.3 & 4.7 & 4.8 & 14.3& 41.4 \\
V12& 0.96 & 1.06 & 1.64 & 2.1 & 3.1 & 6.4 & 4.9 & 14.4 & 34.9 \\
V13& 0.96 & 1.10 & 1.68 & 2.1 & 3.1 & 5.0 & 4.8 & 9.7 & 27.7 \\
V14& 0.96 & 1.10 & 1.70 & 2.0 & 2.9 & 5.4 & 4.8 & 14.6 & 31.8 \\
V15& 0.96 & 1.09 & 1.64 & 2.4 & 3.7 & 4.6 & 4.9 & 15.0 & 34.5 \\
Range& &0.95--2.16  &  &   &1.7--6.4 &  &   &2.4--41.4& \\
\enddata
\label{amch002char}
\tablenotetext{a}{
Same as Table \ref{amch02char} but for $0.1\zsun$. This table corresponds to Fig.~\ref{amch002var}.
}
\end{deluxetable}

\begin{deluxetable}{c c c c c c c c c c}
\tablewidth{390pt}
\tablecaption{Chirp mass characteristics for $0.1\zsun$, submodel B\tablenotemark{a}}
\tablehead{& \multicolumn{3}{c}{NS-NS} & \multicolumn{3}{c}{BH-NS} & \multicolumn{3}{c}{BH-BH}\\
 & min & avg & max & min & avg & max & min & avg & max }
\startdata
S&   0.96 & 1.09 & 1.67  & 2.2 & 3.1 & 4.6 & 4.8 & 9.7 & 18.3 \\ 
V1&  1.08 & 1.11 & 1.56  & 2.9 & 3.6 & 4.4 & 5.9 & 16.1 & 32.3 \\ 
V2&  0.96 & 1.07 & 1.66  & 2.3 & 3.3 & 4.6 & 4.8 & 9.3 & 18.8 \\ 
V3&  0.99 & 1.12 & 1.68  & 2.0 & 2.9 & 4.5 & 4.8 & 6.8 & 14.9 \\ 
V4&  0.98 & 1.20 & 1.60  & 2.1 & 2.7 & 3.6 & 5.0 & 6.7 & 11.5 \\ 
V5&  0.96 & 1.08 & 1.66  & 2.2 & 3.2 & 4.6 & 4.8 & 9.8 & 17.7 \\ 
V6&  0.98 & 1.09 & 1.65  & 1.7 & 3.2 & 4.6 & 4.8 & 9.7 & 18.3 \\ 
V7&  0.96 & 1.08 & 1.60  & 2.1 & 3.1 & 4.6 & 4.9 & 9.7 & 17.7 \\ 
V8&  0.96 & 1.09 & 1.64  & 2.2 & 3.0 & 4.4 & 4.8 & 7.2 & 16.3 \\ 
V9&  0.98 & 1.08 & 1.55  & 2.1 & 3.0 & 4.5 & 4.9 & 9.3 & 18.1 \\ 
V10& 1.10 & 1.20 & 2.13  & 2.0 & 3.4 & 5.0 & 2.4 & 10.2 & 17.5 \\ 
V11& 0.97 & 1.08 & 1.61  & 2.3 & 3.4 & 4.7 & 4.8 & 10.6 & 19.1 \\
V12& 0.96 & 1.18 & 1.65  & 2.1 & 3.2 & 4.6 & 4.9 & 10.1 & 17.5 \\
V13& 0.96 & 1.10 & 1.66  & 2.1 & 3.1 & 5.0 & 4.8 & 9.4 & 21.2 \\
V14& 1.00 & 1.26 & 1.68  & 2.1 & 2.8 & 4.8 & 4.9 & 6.7 & 11.8 \\
V15& 0.96 & 1.06 & 1.64  & 2.5 & 3.7 & 4.6 & 4.9 & 11.8 & 34.5 \\
Range & &0.96--2.13&   &  &1.7--5.0&   &  & 2.4--34.5& \\
\enddata
\label{bmch002char}
\tablenotetext{a}{
Same as Table \ref{amch002char} but for submodel A. This table corresponds to Fig.~\ref{bmch002var}.
}
\end{deluxetable}

\clearpage

\clearpage

\clearpage

\clearpage

\clearpage

\clearpage

\clearpage

\clearpage

\clearpage

\clearpage

\clearpage

\clearpage

\clearpage

\clearpage

\end{document}